\documentclass[longauth]{aa}

\usepackage{graphicx}
\usepackage{txfonts}
\usepackage[svgnames]{xcolor}
\usepackage[colorlinks=true, linkcolor = blue, urlcolor = magenta, hyperfootnotes=false]{hyperref}
\usepackage{multirow}
\usepackage{footmisc}
\newcommand{\astfootnote}[1]{%
\let\oldthefootnote=\thefootnote%
\setcounter{footnote}{0}%
\renewcommand{\thefootnote}{\fnsymbol{footnote}}%
\footnote{#1}
\let\thefootnote=\oldthefootnote%
}
\usepackage{xcolor}
\usepackage{makecell}

\usepackage{tablefootnote}

\hypersetup{
    colorlinks,
    citecolor=blue
}

\usepackage{natbib}
\bibpunct{(}{)}{;}{a}{}{,}
\usepackage{subfigure}
\usepackage{wasysym}
\usepackage{siunitx} 
\usepackage[normalem]{ulem} 

\usepackage{amssymb}
\usepackage{pifont}

\usepackage{placeins}
\usepackage{caption}

\newcommand{\todo}[1]{\textcolor{cyan}{#1}}

\newcommand{\Lfour}{$L_{\mathrm{4}}$}
\newcommand{\Lfive}{$L_{\mathrm{5}}$}

\newcommand{\yes}{[{\color{Green}$\checkmark$}]}
\newcommand{\no}{[{\color{red}$\times$}]}

\begin{document} 

    \title{Two inner dust clumps in PDS~70}
    \subtitle{A third protoplanet traced by trojan material or a substructured inner disk?}
           \author{
           O.~Balsalobre-Ruza\inst{\ref{cab}} 
           ~V.~Christiaens\inst{\ref{leuven}, \ref{liege}, \ref{ups}},~N.~Huélamo\inst{\ref{cab}},~
I.~Hammond\inst{\ref{mpia}},~M.~Benisty\inst{\ref{mpia}},~S.~Lacour\inst{\ref{lira}, \ref{eso_garching}},~D.~Blakely\inst{\ref{ca1}, \ref{ca2}},
           R.G.~van Holstein\inst{\ref{eso_chile}}, 
           J.~Latour\inst{\ref{liege}}, 
           J.~Lillo-Box\inst{\ref{cab}}, 
           Z.~Wahhaj\inst{\ref{eso_chile}},
           D.~Trevascus\inst{\ref{mpia}},
           O.~Absil\inst{\ref{liege}}, 
           J.~Bae\inst{\ref{flusa}}, 
           C.~Charalambous\inst{\ref{catchile}},
           I.~de Gregorio-Monsalvo\inst{\ref{eso_chile}},
           A.~Leleu\inst{\ref{geneve}},
           I.~Mendigutía\inst{\ref{cab}}, 
           C.~Petrovich\inst{\ref{inusa}}, 
           Á.~Ribas\inst{\ref{uk1},\ref{uk2}}
           and
           S.~Juillard\inst{\ref{arizona}}
           }

    \titlerunning{Inner dust clumps in PDS~70: A third protoplanet traced by trojan material or a substructured inner disk?}
    \authorrunning{O. Balsalobre-Ruza et al.}

    \institute{
    Centro de Astrobiolog\'ia (CAB), CSIC-INTA, ESAC campus, Camino Bajo del Castillo s/n, 28692, Villanueva de la Ca\~nada (Madrid), Spain \email{o.balsaruza@gmail.com} \label{cab} 
    \and
    Institute of Astronomy, KU Leuven, Celestijnenlaan 200D, 3001 Leuven, Belgium \label{leuven} 
    \and
    STAR Institute, Université de Liège, Allée du Six Août 19c, 4000 Liège, Belgium\label{liege}
    \and
    Université Paris-Saclay, Université Paris Cité, CEA, CNRS, AIM, F-91191 Gif-sur-Yvette, France\label{ups}
    \and
    Max-Planck-Institut für Astronomie, Königstuhl 17, 69117 Heidelberg, Germany\label{mpia}
    \and 
    LIRA, Observatoire de Paris, Université PSL, Sorbonne Université, Université Paris Cité, CY Cergy Paris Université, CNRS, 92190 Meudon, France\label{lira}
    \and
    European Southern Observatory, Karl-Schwarzschild-Straße 2, 85748 Garching, Germany \label{eso_garching}
    \and
    Department of Physics and Astronomy, University of Victoria, 3800 Finnerty Road, Elliott Building, Victoria, BC, V8P 5C2, Canada\label{ca1}
    \and
    NRC Herzberg Astronomy and Astrophysics, 5071 West Saanich Road, Victoria, BC, V9E 2E7, Canada\label{ca2}
    \and
    European Southern Observatory, Alonso de Córdova 3107, Casilla 19001, Vitacura, Santiago, Chile
    \label{eso_chile}
    \and
    Department of Astronomy, University of Florida, Gainesville, FL 32611, USA\label{flusa}
    \and
    Instituto de Astrofísica, Pontificia Universidad Católica de Chile, Av. Vicuña Mackenna 4860, 782-0436 Macul, Santiago, Chile\label{catchile}
    \and
    Observatoire Astronomique de l’Universit\'{e} de Gen\`{e}ve, Chemin Pegasi 51b, 1290 Versoix, Switzerland\label{geneve}
    \and
    Department of Astronomy, Indiana University, Bloomington, IN 47405, USA\label{inusa}
    \and
    Astronomy Unit, Department of Physics and Astronomy, Queen Mary University of London, Mile End Road, London E1 4NS, UK\label{uk1}
    \and
    Institute of Astronomy, University of Cambridge, Madingley Road, Cambridge, CB3 0HA, UK\label{uk2}
    \and
    Steward Observatory, University of Arizona, Tucson, AZ 85721, USA\label{arizona}
    }
   \date{Received 1 May 2026 / Accepted 4 August 2026}

  \abstract
   {The wide cavity in the PDS~70 protoplanetary disk harbors two directly imaged confirmed protoplanets. 
   Several studies have proposed the existence of a third inner planet candidate at $\sim$13\,au, labeled PDS~70\,d.
   While its motion is consistent with a Keplerian orbit, its unusually blue spectrum challenges a planetary interpretation.
   }
   {
   We further investigate the presence and nature of a third inner planet using new SPHERE and GRAVITY observations.
   }
   {Using the star-hopping strategy, we obtained new coronagraphic SPHERE/IRDIS observations of PDS~70 in dual-polarization imaging mode in the $H$-band, as well as non-coronagraphic observations with SPHERE/IRDIFS covering the $YJHK$-bands. 
   We also searched for a planetary-like signal using GRAVITY+ in astrometric mode with the four-unit-telescope configuration.
   }
   {We consistently detect two inner emission features with SPHERE: one corresponding to the previously proposed planet candidate at $\sim$13\,au,
   and another that appears to share the same orbit while leading it by $\sim$120$^\circ$ (clockwise direction) with a more elongated morphology.
   Both features exhibit dust-scattered light spectra while displaying different colors, which may indicate differences in dust grain sizes.
   We show that such configuration is consistent with co-orbital dust accumulated at the stable Lagrangian regions of a yet undetected planet, distinct from the previously proposed candidate. 
   The analysis of GRAVITY data at the predicted position of this newly proposed planet has resulted in a marginal (3$\sigma$) detection located between the two clumps along the same orbit ($\rho$\,=\,76.2\,$\pm$\,0.29\,mas, PA\,=\,226.50\,$\pm$\,0.21$^\circ$), and consistent with a $\sim$3\,M$_{\rm Jup}$  planet.
   This planet-like signal is aligned with a narrow shadow detected in the outer disk with SPHERE. 
   Additionally, we detect with SPHERE some polarized emission very close to the star likely arising from the inner disk, from which we roughly estimate an inclination of 50$^\circ$ and a PA of $\sim$135$^\circ$. The fact that the two dust clumps appear embedded within this polarized emission motivates an inner disk-related alternative scenario.
   }
   {We conclude that the emission from the previously reported third planet candidate could arise either from a dust clump trailing a yet undetected planet along the same orbit, or from a rotating substructure within the inner disk.
   Additional observations are required to further test these two proposed scenarios.
   In particular, confirming the new GRAVITY planet-like signal would support the use of co-orbital substructures as indirect tracers of embedded planets.}

   \titlerunning{\texttt{}}
   \maketitle
    \nolinenumbers 
%

\section{Introduction} 
\label{sec:intro}

During the past few decades, high-angular resolution imaging  has resolved above a hundred protoplanetary disks, revealing complex substructures in over half of them at sub-millimeter (sub-mm) and Near InfraRed (NIR) wavelengths (e.g., \citealt{andrews18, 
sierra2021, benisty2023, vioque2025}). 
The remaining apparently unstructured disks are likely too compact or faint to reveal further details (e.g., \citealt{jennings2022}).
Consequently, substructures are considered a common property of protoplanetary disks.
Hydrodynamical simulations show that diverse mechanisms can reproduce those disk features, either invoking planetary-mass companions (e.g., \citealt{dong2018}) or other physical phenomena (e.g., turbulence, magnetized winds or the influence of their local environment; \citealt{bethune2017, lesur2023, garufi2026}).

Despite the prevalence of substructures, only a few embedded planets have been robustly confirmed so far (e.g., PDS~70\,b\,and\,c; \citealt{keppler2018,haffert2019}; WISPIT-2\,b\,and\,c; \citealt{vanapelleveen2025,close2025,lawlor2026}) leaving planet-disk interactions poorly constrained observationally.
Their detection is likely hampered by strong extinction from the surrounding material (e.g., \citealt{sanchis2020,quiroz2022,cugno2025}) and/or the low mass of the planet \citep[e.g.][]{ATorres2021}. 
This has motivated indirect approaches to infer the position of planet candidates, like the search of velocity kinks (see e.g., \citealt{perez2015, pinte2020, izquierdo2026}).

An alternative tracer with the potential to efficiently localize embedded planets is the disk material accumulated within their orbits. 
Different models (e.g., \citealt{zhang2018, montesinos2020, garrido2022}) predict that the gas and dust get trapped in 1:1 resonance, arranging co-orbitals either as horseshoe structures or as trojans.
The latter correspond to concentrated clumps at one or both of the stable Lagrangian points of the star-planet system (\Lfour\ and \Lfive), leading and trailing the planet by $\pm60^\circ$.
A couple of dust clumps associated with undetected planets have been proposed based on sub-mm continuum observations, one in LkCa~15 (\citealt{long2022}) and the other in HD~163296 (\citealt{isella2018, rodenkirch2021, garrido2023}).
The TWA~7 debris disk provides the most compelling observational support for this indirect strategy to date.
The \textit{James Webb} Space Telescope (JWST) recently
detected in this system an emission compatible with the lowest-mass planet (0.3\,$M_{\rm Jup}$) ever directly imaged (\citealp{lagrange2025, crotts2025}).
This planet lies within the under-density of a horseshoe-like structure previously detected in polarimetric light (\citealt{ren23}) with the Very Large Telescope (VLT)/Spectro-Polarimetric High contrast imager for Exo-planets REsearch (SPHERE; \citealt{beuzit19}), likely co-orbiting with the planet (\citealt{lacquement2026}).

PDS~70 is one of the most studied protoplanetary disks.
It is a K7 T-Tauri star (5.4~Myr) at a distance of d\,=\,112.39\,$\pm$\,0.24\,pc, belonging to the Sco-Cen star-forming region (\citealt{gregorio1992, riaud2006}) 
The disk consists of (from the inside out) an inner component, a wide gap, and a bright outer ring.
The inner disk was first inferred from the Spectral Energy Distribution (SED; \citealt{hashimoto2012, dong2012}), later from 
polarized light (\citealt{keppler2018}), and then resolved in the sub-mm (\citealt{keppler2019}) with a substructured and potentially variable outer edge (\citealt{casassus2022, fasano2025}; 11-16\,au radius).
The cavity, with a size of $\sim$50\,au (\citealt{keppler2018}), is strongly depleted containing about 1\% of the dust density of the outer ring (\citealt{wahhaj2024}).
The outer disk presents a gas density peak located at $\sim$75\,au (\citealt{portillarevelo2023}).
It also shows substructures in the NIR, including an outer arm in the northeast (\citealt{juillard2022, christiaens2024}).
The origin of this feature remains uncertain, with proposed interpretations including a vortex (\citealt{juillard2022}) or the result of geometrical effects related to disk flaring (\citealt{wahhaj2024}).

Two giant planets have been confirmed within the PDS~70 cavity, PDS~70\,b (20\,au) and c (34\,au).
They have been detected in several epochs at the NIR, as well as in H$\alpha$ emission, indicating ongoing accretion (\citealt{keppler2018, wagner2018, christiaens2019, haffert2019, mesa2019}). 
In the sub-mm, a variable circumplanetary disk has been detected around planet c (\citealt{isella2019, benisty2021, casassus2022}), likely including contribution from free-free emission (\citealt{dominguezjamet2025}).
The sub-mm emission associated to planet b appears offset and elongated toward the trailing part of its orbit (\citealt{isella2018}) and may be accompanied by a clump of material at its \Lfive\ region (\citealt{balsalobre2023}).

PDS~70\,b and c are thought to be near a 2:1 mean-motion resonance (\citealt{bae2019, wang2021}), which could extend to a 4:2:1 resonant chain given the proposed third inner planet at $\sim$13\,au (\citealt{mesa2019, christiaens2024, hammond2025}, which was labeled PDS~70\,d).
This candidate is supported by a decade of NIR observations showing an apparent Keplerian velocity, though its unusually blue spectrum differs markedly from that of the outer planets, raising doubts about its nature.
The existence of a third planet would have major implications for characterizing the system, as demonstrated by  GRAVITY dynamical mass estimates (\citealt{gravity17}).
The mass-upper limits of the confirmed planets would change significantly, for example PDS~70\,c could be $\sim$45\% less massive (\citealt{trevascus2025}). 
If confirmed, PDS~70 would represent a younger dynamical analogue of the HR~8799 system, which exhibits an 8:4:2:1 resonant chain (\citealt{esposito2013, zurlo2022}).

Recent SPHERE observations of PDS~70 in the $YJHK_s$-bands have improved contrast at very close separations (0.1$\arcsec$; \citealt{wahhaj2024}), where the third planet candidate is located. This is possible by using the star-hopping strategy (\citealt{wahhaj2021}), which consists on interleaving observations of the target with a spectro-photometrically similar and close star that is used as a reference
to perform Reference Differential Imaging (RDI; e.g., \citealt{2009ApJ...694L.148L,2019AJ....157..118R}).
This strategy also enables the study of the outer disk emission  avoiding the self-subtraction that affects traditional Angular Differential Imaging (ADI; e.g., \citealt{marois2006, milli2012}) techniques.
Motivated by these results, we aim to further investigate the presence and nature of the third planet candidate using new and deeper star-hopping observations.

In this article, we present new SPHERE 
observations in which we re-detect the emission from the previously proposed protoplanet candidate.
However, based on additional evidence, we propose that it may instead be either a dust clump trailing a yet undetected co-orbital planet (near \Lfive), or a substructure within the inner disk.
To test the first scenario, we conducted a dedicated search for such undetected planet with GRAVITY. 
In Sect.~\ref{sec:obs} we present the observations and data reduction.
Section~\ref{sec:res} describes the main results from all the datasets, while Sect.~\ref{sec:dis} is a discussion of the proposed hypotheses and their caveats.
We summarize our conclusions in Sect.~\ref{sec:concl}.

\section{Observations and data reduction}
\label{sec:obs}

We present new NIR high-contrast and high-angular resolution observations of the PDS~70 system obtained with the SPHERE and GRAVITY instruments. 
The SPHERE data were collected over two consecutive nights in April 2024,
and during three nights in mid-2025 as part of a Director discretionary Time (DDT).
The GRAVITY observations were conducted under the same DDT 
in May 2025.
For completeness, we analyzed three additional SPHERE archival datasets from \cite{wahhaj2024} obtained in July and August 2021 and February 2022 with a similar instrumental setup, which  were not originally designed to study the very close separation emissions that are the focus of our study.
We note that additional star-hopping epochs from \cite{wahhaj2024} were ignored due to poorer observational conditions.
All the studied datasets are detailed below, and summarized in Tables~\ref{tab:obs_sph} and \ref{tab:obs_grav}.

\subsection{Star-hopping IRDIS (H-band) and IRDIFS\_EXT ($Y$--$K$-band) observations}

The SPHERE observations from July 2021 (archival) and April 2024 (new) were performed with the InfraRed Dual-Band Imager and Spectrograph (IRDIS; \citealt{2008SPIE.7014E..3LD}) in Dual-Polarization Imaging mode (DPI; \citealt{deboer2020}) in the $H$-band ($\lambda=1625$~nm, $\Delta \lambda=290$~nm). 
During August 2021, February 2022 (archival), and May-July 2025 (new), PDS~70 was observed using the Integral Field Spectrograph (IFS; \citealt{2008SPIE.7014E..3EC}) in $YJH$-bands (903-1915\,nm) simultaneously with IRDIS (i.e., IRDIFS\_EXT mode). 
The 2025 dataset used IRDIS in the Dual-Band Imaging mode (DBI) with the $K_1$ ($\lambda=2110$~nm, $\Delta \lambda=102$~nm) and $K_2$ ($\lambda=2251$~nm, $\Delta \lambda=109$~nm) filters. 
We note that for the IRDIFS\_EXT archival datasets, we only analyzed the IFS-arm data with the goal of studying the astrometry of the detected features (Sect.~\ref{sec:astr_spec}), 
so that the corresponding IRDIS-arm data are not considered as they do not provide additional constraints to our analysis.

The three archival datasets were obtained with the coronagraphic mask N\_ALC\_YJH\_S with an Inner Working Angle (IWA) of 0.15$\arcsec$. 
For the 2024 observations we used the mask N\_ALC\_YJ\_S with a smaller IWA (0.08$\arcsec$). 
The 2025 observations were conducted without a coronagraph, allowing us to probe emissions down to $\sim$0.05$\arcsec$ from the star.

All observations were taken in pupil-stabilized mode, exploiting the star-hopping technique by using UCAC2~14412811 (2021, 2022, and 2024) and UCAC2~14413562 (2025) as reference stars to perform RDI. 
The former star was chosen in \citet{wahhaj2024} based on its similar brightness and close separation to PDS~70. 
We chose the same reference star in our 2024 run for consistency with previous observations.
However, the Point-Spread-Function (PSF) in Aug 2021 and 2024 was affected by the Wind-Driven Halo (WDH) pattern (\citealt{2020A&A...638A..98C}), which complicates PSF subtraction and can lead to over-subtraction when the WDH orientation differs between the science and reference frames. 
Since the WDH pattern rotates with the parallactic angle, we selected UCAC2~14413562 as the reference star for the 2025 run to reduce the over-subtraction. 
Its right ascension is 5\,min ahead of PDS~70, compensating for the $\sim$10\,min delay between the science and reference sequences and ensuring a closer match of the WDH orientation (see Table~\ref{tab:stars}).

The observational details of all the runs are summarized in Table~\ref{tab:obs_sph}, including exposure times, field rotation, coherence times, and the estimated Strehl ratios. 
We note that the AO loop opened on the April 6 (2024) run, 
forcing us to discard two affected polarimetric cycles (8 frames).
Due to unfavorable conditions during the 2025 run, the observations required three attempts, with the first night (May 28, 2025) aborted and ruled out for our analysis.

\subsection{IRDIS data reduction}

\subsubsection{Total-intensity reduction}
\label{sec:irdis_ti_red}

The IRDIS DPI datasets were processed using the \texttt{star-hopping}\footnote{\url{https://github.com/zwahhaj/starhopping}} \texttt{IDL} code (\citealt{wahhaj2021}) and the Vortex Image Processing (\texttt{VIP})\footnote{ \url{https://github.com/vortex-exoplanet/VIP}} \texttt{Python} package (\citealt{vip1, vip2}, v2.0.0).
The \texttt{IDL} code performed the basic pre-processing reduction steps, including flat-fielding, bad pixels correction, and an initial frame alignment.
Subsequently, \texttt{VIP} was used for Principal Component Analysis (PCA)-based sky subtraction (\citealt{skysubpca, ren23}) and frame centering carried out in two steps. 
First, all frames were aligned using the satellite spot method for coronagraphic observations. 
Then, intra-night drift was corrected by maximizing the cross-correlation of the speckle pattern throughout the image sequence.

In the case of the IRDIS DBI cubes (those obtained simultaneously with the IFS data), they were pre-processed using the \texttt{VCAL-SPHERE}\footnote{\url{https://github.com/VChristiaens/vcal_sphere}} \texttt{Python} pipeline (\citealt{vcal}), which is based on \texttt{esorex} routines and \texttt{VIP}. 
This pipeline perfo\textit{rms} dark subtraction, flat-fielding, PCA-based sky subtraction, bad pixel correction, relative alignment via cross-correlation of the speckle pattern, and absolute centering via 2D Gaussian fits of the PSF as these are non-coronagraphic observations.
Finally, it also includes anamorphism
correction and automatically removes bad frames.

For the post-processing of both the IRDIS DPI and DBI data, we performed an Iterative Principal Component Analysis with a combined Angular and Reference star Differential Imaging strategy (IPCA-ARDI) using \texttt{VIP}, which is particularly appropriate in presence of extended circumstellar signals \citep{juillard2024}.
The main hyper-parameters associated with this algorithm are the number of principal components ($n_{\rm pc}$) and the total number of iterations ($n_{\rm it}$).
An incremental IPCA mode can also be implemented, in which $n_{\rm pc}$ increases progressively as the algorithm iterates. This mode introduces a third parameter, being the number of iterations performed at each value of $n_{\rm pc}$ before incrementing it in the next iteration.

In all our reductions we considered $n_{\rm it} > 100$ iterations, which was sufficient to ensure convergence. The incremental mode was used for datasets affected by WDH, namely the 2021 and 2024 datasets.
The final images for each epoch were obtained as the median  of IPCA-ARDI reductions obtained with different values of $n_{\rm pc}$.
This last step is motivated by the fact that authentic circumstellar signals should be consistently recovered across a range of $n_{\rm pc}$ values, while the residual noise structure is expected to change for different $n_{\rm pc}$ values.
We considered the following $n_{\rm pc}$ ranges: 1\,-\,9, 2\,-\,15, 8\,-\,15, 7\,-\,9, 12\,-\,20, for the 2021 Jul 15, 2024 Apr 6, 2024 Apr 7, 2025 Jul 18, and 2025 Jul 22 datasets, respectively.

For datasets with a field rotation $\leq 50 \deg$, we initialized IPCA-ARDI using PCA-RDI with data imputation (\citealt{ren23}, via \texttt{VIP}) to obtain an initial estimate of the circumstellar flux and improve convergence.
As opposed to regular PCA-RDI, including data imputation computes the projection coefficients of the principal components modeling the stellar PSF using the region of the image containing only speckle noise (referred to as the anchor mask). 
We defined the anchor mask as a combination of 
areas of the disk cavity avoiding known signal locations (i.e., planets b, c, and the inner candidate) and their vicinities.

Finally, and to compare different algorithms, we also post-processed the IRDIS DPI and DBI data using a double-Locally Optimized Combination of Images (LOCI; \citealt{2007ApJ...660..770L}) approach.
The procedure is described in Appendix~\ref{sec:loci_reduction}.

\begin{figure*}[]
\centering\includegraphics[width=0.9\textwidth{}]{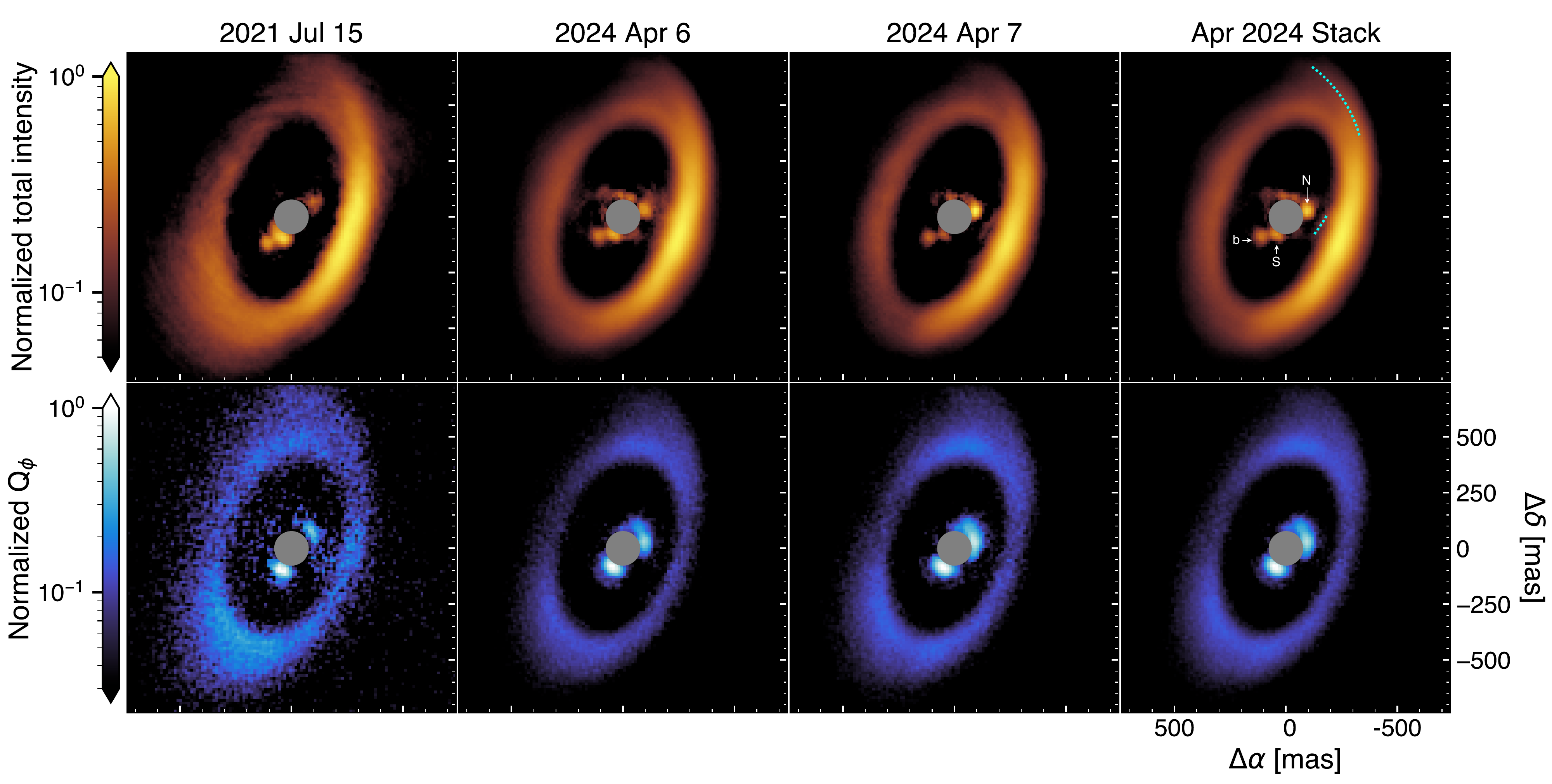}
\caption{Gallery of $H$-band IRDIS DPI images of PDS~70. 
Total-intensity images are shown in the top, 
and polarimetric Q$_{\phi}$ images in the bottom row. 
The three detected emissions are labeled in the 2024 stacked image, where dotted cyan lines indicate the two outer disk substructures reported in the literature. 
In all panels, North is up and East is to the left.
Coronagraphic mask as grey region.}
\label{fig:irdis_dpi}
\end{figure*}

\subsubsection{Polarimetric reduction}
\label{sec:red_pol}

We used the
\texttt{Python} package \texttt{IRDAP}\footnote{\url{https://github.com/robvanholstein/IRDAP}}
(\citealt{irdap}, v1.3.5) to perform Polarimetric Differential Imaging (PDI). 
The \texttt{IRDAP} pipeline pre-processes the data (sky subtraction, flat fielding, bad-pixel correction, and centering) and computes the Stokes parameters from the double difference. Subsequently, the pipeline applies a fully validated Mueller matrix model to correct the data for instrumental polarization and polarization crosstalk of the telescope and the instrument. These steps yield the Stokes $Q$- and $U$-images. 
Next, \texttt{IRDAP} uses a user-defined annulus in the images to measure and subtract the remaining polarization in the stellar halo.
This residual polarization halo can originate from interstellar dust and/or the spatially unresolved inner parts of the circumstellar disk. 
We subtracted the residual polarization halo using an annulus located exterior to the circumstellar disk. 
Finally, \texttt{IRDAP} computes the $Q_\phi$- and $U_\phi$-images following the definitions in \cite{deboer2020}.

After inspecting the resulting $Q$-, $U$-, $Q_\phi$-, and $U_\phi$-images, we found that they still contain a faint residual polarization halo.
To remove it completely, we used the grid-search approach described in \cite{garufi2024}.
In this method, we take the final images from \texttt{IRDAP} as a starting point and create a series of $Q$- and $U$-images where we subtract a range of slightly fainter and stronger halo polarized signals. 
We then compute a grid of $Q_\phi$- and $U_\phi$-images from all possible combination of the previously produced $Q$- and $U$-images, and  we visually inspect them to find the image with the faintest polarization halo emission. 
In case of the $Q$- and $U$-images, any dominating positive or negative signal at the very center of the image must originate from a residual polarization halo.
We therefore selected the images where 
the contribution from the positive and negative signals is equal in the central regions, so that the butterfly pattern of the disk is the sharpest. 
In case of the $Q_\phi$- and $U_\phi$-images, we selected the image with the least polarization signal between the outer and inner parts of the disk since any positive or negative signal there must originate from the polarization halo.

\subsection{IFS data reduction}
\label{sec:ifs_red}

The IFS cubes were calibrated and pre-processed with \texttt{VCAL-SPHERE} following the approach presented in \citet{hammond2025}. The pipeline computed master darks, flats, spectra positions and wavelength calibration with the associated \texttt{esorex} recipes using the day time calibrations proceeding each observation, with the additional step of subtracting the dark from each type of flat. The resulting cubes were sky-subtracted using the sky frame taken closest in time to the science (likewise for the reference star) and any bad pixels identified from the flats by \texttt{esorex} were corrected using the median value of the neighboring 5$\times$5 pixels. After spectral cube building with \texttt{esorex}, we perform a second round of bad pixel correction using a rolling subframe to identify and replace any pixels deviating 6$\sigma$ from the median. We centered the coronagraphic observations using the intersection of the four satellite spots located at 14$\lambda/D$ from the star, while non-coronagraphic data were aligned via cross-correlation of the speckle pattern using discrete Fourier transform upsampling, followed by shifting of frames to the peak of a 2D Gaussian fit to the PSF. Finally, we identified and trimmed bad frames by correlation to the median frame of the entire sequence. In contrast to IRDIS, we opted not to correct the anamorphism since the distortion is much smaller than for IRDIS, and avoids an additional sub-pixel interpolation of pixel intensities.

Using \texttt{VIP}, we applied PCA for PSF subtraction combining RDI and Spectral Differential Imaging (SDI; \citealt{sdi1,sdi2}), hereafter referred to as PCA-RSDI.
We also tested PCA-ARSDI, additionally incorporating ADI, which resulted in images compatible with the former method with lower residual noise but at the expense of introducing geometric artifacts associated with SDI and ADI self-subtraction (see e.g., \citealt{christiaens2019b}).
The scaling factor per spectral channel for SDI was defined as the wavelength ratio with respect to that from the last channel.
We used $n_{pc}$ ranging 20\,-\,500 for 2025 Jul 18, and 30\,-\,500 for 2025 Jul 22. 
We obtained the final images by median-stacking the resulting frames over the channels comprising each band (1\,-\,10, 11\,-\,23, 28\,-\,39, for the $YJH$-bands, respectively) and across the different $n_{pc}$ values.

\subsection{GRAVITY observations and reduction}
\label{sec:grav_obs}
PDS~70 was observed on May 14, 2025, with GRAVITY in the $K$-band using medium spectral resolution (R$\sim$500) and the four 8m-unit telescopes (UT) of the VLTI. 
These observations were obtained with the upgraded adaptive optics system (GRAVITY+; \citealt{gravity+2026}).
The aim was to search for a new inner planet suggested by our 2024 total-intensity IRDIS results (see Sect.~\ref{sec:astr_spec}), different to that previously proposed.

We used the dual-field on-axis mode, simultaneously observing the central star as fringe-tracking reference and the location of the predicted planet. 
The science fiber was alternately centered on two exploratory positions based on our predictions for the putative planet with the corresponding relative coordinates listed in Table~\ref{tab:obs_grav} (see details in Appendix~\ref{sec:grav_point}) and also on the central star for phase referencing.
In total, we integrated for 45\,min at each exploratory location.

All GRAVITY data were reduced with the publicly available consortium \texttt{Python} tools\footnote{\url{https://version-lesia.obspm.fr/repos/DRS_gravity/gravi_tools3/}} which employ the official \texttt{EsoReflex} scripts. 
We first used the \texttt{run\_gravi\_reduce} \texttt{Python} script to reduce the raw images to the so-called \texttt{astroreduced} files using ESO GRAVITY pipeline (v1.6.4b1). 
These files were then processed with the \texttt{run\_gravi\_astrored\_astrometry} script to search for an off-axis companion by fitting, at each position in the field of view, a model including both the interferometric signature of a point source and the residual stellar leakage (see Appendix~A of \citealt{gravity2020}).
For each astrometric position, the model parameters are optimized to minimize the $\chi^2$, yielding a two-dimensional $\chi^2$ map across the field of view.
Since the two science-fiber pointings are separated by only 36\,mas (fiber transmission of $\sim55\%$ at 30\,mas; \citealt{wang2021}), both datasets were combined to increase the sensitivity within their overlapping region of the field of view.

\section{Results}
\label{sec:res}

\subsection{Total-intensity imaging}
\label{sec:im}

\begin{figure}[]
\centering\includegraphics[width=0.45\textwidth{}]{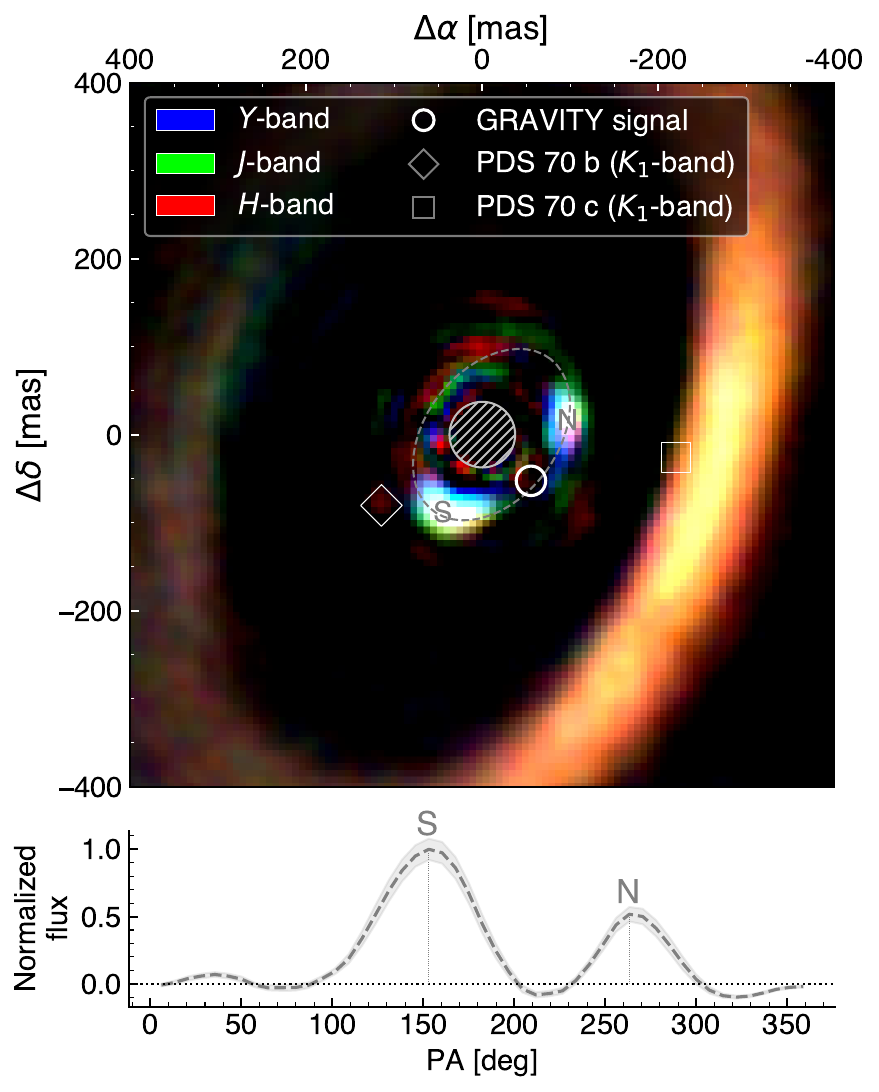}
\caption{\textit{Top:} RGB composition image of PDS~70 for the July 2025 IFS observation showing the $YJH$-bands.
Mask corresponds to saturated pixels.
\textit{Bottom:} Azimuthal profile for the elliptical ring represented in dashed grey in the top image, where two prominent emissions are detected corresponding to the N and S blobs.}
\label{fig:ifs_n2_rgb}
\end{figure}

The complete gallery of images obtained with IPCA-ARDI for IRDIS and with PCA-RSDI for IFS are presented in Fig.~\ref{fig:irdis_dpi} and in Fig.~\ref{fig:irdifs}.
For comparison, the IRDIS images reduced with the double-LOCI algorithm are shown in Fig.~\ref{fig:irdis_ti_2loci}. 
In Fig.~\ref{fig:ifs_n2_rgb}, we show an RGB composite image (combining the 2025 observations) built from the three SPHERE/IFS bands ($YJH$ assigned to blue, green, and red, respectively), together with the azimuthal profile measured along an ellipse drawn in the innermost regions.
For the archival 2021 and 2022 IFS images, we refer the reader to \citet{hammond2025}.
In the next subsections, we describe the main findings.

\subsubsection{Inner regions}
\label{sec:im_cav}

We consistently detect the same features across all observing epochs. 
In the three IRDIS $H$-band datasets (2021 and 2024), we identify planet b together with two additional inner emissions located to the northwest and southeast of the host star, hereafter referred to as N and S, respectively (see Fig.~\ref{fig:irdis_dpi}), with N corresponding to the previously proposed planet candidate.
In the two nights from July 2025 (non-coronagraphic), the N and S features are recovered in all three IFS bands ($YJH$), while the simultaneous IRDIS $K_{1}K_{2}$-bands reveal planets b, c, and the feature N, while S is absent (see Fig.~\ref{fig:irdifs}).
In the archival IFS data from 2021 and 2022, both inner features are also detected (see Fig.~1 from \citealt{hammond2025}).

The N and S features are coincident in position in the three IFS bands (Fig.~\ref{fig:ifs_n2_rgb}) therefore ruling out a chromatic origin. 
Apart from N and S, the azimuthal profile computed along an ellipse in the inner regions of the IFS image, shows no other significant emissions (bottom panel in Fig.~\ref{fig:ifs_n2_rgb}).
While S is clearly extended in all three bands, the morphology of N appears wavelength-dependent, being more extended in the blue ($Y$-band) and point-like in the red ($K$-band). 
\cite{hammond2025} already reported an extended shape for N in their RDI observations.

The double-LOCI reductions are consistent with the IPCA-ARDI results recovering both the N and S features.
We note that an elongated feature trailing N is visible in the 7 April 2024 (third panel of Fig.~\ref{fig:irdis_ti_2loci}).
As such structure is absent from all other datasets (although also present in some PCA components of the ARDI algorithm for the same epoch), we consider it is most likely an artifact.
In the 2021 dataset, the incremental IPCA mode proves essential to recover both N and S emissions, as it is severely affected by the WDH effect.
Consequently, double-LOCI only retrieves N in that epoch.

\begin{figure}[]
\centering\includegraphics[width=0.44\textwidth{}]{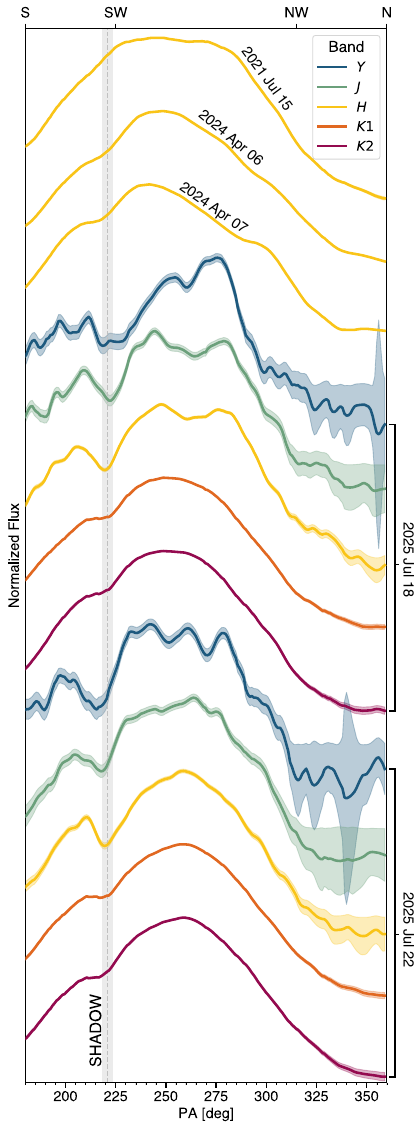}
\caption{Relative azimuthal profiles of the bright side (PA\,$>$\,180$^\circ$) of the outer disk in the SPHERE images. 
The colored lines indicate the band (see legend), and the epoch is labeled in the right $y$-axis. 
Each profile is normalized and offseted.
Color-shaded regions show the 1$\sigma$ uncertainty of the aperture photometry.
The 2$\sigma$ inferred angle of the shadow is shown with a gray vertical area. 
The top $x-$axis indicates the corresponding side of the disk.}
\label{fig:az_prof_shadow}
\end{figure}

Given that N and S are detected in multiple star-hopping epochs at different wavelengths, and using independent post-processing algorithms, we claim these are robustly detected emissions rather than artifacts.
As it will be shown in Sect.~\ref{sec:astr_spec}, the astrometry of N is consistent with that reported for the previous inner planet candidate\footnote{Hereafter, we refer to the emission as N instead of PDS~70\,d.}.
Meanwhile, this is the first time that the S emission is reported.
Despite its small projected separation from PDS~70\,b, we find no evidence of a direct physical connection between the two features as the intensity decreases between them.
The non-detection of S in the datasets previous to 2021 (\citealt{mesa2019}) is discussed in Sect.~\ref{sec:dis}.

\subsubsection{Outer disk}
\label{sec:shadow}

Taking advantage of the improved recovery of extended emission provided by RDI, we extracted the outer-disk azimuthal profiles for each epoch and band from our total-intensity images.
We excluded here the archival Aug 2021 and Feb 2022 epochs since they yield lower-quality images. 
We computed these profiles using aperture photometry, placing circular apertures with a diameter of 2.5\,pixels along the curve tracing the maxima of the signal in 1$^\circ$ steps.
The considered uncertainty per pixel is the standard deviation in the background of the image, which was quadratically propagated by \texttt{photutils}\footnote{\url{https://github.com/astropy/photutils}} (\citealt{bradley2024}, v2.0.2).
The ``elliptic'' curve to trace the disk maxima was derived with \texttt{spifit}\footnote{\url{https://github.com/VChristiaens/spifit}} (\citealt{casassus2021}),
configuring it to fit a single trace over 360$^\circ$, which allows the solution to converge to a closed ellipse-like curve.

Figure~\ref{fig:az_prof_shadow} shows the resulting normalized profiles for the bright (near) side of the outer disk, sorted by epoch and color-coded by spectral band.
We identified a local minimum that appears at comparable azimuths across all bands and epochs,
with the only exception of 2021 July 15, where this minimum is not detected.
This may be related to sensitivity limitations, as both the image quality and the on-source exposure time were lower in the 2021 dataset than in the 2024 and 2025 ones.
The persistence of this localized minimum across epochs and wavelengths is consistent with a shadow cast onto the outer disk, as commonly observed in protoplanetary disks (e.g., \citealt{marino2015, stolker2016}).
We inferred the azimuthal location of this feature by using a hierarchical model for these local minima (see Appendix~\ref{sec:shadow_loc} for details), which considers that all dimmings share the PA and that the deviations arise from uncertainties in the extraction of the azimuthal profile.
In particular, we modeled the multi-wavelength (excluding the $Y$-band given the low signal-to-noise ratio, SNR) data from the two 2025 July epochs (deeper than those from 2024), previously isolating the local minima. 
From this analysis, we constrained the location of the shadow to be at PA~$= 220.74 ^{+0.52^\circ}_{-0.49^\circ}$.

We note that the northern arm-like structure reported by \cite{juillard2022} is recovered in all bands and epochs.
Additionally, in the 2024 $H$-band and 2025 $K_1$-band images, we find a similar structure to the spiral accretion stream of planet c reported by \cite{christiaens2024}.
Both substructures are indicated in Fig.~\ref{fig:irdis_dpi}.
An analysis of these structures is beyond the scope of this work.

\subsection{Polarimetry}
\label{sec:pol}

\begin{figure*}
\sidecaption
\centering\includegraphics[width=12cm]
{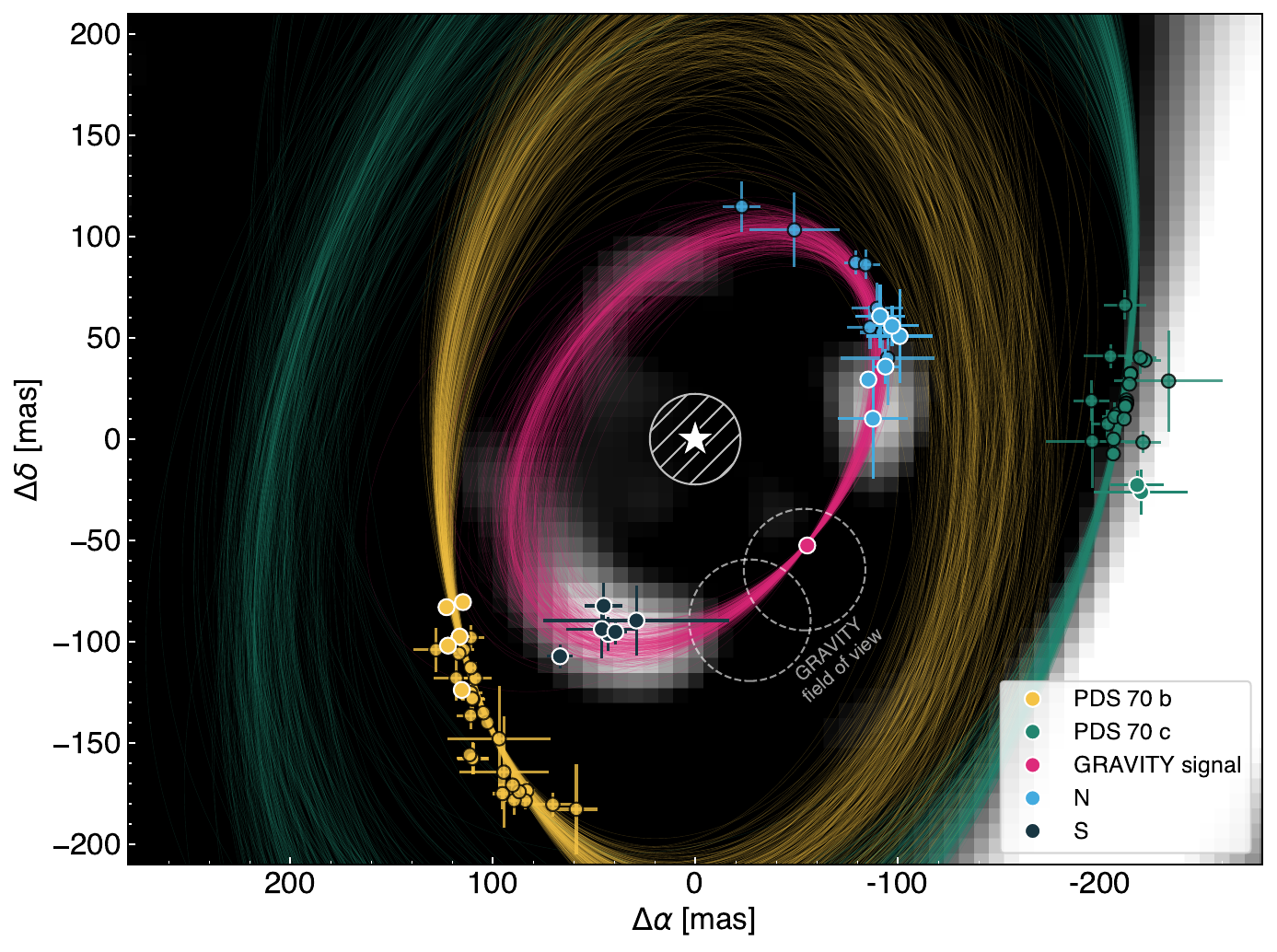}
\caption{Astrometric measurements and inferred orbits for the \textit{3p$_{\rm troj}$} model (see Sect.~\ref{sec:orb}) overlaid in the PDS~70 SPHERE/IFS $H$-band image of 18 July 2025.
Literature astrometry is displayed with black-edged  symbols. 
From this work (white-edged symbols), only astrometric measurements derived in the $H$-band for PDS~70\,b, N and S features, and $K_1$-band for PDS~70\,b and c are included.
Dashed-white circles show the 30\,mas radius of each GRAVITY fiber pointing.}
\label{fig:astrometry}
\end{figure*}

The reduced Q$_{\phi}$ images from the 2021 and 2024 observations are presented in the bottom row of Fig.~\ref{fig:irdis_dpi}.
For the combined 2024 dataset, Fig.~\ref{fig:qp_ti} compares the inner region of the Q$_{\phi}$ image with its total-intensity counterpart, both before and after applying a correction for the coronagraphic transmission.
This correction was computed by dividing each pixel signal by the transmission as provided in \cite{guerri2011}.
We note that this approach may increase the noise very close to the IWA and assumes azimuthal isotropy, thus not preserving the true morphology of the emission.

The N and S features identified in total intensity are recovered in the Q$_{\phi}$ images with similar position and morphology (see Fig.~\ref{fig:qp_ti}, left panel).
Their detection in polarized intensity hence suggests they are scattered-light structures associated with dust.
When applying the coronagraph transmission correction, additional emission is recovered towards the innermost regions. 
This emission is likely tracing the inner disk, previously detected in the NIR \citep{keppler2018}, and the fact that the N and S blobs are embedded in this polarized emission indicates that they may be part of it. 

In the right panel from Fig.~\ref{fig:qp_ti}, we show an elliptical fit (dashed blue line) to the Q$_{\phi}$ signal, inferring approximate values of semi-major (minor) axis of $a_{\rm in}$\,$\approx$\,14\,au ($b_{\rm in}$\,$\approx$\,9\,au). Under the assumption of an intrinsically circular structure projected onto the sky, this translates into $i_{\rm in} \sim 51^{\circ}$ and PA$_{\rm in}$\,$\sim$\,136$^\circ$.
There is an offset with respect to the stellar centroid of $\Delta\alpha_{\rm off}$\,$\sim$\,4\,mas and $\Delta\delta_{\rm off}$\,$\sim$\,--16\,mas.
These estimations are consistent with those derived from GRAVITY dedicated observations, that inferred $i_{\rm in} = 66^{+18^\circ}_{-28^\circ}$ and PA$_{\rm in} = 166^{+50^\circ}_{-51^\circ}$ (\citealt{bohn2022}). 
However, we emphasize that both approaches are subject to large uncertainties.

\subsection{N \& S astrometry and spectra}
\label{sec:astr_spec}

We retrieved the astrometry and photometry of PDS~70\,b, PDS~70\,c, N and S (hereafter, objects) from the total-intensity IRDIS and IFS images.
The full procedure to derive the astrometric and photometric measurements is described in Appendix~\ref{app:astro_phot}. 
The results for the final astrometry and the calibrated fluxes are collected in Table~\ref{tab:astr_spec} for all IRDIS datasets and the wavelength-collapsed IFS cubes, and in Table~\ref{tab:astr_spec_ch} for the 2025 July 18 IFS cube on a channel-by-channel basis. 

Figure~\ref{fig:astrometry} shows our derived astrometry together with bibliographic measurements of the different objects (\citealt{keppler2018, muller2018, christiaens2019, haffert2019, wang2020, wang2021, christiaens2024, wahhaj2024, close2025pds, hammond2025, trevascus2025}).
As explained above, the N feature corresponds to the candidate inner planet previously reported in the literature, with our inferred astrometric measurements for the archival 2021 and 2022 epochs being compatible with those reported in \cite{hammond2025}.
Additionally, we found that the S feature not only shares similar separation to the host star as N, but also lies within its inferred 1$\sigma$ orbital solutions extracted in \citet{hammond2025}.
However, we note that the derived S position is very similar within uncertainties in all epochs (2021-2025). Its compatibility with Keplerian motion is discussed in Sect.~\ref{sec:dis}.

\begin{figure}[t!]
\centering\includegraphics[width=0.4\textwidth{}]{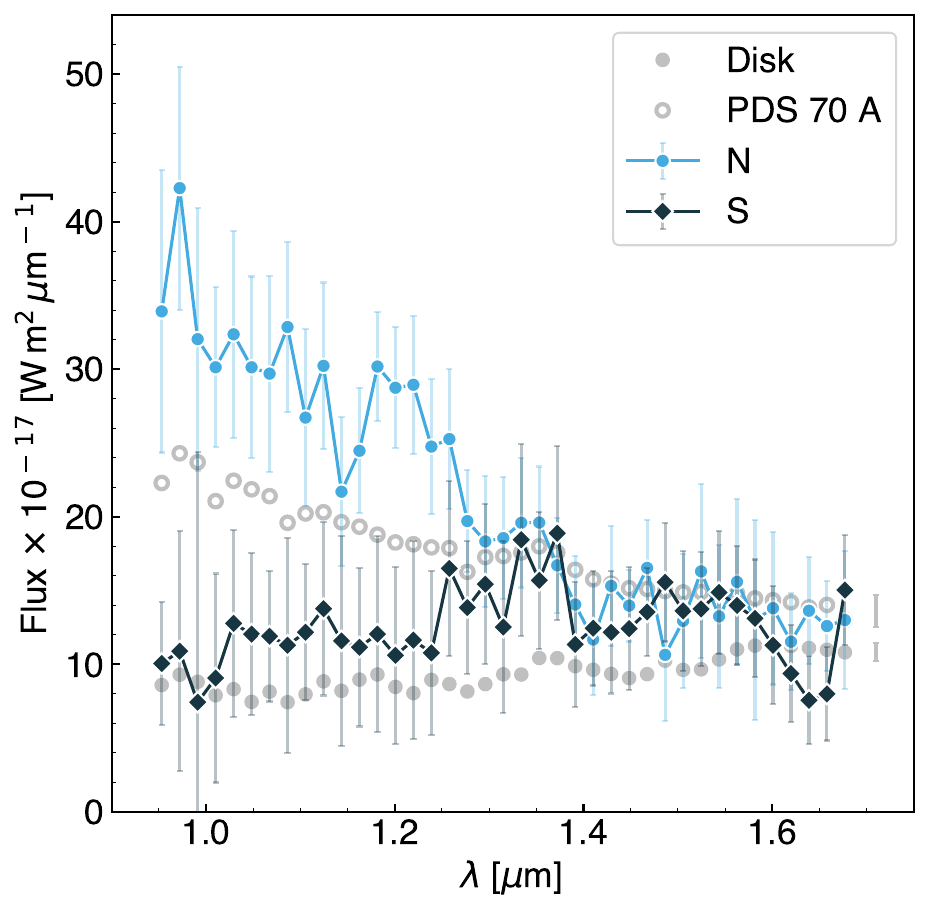}
\caption{Spectra for the two NIR inner features, N and S, inferred from SPHERE/IFS taken in 18 Jul 2025. 
Disk and stellar spectra are shown for comparison, with their associated mean uncertainties shown as grey error bars at the right.}
\vspace{-1cm}
\label{fig:spectra}
\end{figure}

The spectra for the N and S features obtained from July 2025 IFS channels are displayed in Fig.~\ref{fig:spectra}.
The disk and stellar spectra are shown for comparison, both extracted via aperture photometry and later scaled with the NIR features.  For the disk, the flux was averaged over three separate locations along the bright side, using an aperture radius equal to the FWHM of the PSF.
The N spectrum is compatible with that computed by \cite{mesa2019} and \cite{hammond2025}, being brighter at shorter wavelengths and showing a slope similar to that of the host star.
On the other hand, the S spectrum is very similar to that of the outer disk indicating a grey color.
Both N and S are compatible with a dusty nature rather than self-luminous planetary objects which have redder spectra (see \citealt{hammond2025} for a comparison of N spectrum with those from b and c planets).
In fact, N has been recently followed up with GRAVITY observations placing the fiber on the source, resulting in a non-detection (\citealt{trevascus26}, in press). 
This result supports its dusty and extended 
nature. 

Putting together both the astrometric and spectral evidence, we found that N and S are consistent with two dusty clumps lying along the same orbital path, with a preliminary angular separation compatible with $\sim$120$^\circ$. 
This configuration is suggestive of a Trojan scenario, in which circumstellar material is trapped at the \Lfour\ and \Lfive\ Lagrangian points (corresponding to S and N, respectively, given the direction of orbital motion) of a yet undetected planet. 
To test this hypothesis, we conducted a dedicated search with GRAVITY aimed at detecting such a putative planet within the orbit.

\subsection{GRAVITY marginal signal}
\label{sec:grav_res}

Following \citet{nowak20}, we converted the $\chi^2$ map produced by the GRAVITY astrometric fit (Sect.~\ref{sec:grav_obs}) into 
\begin{equation}
    z = \chi^2_{\rm no\,planet} - \chi^2_{\rm planet}(\Delta\alpha,\,\Delta\delta),
\end{equation}
where $\chi^2_{\rm no\,planet}$ corresponds to the value measured at the position of the star.
The $z$ metric therefore measures the preference for the planet model over the null hypothesis.

\begin{figure}[t!]
\centering\includegraphics[width=0.45\textwidth{}]{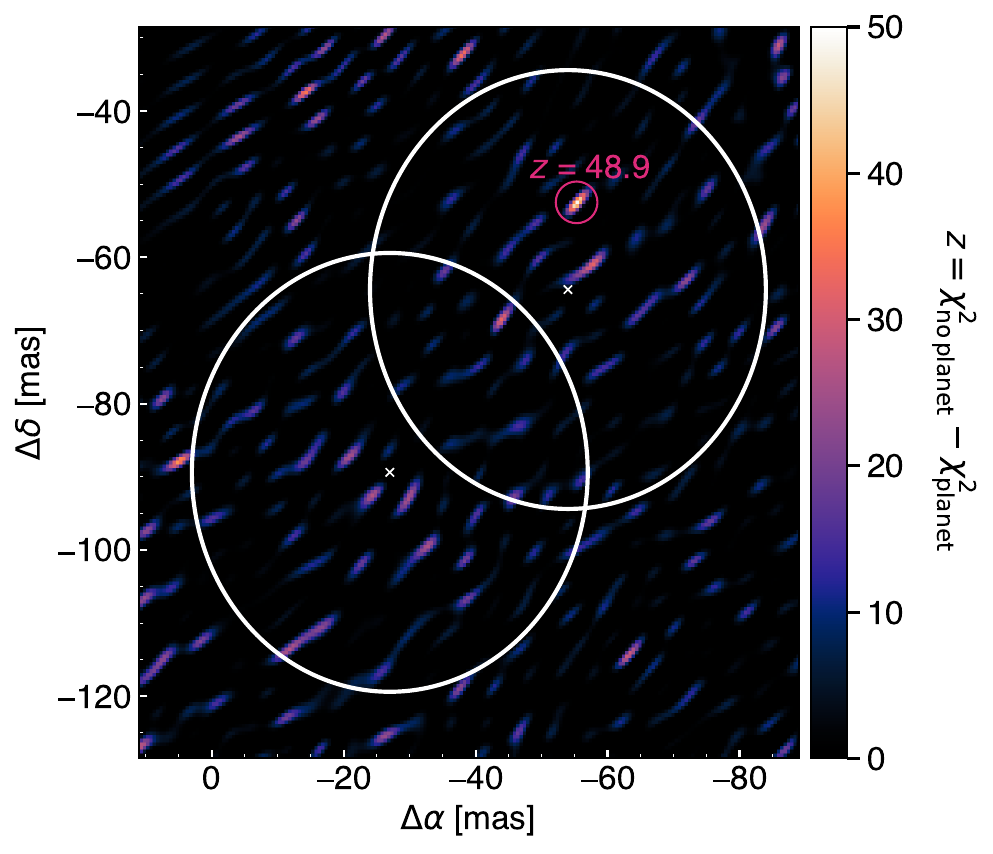}
\caption{$z$ map from the GRAVITY analysis.
The magenta circle marks the global maximum of the point-source model. White circles show the 30\,mas radius of each GRAVITY fiber pointing, and white crosses their centers.}
\label{fig:chi2_gravity}
\end{figure}

The resulting map exhibits a global maximum with $z \approx 50$ (see Fig.~\ref{fig:chi2_gravity}).
This value is well separated from the distribution of the remaining local maxima (see Fig.~\ref{fig:distr_grav}), whose median is $z\sim10$, and the next highest peaks reach $z\sim$30\,-\,38.
The global maximum is located at $\Delta\alpha = -55.31 \pm 0.29$\,mas, $\Delta\delta= -52.48 \pm 0.28$\,mas (separation $\rho = 76.2 \pm 0.29$\,mas and position angle PA = $226.50 \pm 0.21^\circ$; see Fig.~\ref{fig:ifs_n2_rgb} and \ref{fig:astrometry}), with a Pearson correlation coefficient $-0.3$ between $\Delta\alpha$ and $\Delta\delta$. 
The inferred contrast with respect to the star is $(4.4 \pm 1.3) \times 10^{-5}$, corresponding with a $K$-band magnitude
of $19.4 \pm 0.4$\,mag. 
This magnitude is translated into an approximate mass of 3\,M$_{\rm Jup}$ using \texttt{species}\footnote{\url{https://github.com/tomasstolker/species}} (\citealt{stolker2020}, v0.10.0) according to the \texttt{AMES-DUSTY} and \texttt{ATMO} models (\citealt{chabrier2000, tremblin2015}).

The reported signal shows a $\sim$3$\sigma$ significance, estimated as the ratio between the signal contrast and its associated \textit{rms} measured within an elliptical aperture (3.5\,mas\,$\times$\,1.5\,mas, 140$^\circ$) that encloses the signal.
We note that the emission is too faint for a reliable spectral extraction.
Thus, this marginal signal requires future observations to be confirmed and if so, to test its planetary nature (i.e., to check if the emission is compatible with a young planet in a Keplerian orbit).
Finally, we have confirmed that this GRAVITY signal is consistent with a non-detection in the $K_1K_2$ SPHERE IRDIS data, by computing the corresponding contrast curves (see Appendix~\ref{app:contr_curve} and Fig.~\ref{fig:contr_curve}).

\subsection{Orbital fitting}
\label{sec:orb}

We revisited the orbital solution of the PDS~70 planetary system exploiting the information from our new astrometric measurements.
These include the two confirmed planets (PDS~70\,b and c), the N emission (previously proposed as a planet candidate), 
and the newly identified S feature and the planet candidate marginally detected with GRAVITY.

To this end, we considered two models. 
The first (labeled as \textit{3p}) considers the three planetary orbits fitting the measurements from PDS~70\,b, c and N (equivalent to the model of \citealt{hammond2025}, as N is moving in Keplerian motion).
The second model (\textit{3p}$_{\rm troj}$) assumes the GRAVITY planet-like signal to be the innermost planet, and additionally constrains the solution to be consistent with the S and N astrometry under a co-orbital configuration.
We note that for N and S, we only considered our astrometric measurements from the $H$-band collapsed datasets to avoid possible wavelength-dependent effects given their non-compact shape and dusty-like spectra.

We sampled the posterior distributions using the \texttt{Julia} package \texttt{Octofitter}\footnote{\url{https://github.com/sefffal/Octofitter.jl}} (\citealt{thompson2023}, v8.1.2).  
In particular, we used the parallel-tempered sampler \texttt{pigeons}\footnote{\url{http://pigeons.run/stable/}} (\citealt{surjanovic2025}), which allows to efficiently explore multi-modal distributions.
We run it for 17 rounds and 48 chains to ensure the convergence of the models.
We set mostly the same priors as in previous works (e.g.,
\citealt{wang2021, hammond2025, trevascus2025}), which are uninformative for all parameters but the system parallax and stellar mass ($M_\star$). 
As additional priors, we included a non-crossing orbits condition and near co-planarity ($\sim10^\circ$) between the b and c planets and the outer disk (see \citealt{wang2020, wang2021}).
The \textit{3p}$_{\rm troj}$ model was implemented by including an additional likelihood term for the N and S features. 
These are assumed to share the same Keplerian solution as the inner GRAVITY planet-like signal, while accounting for mean longitude offset relative to the proposed planet, introduced as two additional free parameters ($\Delta\lambda_{\rm N/S}$).
To compute it, we assumed the circular case (preferred solution found in \citealt{hammond2025}) in which a shift in the mean longitude is equivalent to a mean anomaly shift ($\Delta\lambda = \Delta M$).
Two jitter terms ($J_{\rm N/S}$) were further included to account for the wavelength-dependent effects and for the increased uncertainty in the astrometric measurements of these two features, given their elongated morphology.

\begin{figure*}[h!]
\sidecaption
\centering\includegraphics[width=12cm]{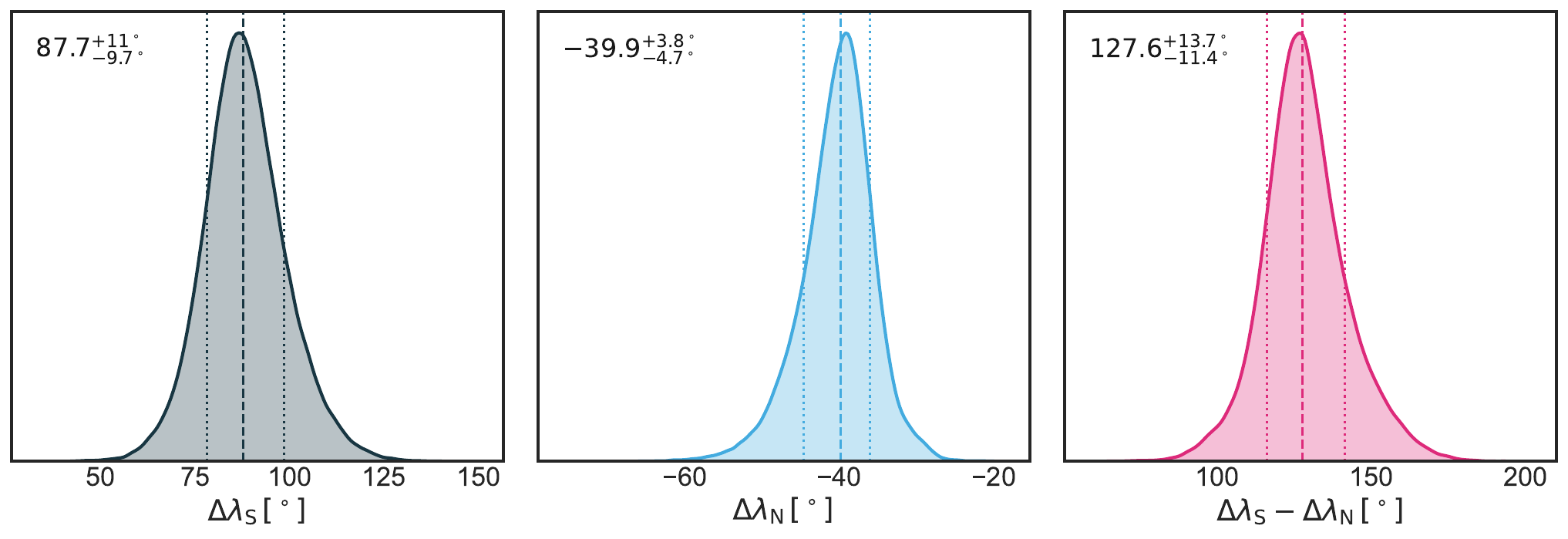}
\caption{Posterior distributions of the mean longitude offset of S (\textit{left}) and N (\textit{center}) relative to the GRAVITY planet-like candidate emission, and their orbital separation (\textit{right}).}
\label{fig:delta_nu}
\end{figure*}

Table~\ref{tab:post_orb} shows the priors and posterior distributions for both models.
The inferred solutions are consistent with those reported by \cite{hammond2025}, and therefore support the possibility of a near 4:2:1 Laplace resonance.
We note that the posterior distributions of the planetary masses remain largely unconstrained, with median values higher than in previous works, likely reflecting the broader upper limits adopted in our prior distributions. 
Figure~\ref{fig:corner} shows the corner plot for the inner planet under both models.
As expected, the inclusion of measurements covering a larger fraction of the orbital path in the \textit{3p}$_{\rm troj}$ model leads to tighter posteriors on the semi-major axis ($a$), eccentricity ($e$), and inclination ($i$), favoring a nearly circular orbit with $a_d = 12.87^{+0.50}_{-0.47}$\,au.
The posterior distributions of the mean longitude offsets for the clumps are shown in Fig.~\ref{fig:delta_nu}. 
While their offsets relative to the GRAVITY planet-like signal do not correspond to the expected $\pm 60^\circ$ configuration for Trojan dust, their mutual separation is consistent with $\sim$120$^\circ$.

\section{Discussion}
\label{sec:dis}

\setlength{\tabcolsep}{6pt}
\begin{table*}[h!]
\caption[]{Summary of the observational evidence for the competing scenarios to explain N and S NIR features.}
\label{tab:scenarios}

\begin{tabular}{llll}

\hline
\hline \noalign{\smallskip}
 & Observational evidence & (S1) Planet with trojan dust & (S2) Inner disk features \\ \noalign{\smallskip} \hline \noalign{\smallskip}
N feature  & Keplerian motion  & \yes\ Clump at the \Lfive\ region of a planet & [?] Rotating substructure \\ \noalign{\smallskip} \hline \noalign{\smallskip}
\multirow{3}{*}{S feature}  &  Detectability  &  \yes\ Behind the coronagraph until 2021  &  [?] Self and over-subtraction  \\  \noalign{\smallskip}
 &  \multirow{2}{*}{Static}  &  \multirow{2}{*}{[?]} Slow rotation as near the aphelion and  &  \multirow{2}{*}{\yes\ Southern ansa} \\ 
  &    &  \hspace{0.42cm} challenging astrometric measurement  &    \\ \noalign{\smallskip} \hline \noalign{\smallskip}
N $\&$ S &  \multirow{2}{*}{Differing slopes}  &   \multirow{2}{*}{\yes} Dust population asymmetry for &   \multirow{2}{*}{[?] N tracing a planetesimal collision}  \\
spectra & & \hspace{0.52cm} \Lfour\ $\&$ \Lfive\ predicted by models &  \\ \noalign{\smallskip} \hline \noalign{\smallskip}
Outer disk & Narrow shadow & [?] Planet blocking the light ($\Delta$PA$\sim$6$^\circ$) & [?] Inclined inner disk \\ \noalign{\smallskip} \hline \noalign{\smallskip}
Polarimetry & Inner disk emission & [?] Inner disk with co-orbitals & \yes\ N \& S embedded in the inner disk \\ \noalign{\smallskip}  \hline \noalign{\smallskip} 
GRAVITY & Planet-like & \multirow{2}{*}{\yes} Compatible with the expected & \multirow{2}{*}{\no\ But signals yet to be confirmed}\\ 
\& JWST & tentative emission & \hspace{0.5cm} location from a trojan configuration &  \\ \noalign{\smallskip}  \hline
\end{tabular}
\\[1ex]
\footnotesize{\textbf{Note.} Symbols \yes, [?], and \no\ indicate if the observational evidence supports, does not constrain, or is inconsistent with the scenario.}
\end{table*}

We propose two hypotheses to explain the inner features detected with SPHERE, which are consistently observed across multiple wavelengths and epochs.
Scenario 1 (S1): the N and S features could correspond to circumstellar material trapped at the stable Lagrange points of an inner protoplanet.
Scenario 2 (S2): N and S could be substructures of the inner disk not related with a co-orbital planet. For instance, S may be the southern ansa of the inner disk, while N could correspond to a rotating substructure within the same disk (e.g., an embedded protoplanet enshrouded in dust, an outer spiral arm, or a dust clump from a recent planetesimal collision),
or they could be vortices triggered by planet b (e.g., \citealt{hammerz2019}).
In the following, we review the observational evidence and discuss the possible interpretations for each case. 
A summary is presented in Table~\ref{tab:scenarios}.

\paragraph{Northern feature in Keplerian motion.}
The N feature is unambiguously moving across the 13 available epochs spanning an 11-year baseline, indicating that it is an orbiting structure around PDS~70. 
Taking advantage of this behavior, in Sect.~\ref{sec:astr_spec} we assumed that N follows the same orbital motion as the hypothetical planet it might trail, on a 1:1 mean-motion resonance (i.e., S1). 
Based on the \textit{3p}$_{\rm troj}$ solution, the orbital separation (i.e., mean-longitude difference) between N and S is $127.6^{+14^\circ}_{-11^\circ}$ (see Fig.~\ref{fig:delta_nu}). 
This is compatible with the 120$^\circ$ separation expected for a tadpole co-orbital configuration (i.e., S-N being \Lfour-\Lfive, respectively).
The fact that both structures are offset from the nominal $\pm60^\circ$ from the planet in the same orbital direction may be explained by the presence of non-negligible dissipative forces (\citealt{2019A&A...631A...6L}).
On the other hand, if N were physically linked to the inner disk instead (S2), its Keplerian motion 
would suggest it could be a partially resolved rotating substructure, such as a spiral arm (e.g., \citealt{deboer2021, ginski2025}).
An inner-disk arm could be excited by a planetary companion (\citealt{dong2017}) but located at a different position than that explored in this work.

\paragraph{Southern feature detectability.}
Conversely to the N feature, S has not been reported previously in the literature. 
We identify two reasons that may explain these non-detections for each proposed scenario. 
In the case where S is co-orbiting with N (S1), the derived orbital solutions (Sect.~\ref{sec:orb}) predict that S was mostly obscured by the coronagraphic masks in all previous observing epochs, with separations below 100\,mas until 2020.
But even if S is part of the inner disk (S2), in non-star-hopping datasets it is susceptible to be severely affected by self- and over-subtraction effects given its particularly elongated morphology.
This effect could also explain why classical ADI strategies failed to recover the elongated shape of N that we found at bluer wavelengths, appearing as a point-like source instead (e.g., \citealt{mesa2019}).

\paragraph{Static southern feature.}
No significant motion within the uncertainties is detected for S across the epochs in which it is detected (2021, 2024, and 2025).
However, there are several caveats to consider which make this result still compatible with the co-orbital scenario (S1).
First, given its current location is near the aphelion of the inferred orbit, the motion of S is expected to be slower than that of N: we estimated that S should have undergone a projected displacement of $\Delta\alpha$\,$\sim$36\,mas and $\Delta\delta$\,$\sim$7\,mas between 2021 and 2025. 
Second, its substantially more extended morphology (elongated in the $\alpha$ direction) as compared to N makes an accurate astrometric extraction technically challenging. 
Third, if S is indeed a dust cloud, its apparent position and morphology may be strongly affected by illumination effects and by the known variability of the host star.
For instance, during dipper events the stellar flux is partially blocked (e.g., \citealt{gaidos2024}), potentially in an anisotropic manner, which can modulate the illumination of the surrounding circumstellar material. 
Hence, additional epochs employing star-hopping observations sufficiently separated in time from the existing data are required to test the co-orbital motion of S.
In case that no motion is confirmed with future observations, the scenario in which S corresponds to the ansa of the inner disk (S2) would be further supported.

\vspace{-0.35cm}
\paragraph{N \& S spectra.}
The spectra recovered for N and S resemble those of the star (blue) and the outer circumstellar disk (grey), respectively (Sect.~\ref{sec:astr_spec}). 
Both are consistent with scattered light emission from dust around the source.
The observed discrepancy in their spectral behavior might be attributed to differences in their dust populations: N being dominated by smaller grains, which enhance Rayleigh scattering, whereas S being composed of larger grains.
Such difference is indeed expected in the co-orbital scenario (S1).
Hydrodynamical simulations predict gas and dust population asymmetries between \Lfour\ and \Lfive\ (e.g., \citealt{montesinos2020}) as a result of local disk temperature gradients (\citealt{heron2026}).
On the other hand, in the speculative case in which N is tracing a collision (S2), a local enhancement in smaller dust grains would be expected.

\begin{figure}[]
\centering
\includegraphics[width=0.42\textwidth{}]{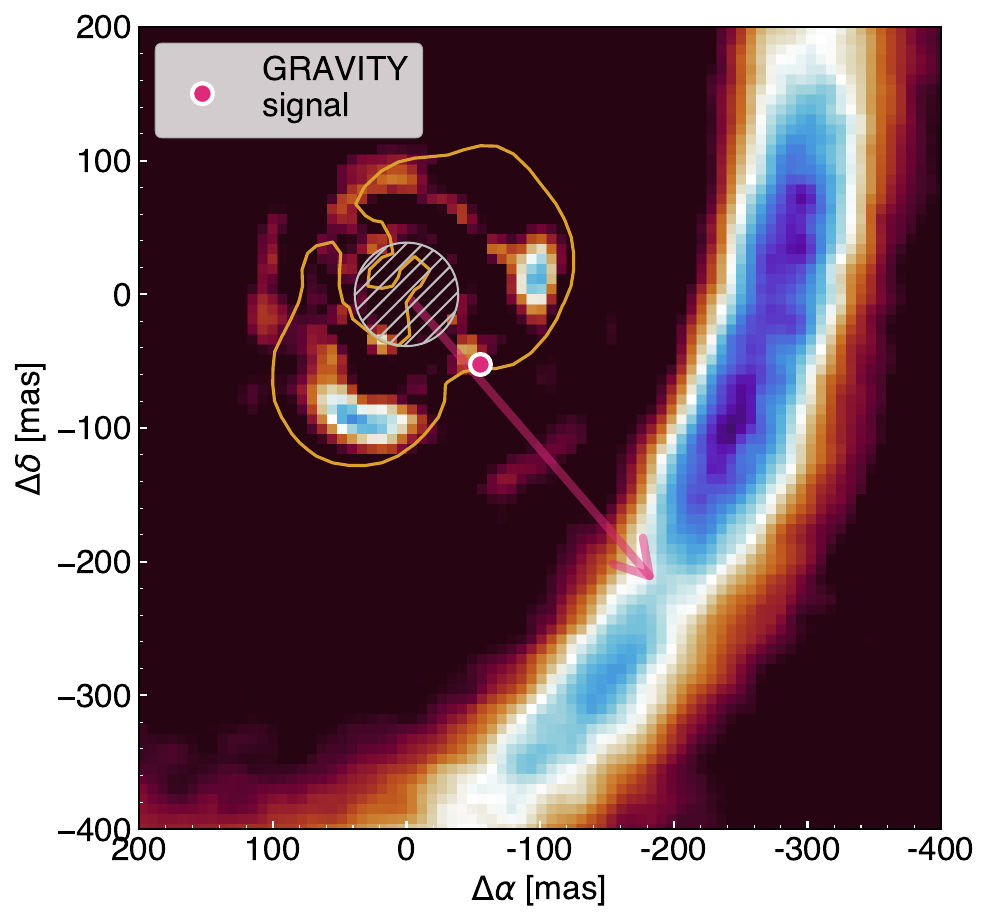}
\caption{SPHERE/IFS image $H$-band (18 July 2025) in the background with yellow contour showing the inner emission from Q$_{\phi}$ in the 2024 combined dataset.
The magenta arrow and dot indicate the direction from the host star to the shadow, and the location of the putative planet, respectively.}
\label{fig:shadow_pl}
\end{figure}

\paragraph{Outer disk narrow shadow.}
Localized brightness drops in outer disks are generally interpreted as shadows cast by misaligned inner disks (e.g., \citealt{canovas2013, 
stolker2016, benisty17}).
As mentioned before (Sect.~\ref{sec:pol}), the inner disk inclination is currently poorly constrained (inner-outer disk misalignment of $15.6\pm18.5^\circ$; \citealt{bohn2022}), but these measurements do not support a strong misalignment that could produce the narrow shadow we detected in 2024 and 2025. 

Alternatively, the presence of the single shadow in the outer disk could also be caused by a planet partially blocking the stellar light (e.g., \citealt{montesinos2021}).
\cite{akansoy2025} noted that the larger the planet-disk separation, the more extended and deeper the shadow becomes.
They emphasized that PDS~70 is a particularly favorable system for detecting planet-induced shadows given its wide cavity, however, no shadow has been found for PDS~70\,b. 
Given that the PA of the observed shadow is close to that of the GRAVITY marginal signal (difference of $\Delta$PA\,$\sim$\,6$^\circ$; see magenta arrow from Fig.~\ref{fig:shadow_pl}),
this partial alignment might support the planetary scenario (S1) if confirmed.
Testing this hypothesis requires re-detecting the shadow using an RDI approach, and verifying that it shares the Keplerian motion of the proposed planet.
At present, this test cannot be performed with the data in hand, as the shadow is detected only in the 2024 and 2025 epochs.
The expected variation in PA over this time baseline, assuming the \textit{3p}$_{\rm troj}$ solution, could be as small as 10$^\circ$. 
This shift is $\sim$2 times the PA deviation found for the shadow between July 18$^{\mathrm{th}}$ and 22$^{\mathrm{th}}$ in the $J$-band.
Given the low contrast of the shadow in the 2024 data, an accurate position constraint is further limited.
We discuss a possible third interpretation for the outer disk shadow in Appendix~\ref{sec:self_shadow}.

\vspace{-0.5cm}
\paragraph{Inner-disk emission.}
The polarized emission detected in the inner regions of the Q$_{\phi}$ image is likely tracing the inner disk. 
The fact that N and S appear embedded in this emission can be interpreted as both being part of it (S2).
A comparison of our SPHERE images with ALMA data resolving the inner disk (\citealt{fasano2025}; Zagaria et al. in prep.) does not put a strong constraint on whether the NIR emissions are embedded or not in the disk given the ALMA beam size ($\sim$50-70\,mas), sensitivity, and different dust populations.
We note, however, that the clumps could still be part of a substructure within the inner disk (e.g., a ring), in which a planet with co-orbitals could be forming. 
Hence, although S2 provides a simpler explanation for the polarimetric observations, S1 would still hold.

\vspace{-0.35cm}
\paragraph{Planet-like marginal detections.}
The GRAVITY search resulted in a marginal signal detection compatible with a $\sim$3\,M$_{\rm Jup}$ planet at the predicted position for a co-orbital planet. 
We note that previous observations with the JWST/Near Infrared Imager and Slitless Spectrograph (NIRISS) using the Aperture Masking Interferometric (AMI) mode at $4.8\, \mu m$, reported a 4$\sigma$ detection at PA~=~$220^{+10^\circ}_{-14^\circ}$, with an unconstrained separation in the 50-200\,mas range (\citealt{blakely2025}). 
To assess whether the GRAVITY and \textit{JWST} data could be tracing the same emission, we propagated the astrometric position of the GRAVITY signal to the epoch of the \textit{JWST} observations (2023 February 24) using the orbital solution from our \textit{3p${\rm troj}$} model and Eq.~\ref{eq:pred_M}. 
We predict a separation of $76.8^{+1.1}_{-1.2}$\,mas and a PA of $249.2^{+1.4^\circ}_{-1.0^\circ}$, which is marginally compatible with the \textit{JWST} detection.

\section{Conclusion}
\label{sec:concl}

The multi-epoch and multi-wavelength SPHERE observations, obtained with the star-hopping strategy, have revealed two elongated inner emissions in the PDS~70 disk.
The northern one corresponds to the previously proposed inner planet candidate.
Their spectra suggest a dusty nature, potentially dominated by different populations of grains, with the northern clump likely composed of smaller grains than the southern counterpart.
The resolved Keplerian motion of N, tracked over more than a decade, together with S laying along its orbital path with a separation compatible with 120$^\circ$, motivated a direct search with GRAVITY for a planet between them. 
This search resulted in a marginal detection setting an approximate mass of 3\,$M_{\rm Jup}$ for the putative planet.

The observed angular separation between the two inner clumps is consistent with material trapped in a tadpole configuration with a planet located between them, corresponding to the \Lfour\ and \Lfive\ points, respectively.
The substantial deviation of both features from the expected 60$^\circ$ ($\sim$25$^\circ$ in the direction of movement) could be caused by dissipative forces, which requires further investigation.
Several but tentative pieces of evidence support the presence of this proposed planet: 
(i) the GRAVITY marginal signal at the expected location; 
(ii) a JWST/NIRISS AMI 4.8\,$\mu$m detection of a 4$\sigma$ excess at a marginally compatible position (\citealt{blakely2025}); 
and (iii) a narrow shadow in the outer disk at a similar PA, which could suggest the putative planet is partially blocking stellar light.

We emphasize that confirming the co-orbital scenario requires additional observations. In particular: 
(i) deeper GRAVITY observations at the proposed location to achieve a direct detection above the 5$\sigma$ level and verify its Keplerian motion; 
(ii) follow-up star-hopping observations to check whether S is following co-orbital motion with N or remains static; 
and (iii) same SPHERE observations monitoring the narrow shadow to determine if it follows the same Keplerian motion.

If this newly proposed planet is confirmed with future observations, this result (together with TWA~7\,b), would showcase the use of co-orbital substructures as an indirect tracer to pinpoint the location of new protoplanets.
Moreover, confirming the co-orbital scenario would constitute the first confirmed detection of exotrojan material, shedding light both on the in-situ formation of trojans (e.g., \citealt{beauge2007, lyra2009, bottke2023}) and the potential precursors of yet-to-be-confirmed exotrojan planets currently under search (e.g., \citealt{2002AJ....124..592L, 2017A&A...599L...7L, 2018A&A...609A..96L, 2018A&A...618A..42L, balsalobre2024, zhang2026}; Haukes et al., in prep.).

However, the embedding of N and S within the inner disk in polarimetric observations, and the apparent lack of motion of S between 2021 and 2025, are important caveats for the co-orbital hypothesis; nevertheless, these do not exclude the trojan scenario given the current uncertainties.
Therefore, we propose an alternative interpretation in which S corresponds to the southern ansa of the inner disk, and N could represent a rotating substructure from this inner disk.
Future high-contrast infrared imagers on next-generation extremely large telescopes (e.g., ELT/METIS; \citealt{metis}), capable of probing smaller angular separations, achieving higher spatial resolution, and covering longer wavelength ranges, will be key to determining whether the N and S features correspond to two dust clumps detached from the inner disk, or if they are indeed associated with it. 
Dedicated hydrodynamical simulations and radiative transfer models will also be essential to assess the stability and expected observational signatures of these scenarios.

\begin{acknowledgements}
We thank the referee for their constructive comments, which helped us improve the manuscript.
We thank Faustine Cantalloube, Laird Close, Daniele Fasano, Johan Olofsson, Jason Wang and Francesco Zagaria for useful discussions.
This work is based on observations collected at the European Southern Observatory under ESO programmes 113.26PM.001, 115.29EH.001 and 115.29EH.002.
This research is funded by PID2019-107061GB-C61, PID2023-150468NB-I00 and MDM-2017-0737 by the Spanish Ministry of Science and Innovation/State Agency of Research MCIN/AEI/10.13039/501100011033.
J.L.-B. is also supported by MICIU/AEI/10.13039/501100011033 and NextGenerationEU/PRTR, grant CNS2023-144309.
I.H. acknowledges funding from the European Research Council (ERC) under the European Union’s Horizon 2020 programme (PROTOPLANETS, grant No. 101002188).
C.C. acknowledges support from Agencia Nacional de Investigación y Desarrollo (ANID) through FONDECYT Postdoctoral grant n3230283 and CAS250005.
I.M. is funded by grant PID2022-138366NA-I00 (MCIN/AEI/10.13039/501100011033) and by the European Union.
I.d~G. acknowledges funding from grant PID2023-146295NB-I00 (MCIN/AEI/10.13039/501100011033).
A.R. acknowledges funding from the Royal Society through a University Research Fellowship grant URF\textbackslash R1\textbackslash 241791.
OBR acknowledges the University of Liège for hosting a two-week research stay that benefited the results and scientific discussion of this work.
\end{acknowledgements}
\vspace{-1 cm}

\bibliography{references}

@ARTICLE{ATorres2021,
       author = {{Asensio-Torres}, R. and {Henning}, Th. and {Cantalloube}, F. and {Pinilla}, P. and {Mesa}, D. and {Garufi}, A. and {Jorquera}, S. and {Gratton}, R. and {Chauvin}, G. and {Szul{\'a}gyi}, J. and {van Boekel}, R. and {Dong}, R. and {Marleau}, G. -D. and {Benisty}, M. and {Villenave}, M. and {Bergez-Casalou}, C. and {Desgrange}, C. and {Janson}, M. and {Keppler}, M. and {Langlois}, M. and {M{\'e}nard}, F. and {Rickman}, E. and {Stolker}, T. and {Feldt}, M. and {Fusco}, T. and {Gluck}, L. and {Pavlov}, A. and {Ramos}, J.},
        title = "{Perturbers: SPHERE detection limits to planetary-mass companions in protoplanetary disks}",
      journal = {\aap},
         year = 2021,
        month = aug,
       volume = {652},
          eid = {A101},
        pages = {A101},
          doi = {10.1051/0004-6361/202140325},
archivePrefix = {arXiv},
       eprint = {2103.05377},
 primaryClass = {astro-ph.EP},
       adsurl = {https://ui.adsabs.harvard.edu/abs/2021A&A...652A.101A}
}

@ARTICLE{sanchis2020,
       author = {{Sanchis}, E. and {Picogna}, G. and {Ercolano}, B. and {Testi}, L. and {Rosotti}, G.},
        title = "{Detectability of embedded protoplanets from hydrodynamical simulations}",
      journal = {\mnras},
         year = 2020,
        month = mar,
       volume = {492},
       number = {3},
        pages = {3440-3458},
          doi = {10.1093/mnras/staa074},
archivePrefix = {arXiv},
       eprint = {2001.03565},
 primaryClass = {astro-ph.EP},
       adsurl = {https://ui.adsabs.harvard.edu/abs/2020MNRAS.492.3440S}
}

@ARTICLE{swastik2025,
       author = {{Swastik}, C. and {Wahhaj}, Z. and {Benisty}, M. and {Arora}, S. and {Ginski}, C. and {Ren}, B.~B. and {van Holstein}, R.~G. and {de Rosa}, R. and {Banyal}, R.~K. and {Tazaki}, R.},
        title = "{Imaging the LkCa 15 system in polarimetry and total intensity without self-subtraction artefacts}",
      journal = {\aap},
         year = 2026,
        month = feb,
       volume = {706},
          eid = {A312},
        pages = {A312},
          doi = {10.1051/0004-6361/202449743},
archivePrefix = {arXiv},
       eprint = {2512.18439},
 primaryClass = {astro-ph.EP},
       adsurl = {https://ui.adsabs.harvard.edu/abs/2026A&A...706A.312S}
}

@ARTICLE{wang2020,
       author = {{Wang}, Jason J. and {Ginzburg}, Sivan and {Ren}, Bin and {Wallack}, Nicole and {Gao}, Peter and {Mawet}, Dimitri and {Bond}, Charlotte Z. and {Cetre}, Sylvain and {Wizinowich}, Peter and {De Rosa}, Robert J. and {Ruane}, Garreth and {Liu}, Michael C. and {Absil}, Olivier and {Alvarez}, Carlos and {Baranec}, Christoph and {Choquet}, {\'E}lodie and {Chun}, Mark and {Defr{\`e}re}, Denis and {Delorme}, Jacques-Robert and {Duch{\^e}ne}, Gaspard and {Forsberg}, Pontus and {Ghez}, Andrea and {Guyon}, Olivier and {Hall}, Donald N.~B. and {Huby}, Elsa and {Jolivet}, A{\"\i}ssa and {Jensen-Clem}, Rebecca and {Jovanovic}, Nemanja and {Karlsson}, Mikael and {Lilley}, Scott and {Matthews}, Keith and {M{\'e}nard}, Fran{\c{c}}ois and {Meshkat}, Tiffany and {Millar-Blanchaer}, Maxwell and {Ngo}, Henry and {Orban de Xivry}, Gilles and {Pinte}, Christophe and {Ragland}, Sam and {Serabyn}, Eugene and {Catal{\'a}n}, Ernesto Vargas and {Wang}, Ji and {Wetherell}, Ed and {Williams}, Jonathan P. and {Ygouf}, Marie and {Zuckerman}, Ben},
        title = "{Keck/NIRC2 L'-band Imaging of Jovian-mass Accreting Protoplanets around PDS 70}",
      journal = {\aj},
         year = 2020,
        month = jun,
       volume = {159},
       number = {6},
          eid = {263},
        pages = {263},
          doi = {10.3847/1538-3881/ab8aef},
archivePrefix = {arXiv},
       eprint = {2004.09597},
 primaryClass = {astro-ph.EP},
       adsurl = {https://ui.adsabs.harvard.edu/abs/2020AJ....159..263W}
}

@ARTICLE{gravity2020,
       author = {{GRAVITY Collaboration} and {Nowak}, M. and {Lacour}, S. and {Molli{\`e}re}, P. and {Wang}, J. and {Charnay}, B. and {van Dishoeck}, E.~F. and {Abuter}, R. and {Amorim}, A. and {Berger}, J.~P. and {Beust}, H. and {Bonnefoy}, M. and {Bonnet}, H. and {Brandner}, W. and {Buron}, A. and {Cantalloube}, F. and {Collin}, C. and {Chapron}, F. and {Cl{\'e}net}, Y. and {Coud{\'e} Du Foresto}, V. and {de Zeeuw}, P.~T. and {Dembet}, R. and {Dexter}, J. and {Duvert}, G. and {Eckart}, A. and {Eisenhauer}, F. and {F{\"o}rster Schreiber}, N.~M. and {F{\'e}dou}, P. and {Garcia Lopez}, R. and {Gao}, F. and {Gendron}, E. and {Genzel}, R. and {Gillessen}, S. and {Hau{\ss}mann}, F. and {Henning}, T. and {Hippler}, S. and {Hubert}, Z. and {Jocou}, L. and {Kervella}, P. and {Lagrange}, A. -M. and {Lapeyr{\`e}re}, V. and {Le Bouquin}, J. -B. and {L{\'e}na}, P. and {Maire}, A. -L. and {Ott}, T. and {Paumard}, T. and {Paladini}, C. and {Perraut}, K. and {Perrin}, G. and {Pueyo}, L. and {Pfuhl}, O. and {Rabien}, S. and {Rau}, C. and {Rodr{\'\i}guez-Coira}, G. and {Rousset}, G. and {Scheithauer}, S. and {Shangguan}, J. and {Straub}, O. and {Straubmeier}, C. and {Sturm}, E. and {Tacconi}, L.~J. and {Vincent}, F. and {Widmann}, F. and {Wieprecht}, E. and {Wiezorrek}, E. and {Woillez}, J. and {Yazici}, S. and {Ziegler}, D.},
        title = "{Peering into the formation history of {\ensuremath{\beta}} Pictoris b with VLTI/GRAVITY long-baseline interferometry}",
      journal = {\aap},
         year = 2020,
        month = jan,
       volume = {633},
          eid = {A110},
        pages = {A110},
          doi = {10.1051/0004-6361/201936898},
archivePrefix = {arXiv},
       eprint = {1912.04651},
 primaryClass = {astro-ph.EP},
       adsurl = {https://ui.adsabs.harvard.edu/abs/2020A&A...633A.110G}
}

@ARTICLE{ren23,
       author = {{Ren}, Bin B.},
        title = "{Karhunen-Lo{\`e}ve data imputation in high-contrast imaging}",
      journal = {\aap},
         year = 2023,
        month = nov,
       volume = {679},
          eid = {A18},
        pages = {A18},
          doi = {10.1051/0004-6361/202347354},
archivePrefix = {arXiv},
       eprint = {2308.16912},
 primaryClass = {astro-ph.IM},
       adsurl = {https://ui.adsabs.harvard.edu/abs/2023A&A...679A..18R}
}

@ARTICLE{2007ApJ...660..770L,
       author = {{Lafreni{\`e}re}, David and {Marois}, Christian and {Doyon}, Ren{\'e} and {Nadeau}, Daniel and {Artigau}, {\'E}tienne},
        title = "{A New Algorithm for Point-Spread Function Subtraction in High-Contrast Imaging: A Demonstration with Angular Differential Imaging}",
      journal = {\apj},
         year = 2007,
        month = may,
       volume = {660},
       number = {1},
        pages = {770-780},
          doi = {10.1086/513180},
archivePrefix = {arXiv},
       eprint = {astro-ph/0702697},
 primaryClass = {astro-ph},
       adsurl = {https://ui.adsabs.harvard.edu/abs/2007ApJ...660..770L}
}

@ARTICLE{2020A&A...638A..98C,
       author = {{Cantalloube}, F. and {Farley}, O.~J.~D. and {Milli}, J. and {Bharmal}, N. and {Brandner}, W. and {Correia}, C. and {Dohlen}, K. and {Henning}, Th. and {Osborn}, J. and {Por}, E. and {Su{\'a}rez Valles}, M. and {Vigan}, A.},
        title = "{Wind-driven halo in high-contrast images. I. Analysis of the focal-plane images of SPHERE}",
      journal = {\aap},
         year = 2020,
        month = jun,
       volume = {638},
          eid = {A98},
        pages = {A98},
          doi = {10.1051/0004-6361/201937397},
archivePrefix = {arXiv},
       eprint = {2003.05794},
 primaryClass = {astro-ph.IM},
       adsurl = {https://ui.adsabs.harvard.edu/abs/2020A&A...638A..98C}
}

@ARTICLE{2019AJ....157..118R,
       author = {{Ruane}, Garreth and {Ngo}, Henry and {Mawet}, Dimitri and {Absil}, Olivier and {Choquet}, {\'E}lodie and {Cook}, Therese and {Gomez Gonzalez}, Carlos and {Huby}, Elsa and {Matthews}, Keith and {Meshkat}, Tiffany and {Reggiani}, Maddalena and {Serabyn}, Eugene and {Wallack}, Nicole and {Xuan}, W. Jerry},
        title = "{Reference Star Differential Imaging of Close-in Companions and Circumstellar Disks with the NIRC2 Vortex Coronagraph at the W. M. Keck Observatory}",
      journal = {\aj},
         year = 2019,
        month = mar,
       volume = {157},
       number = {3},
          eid = {118},
        pages = {118},
          doi = {10.3847/1538-3881/aafee2},
archivePrefix = {arXiv},
       eprint = {1901.04090},
 primaryClass = {astro-ph.IM},
       adsurl = {https://ui.adsabs.harvard.edu/abs/2019AJ....157..118R}
}

@ARTICLE{2009ApJ...694L.148L,
       author = {{Lafreni{\`e}re}, David and {Marois}, Christian and {Doyon}, Ren{\'e} and {Barman}, Travis},
        title = "{HST/NICMOS Detection of HR 8799 b in 1998}",
      journal = {\apjl},
         year = 2009,
        month = apr,
       volume = {694},
       number = {2},
        pages = {L148-L152},
          doi = {10.1088/0004-637X/694/2/L148},
archivePrefix = {arXiv},
       eprint = {0902.3247},
 primaryClass = {astro-ph.EP},
       adsurl = {https://ui.adsabs.harvard.edu/abs/2009ApJ...694L.148L}
}

@ARTICLE{wahhaj2021,
       author = {{Wahhaj}, Z. and {Milli}, J. and {Romero}, C. and {Cieza}, L. and {Zurlo}, A. and {Vigan}, A. and {Pe{\~n}a}, E. and {Valdes}, G. and {Cantalloube}, F. and {Girard}, J. and {Pantoja}, B.},
        title = "{A search for a fifth planet around HR 8799 using the star-hopping RDI technique at VLT/SPHERE}",
      journal = {\aap},
         year = 2021,
        month = apr,
       volume = {648},
          eid = {A26},
        pages = {A26},
          doi = {10.1051/0004-6361/202038794},
archivePrefix = {arXiv},
       eprint = {2101.08268},
 primaryClass = {astro-ph.EP},
       adsurl = {https://ui.adsabs.harvard.edu/abs/2021A&A...648A..26W}
}

@ARTICLE{beuzit19,
       author = {{Beuzit}, J. -L. and {Vigan}, A. and {Mouillet}, D. and {Dohlen}, K. and {Gratton}, R. and {Boccaletti}, A. and {Sauvage}, J. -F. and {Schmid}, H.~M. and {Langlois}, M. and {Petit}, C. and {Baruffolo}, A. and {Feldt}, M. and {Milli}, J. and {Wahhaj}, Z. and {Abe}, L. and {Anselmi}, U. and {Antichi}, J. and {Barette}, R. and {Baudrand}, J. and {Baudoz}, P. and {Bazzon}, A. and {Bernardi}, P. and {Blanchard}, P. and {Brast}, R. and {Bruno}, P. and {Buey}, T. and {Carbillet}, M. and {Carle}, M. and {Cascone}, E. and {Chapron}, F. and {Charton}, J. and {Chauvin}, G. and {Claudi}, R. and {Costille}, A. and {De Caprio}, V. and {de Boer}, J. and {Delboulb{\'e}}, A. and {Desidera}, S. and {Dominik}, C. and {Downing}, M. and {Dupuis}, O. and {Fabron}, C. and {Fantinel}, D. and {Farisato}, G. and {Feautrier}, P. and {Fedrigo}, E. and {Fusco}, T. and {Gigan}, P. and {Ginski}, C. and {Girard}, J. and {Giro}, E. and {Gisler}, D. and {Gluck}, L. and {Gry}, C. and {Henning}, T. and {Hubin}, N. and {Hugot}, E. and {Incorvaia}, S. and {Jaquet}, M. and {Kasper}, M. and {Lagadec}, E. and {Lagrange}, A. -M. and {Le Coroller}, H. and {Le Mignant}, D. and {Le Ruyet}, B. and {Lessio}, G. and {Lizon}, J. -L. and {Llored}, M. and {Lundin}, L. and {Madec}, F. and {Magnard}, Y. and {Marteaud}, M. and {Martinez}, P. and {Maurel}, D. and {M{\'e}nard}, F. and {Mesa}, D. and {M{\"o}ller-Nilsson}, O. and {Moulin}, T. and {Moutou}, C. and {Orign{\'e}}, A. and {Parisot}, J. and {Pavlov}, A. and {Perret}, D. and {Pragt}, J. and {Puget}, P. and {Rabou}, P. and {Ramos}, J. and {Reess}, J. -M. and {Rigal}, F. and {Rochat}, S. and {Roelfsema}, R. and {Rousset}, G. and {Roux}, A. and {Saisse}, M. and {Salasnich}, B. and {Santambrogio}, E. and {Scuderi}, S. and {Segransan}, D. and {Sevin}, A. and {Siebenmorgen}, R. and {Soenke}, C. and {Stadler}, E. and {Suarez}, M. and {Tiph{\`e}ne}, D. and {Turatto}, M. and {Udry}, S. and {Vakili}, F. and {Waters}, L.~B.~F.~M. and {Weber}, L. and {Wildi}, F. and {Zins}, G. and {Zurlo}, A.},
        title = "{SPHERE: the exoplanet imager for the Very Large Telescope}",
      journal = {\aap},
         year = 2019,
        month = nov,
       volume = {631},
          eid = {A155},
        pages = {A155},
          doi = {10.1051/0004-6361/201935251},
archivePrefix = {arXiv},
       eprint = {1902.04080},
 primaryClass = {astro-ph.IM},
       adsurl = {https://ui.adsabs.harvard.edu/abs/2019A&A...631A.155B}
}

@ARTICLE{gravity17,
       author = {{GRAVITY Collaboration} and {Abuter}, R. and {Accardo}, M. and {Amorim}, A. and {Anugu}, N. and {{\'A}vila}, G. and {Azouaoui}, N. and {Benisty}, M. and {Berger}, J.~P. and {Blind}, N. and {Bonnet}, H. and {Bourget}, P. and {Brandner}, W. and {Brast}, R. and {Buron}, A. and {Burtscher}, L. and {Cassaing}, F. and {Chapron}, F. and {Choquet}, {\'E}. and {Cl{\'e}net}, Y. and {Collin}, C. and {Coud{\'e} Du Foresto}, V. and {de Wit}, W. and {de Zeeuw}, P.~T. and {Deen}, C. and {Delplancke-Str{\"o}bele}, F. and {Dembet}, R. and {Derie}, F. and {Dexter}, J. and {Duvert}, G. and {Ebert}, M. and {Eckart}, A. and {Eisenhauer}, F. and {Esselborn}, M. and {F{\'e}dou}, P. and {Finger}, G. and {Garcia}, P. and {Garcia Dabo}, C.~E. and {Garcia Lopez}, R. and {Gendron}, E. and {Genzel}, R. and {Gillessen}, S. and {Gonte}, F. and {Gordo}, P. and {Grould}, M. and {Gr{\"o}zinger}, U. and {Guieu}, S. and {Haguenauer}, P. and {Hans}, O. and {Haubois}, X. and {Haug}, M. and {Haussmann}, F. and {Henning}, Th. and {Hippler}, S. and {Horrobin}, M. and {Huber}, A. and {Hubert}, Z. and {Hubin}, N. and {Hummel}, C.~A. and {Jakob}, G. and {Janssen}, A. and {Jochum}, L. and {Jocou}, L. and {Kaufer}, A. and {Kellner}, S. and {Kendrew}, S. and {Kern}, L. and {Kervella}, P. and {Kiekebusch}, M. and {Klein}, R. and {Kok}, Y. and {Kolb}, J. and {Kulas}, M. and {Lacour}, S. and {Lapeyr{\`e}re}, V. and {Lazareff}, B. and {Le Bouquin}, J. -B. and {L{\`e}na}, P. and {Lenzen}, R. and {L{\'e}v{\^e}que}, S. and {Lippa}, M. and {Magnard}, Y. and {Mehrgan}, L. and {Mellein}, M. and {M{\'e}rand}, A. and {Moreno-Ventas}, J. and {Moulin}, T. and {M{\"u}ller}, E. and {M{\"u}ller}, F. and {Neumann}, U. and {Oberti}, S. and {Ott}, T. and {Pallanca}, L. and {Panduro}, J. and {Pasquini}, L. and {Paumard}, T. and {Percheron}, I. and {Perraut}, K. and {Perrin}, G. and {Pfl{\"u}ger}, A. and {Pfuhl}, O. and {Phan Duc}, T. and {Plewa}, P.~M. and {Popovic}, D. and {Rabien}, S. and {Ram{\'\i}rez}, A. and {Ramos}, J. and {Rau}, C. and {Riquelme}, M. and {Rohloff}, R. -R. and {Rousset}, G. and {Sanchez-Bermudez}, J. and {Scheithauer}, S. and {Sch{\"o}ller}, M. and {Schuhler}, N. and {Spyromilio}, J. and {Straubmeier}, C. and {Sturm}, E. and {Suarez}, M. and {Tristram}, K.~R.~W. and {Ventura}, N. and {Vincent}, F. and {Waisberg}, I. and {Wank}, I. and {Weber}, J. and {Wieprecht}, E. and {Wiest}, M. and {Wiezorrek}, E. and {Wittkowski}, M. and {Woillez}, J. and {Wolff}, B. and {Yazici}, S. and {Ziegler}, D. and {Zins}, G.},
        title = "{First light for GRAVITY: Phase referencing optical interferometry for the Very Large Telescope Interferometer}",
      journal = {\aap},
         year = 2017,
        month = jun,
       volume = {602},
          eid = {A94},
        pages = {A94},
          doi = {10.1051/0004-6361/201730838},
archivePrefix = {arXiv},
       eprint = {1705.02345},
 primaryClass = {astro-ph.IM},
       adsurl = {https://ui.adsabs.harvard.edu/abs/2017A&A...602A..94G}
}

@ARTICLE{mesa2019,
       author = {{Mesa}, D. and {Keppler}, M. and {Cantalloube}, F. and {Rodet}, L. and {Charnay}, B. and {Gratton}, R. and {Langlois}, M. and {Boccaletti}, A. and {Bonnefoy}, M. and {Vigan}, A. and {Flasseur}, O. and {Bae}, J. and {Benisty}, M. and {Chauvin}, G. and {de Boer}, J. and {Desidera}, S. and {Henning}, T. and {Lagrange}, A. -M. and {Meyer}, M. and {Milli}, J. and {M{\"u}ller}, A. and {Pairet}, B. and {Zurlo}, A. and {Antoniucci}, S. and {Baudino}, J. -L. and {Brown Sevilla}, S. and {Cascone}, E. and {Cheetham}, A. and {Claudi}, R.~U. and {Delorme}, P. and {D'Orazi}, V. and {Feldt}, M. and {Hagelberg}, J. and {Janson}, M. and {Kral}, Q. and {Lagadec}, E. and {Lazzoni}, C. and {Ligi}, R. and {Maire}, A. -L. and {Martinez}, P. and {Menard}, F. and {Meunier}, N. and {Perrot}, C. and {Petrus}, S. and {Pinte}, C. and {Rickman}, E.~L. and {Rochat}, S. and {Rouan}, D. and {Samland}, M. and {Sauvage}, J. -F. and {Schmidt}, T. and {Udry}, S. and {Weber}, L. and {Wildi}, F.},
        title = "{VLT/SPHERE exploration of the young multiplanetary system PDS70}",
      journal = {\aap},
         year = 2019,
        month = dec,
       volume = {632},
          eid = {A25},
        pages = {A25},
          doi = {10.1051/0004-6361/201936764},
archivePrefix = {arXiv},
       eprint = {1910.11169},
 primaryClass = {astro-ph.EP},
       adsurl = {https://ui.adsabs.harvard.edu/abs/2019A&A...632A..25M}
}

@ARTICLE{deboer2020,
       author = {{de Boer}, J. and {Langlois}, M. and {van Holstein}, R.~G. and {Girard}, J.~H. and {Mouillet}, D. and {Vigan}, A. and {Dohlen}, K. and {Snik}, F. and {Keller}, C.~U. and {Ginski}, C. and {Stam}, D.~M. and {Milli}, J. and {Wahhaj}, Z. and {Kasper}, M. and {Schmid}, H.~M. and {Rabou}, P. and {Gluck}, L. and {Hugot}, E. and {Perret}, D. and {Martinez}, P. and {Weber}, L. and {Pragt}, J. and {Sauvage}, J. -F. and {Boccaletti}, A. and {Le Coroller}, H. and {Dominik}, C. and {Henning}, T. and {Lagadec}, E. and {M{\'e}nard}, F. and {Turatto}, M. and {Udry}, S. and {Chauvin}, G. and {Feldt}, M. and {Beuzit}, J. -L.},
        title = "{Polarimetric imaging mode of VLT/SPHERE/IRDIS. I. Description, data reduction, and observing strategy}",
      journal = {\aap},
         year = 2020,
        month = jan,
       volume = {633},
          eid = {A63},
        pages = {A63},
          doi = {10.1051/0004-6361/201834989},
archivePrefix = {arXiv},
       eprint = {1909.13107},
 primaryClass = {astro-ph.IM},
       adsurl = {https://ui.adsabs.harvard.edu/abs/2020A&A...633A..63D}
}

@ARTICLE{marois2006,
       author = {{Marois}, Christian and {Lafreni{\`e}re}, David and {Doyon}, Ren{\'e} and {Macintosh}, Bruce and {Nadeau}, Daniel},
        title = "{Angular Differential Imaging: A Powerful High-Contrast Imaging Technique}",
      journal = {\apj},
         year = 2006,
        month = apr,
       volume = {641},
       number = {1},
        pages = {556-564},
          doi = {10.1086/500401},
archivePrefix = {arXiv},
       eprint = {astro-ph/0512335},
 primaryClass = {astro-ph},
       adsurl = {https://ui.adsabs.harvard.edu/abs/2006ApJ...641..556M}
}

@INPROCEEDINGS{2008SPIE.7014E..3EC,
       author = {{Claudi}, R.~U. and {Turatto}, M. and {Gratton}, R.~G. and {Antichi}, J. and {Bonavita}, M. and {Bruno}, P. and {Cascone}, E. and {De Caprio}, V. and {Desidera}, S. and {Giro}, E. and {Mesa}, D. and {Scuderi}, S. and {Dohlen}, K. and {Beuzit}, J.~L. and {Puget}, P.},
        title = "{SPHERE IFS: the spectro differential imager of the VLT for exoplanets search}",
    booktitle = {Ground-based and Airborne Instrumentation for Astronomy II},
         year = 2008,
       editor = {{McLean}, Ian S. and {Casali}, Mark M.},
       series = {Society of Photo-Optical Instrumentation Engineers (SPIE) Conference Series},
       volume = {7014},
        month = jul,
          eid = {70143E},
        pages = {70143E},
          doi = {10.1117/12.788366},
       adsurl = {https://ui.adsabs.harvard.edu/abs/2008SPIE.7014E..3EC}
}

@INPROCEEDINGS{2008SPIE.7014E..3LD,
       author = {{Dohlen}, Kjetil and {Langlois}, Maud and {Saisse}, Michel and {Hill}, Lucien and {Origne}, Alain and {Jacquet}, Marc and {Fabron}, Christophe and {Blanc}, Jean-Claude and {Llored}, Marc and {Carle}, Michael and {Moutou}, Claire and {Vigan}, Arthur and {Boccaletti}, Anthony and {Carbillet}, Marcel and {Mouillet}, David and {Beuzit}, Jean-Luc},
        title = "{The infra-red dual imaging and spectrograph for SPHERE: design and performance}",
    booktitle = {Ground-based and Airborne Instrumentation for Astronomy II},
         year = 2008,
       editor = {{McLean}, Ian S. and {Casali}, Mark M.},
       series = {Society of Photo-Optical Instrumentation Engineers (SPIE) Conference Series},
       volume = {7014},
        month = jul,
          eid = {70143L},
        pages = {70143L},
          doi = {10.1117/12.789786},
       adsurl = {https://ui.adsabs.harvard.edu/abs/2008SPIE.7014E..3LD}
}

@ARTICLE{wahhaj2024,
       author = {{Wahhaj}, Z. and {Benisty}, M. and {Ginski}, C. and {Swastik}, C. and {Arora}, S. and {van Holstein}, R.~G. and {De Rosa}, R. and {Yang}, B. and {Bae}, J. and {Ren}, B.},
        title = "{PDS 70 unveiled by star-hopping: Total intensity, polarimetry, and millimeter imaging modeled in concert}",
      journal = {\aap},
         year = 2024,
        month = jul,
       volume = {687},
          eid = {A257},
        pages = {A257},
          doi = {10.1051/0004-6361/202349018},
archivePrefix = {arXiv},
       eprint = {2404.11641},
 primaryClass = {astro-ph.EP},
       adsurl = {https://ui.adsabs.harvard.edu/abs/2024A&A...687A.257W}
}

@ARTICLE{bae2019,
       author = {{Bae}, Jaehan and {Zhu}, Zhaohuan and {Baruteau}, Cl{\'e}ment and {Benisty}, Myriam and {Dullemond}, Cornelis P. and {Facchini}, Stefano and {Isella}, Andrea and {Keppler}, Miriam and {P{\'e}rez}, Laura M. and {Teague}, Richard},
        title = "{An Ideal Testbed for Planet-Disk Interaction: Two Giant Protoplanets in Resonance Shaping the PDS 70 Protoplanetary Disk}",
      journal = {\apjl},
         year = 2019,
        month = oct,
       volume = {884},
       number = {2},
          eid = {L41},
        pages = {L41},
          doi = {10.3847/2041-8213/ab46b0},
archivePrefix = {arXiv},
       eprint = {1909.09476},
 primaryClass = {astro-ph.EP},
       adsurl = {https://ui.adsabs.harvard.edu/abs/2019ApJ...884L..41B}
}

@ARTICLE{garrido2023,
       author = {{Garrido-Deutelmoser}, Juan and {Petrovich}, Cristobal and {Charalambous}, Carolina and {Guzm{\'a}n}, Viviana V. and {Zhang}, Ke},
        title = "{A Gap-sharing Planet Pair Shaping the Crescent in HD 163296: A Disk Sculpted by a Resonant Chain}",
      journal = {\apjl},
         year = 2023,
        month = mar,
       volume = {945},
       number = {2},
          eid = {L37},
        pages = {L37},
          doi = {10.3847/2041-8213/acbea8},
archivePrefix = {arXiv},
       eprint = {2301.13260},
 primaryClass = {astro-ph.EP},
       adsurl = {https://ui.adsabs.harvard.edu/abs/2023ApJ...945L..37G}
}

@ARTICLE{keppler2019,
       author = {{Keppler}, M. and {Teague}, R. and {Bae}, J. and {Benisty}, M. and {Henning}, T. and {van Boekel}, R. and {Chapillon}, E. and {Pinilla}, P. and {Williams}, J.~P. and {Bertrang}, G.~H. -M. and {Facchini}, S. and {Flock}, M. and {Ginski}, Ch. and {Juhasz}, A. and {Klahr}, H. and {Liu}, Y. and {M{\"u}ller}, A. and {P{\'e}rez}, L.~M. and {Pohl}, A. and {Rosotti}, G. and {Samland}, M. and {Semenov}, D.},
        title = "{Highly structured disk around the planet host PDS 70 revealed by high-angular resolution observations with ALMA}",
      journal = {\aap},
         year = 2019,
        month = may,
       volume = {625},
          eid = {A118},
        pages = {A118},
          doi = {10.1051/0004-6361/201935034},
archivePrefix = {arXiv},
       eprint = {1902.07639},
 primaryClass = {astro-ph.EP},
       adsurl = {https://ui.adsabs.harvard.edu/abs/2019A&A...625A.118K}
}

@BOOK{meeus1998,
       author = {{Meeus}, J.},
        title = "{Astronomical algorithms}",
         year = 1998,
       adsurl = {https://ui.adsabs.harvard.edu/abs/1998aalg.book.....M}
}

@INPROCEEDINGS{rodrigo2020,
       author = {{Rodrigo}, C. and {Solano}, E.},
        title = "{The SVO Filter Profile Service}",
    booktitle = {XIV.0 Scientific Meeting (virtual) of the Spanish Astronomical Society},
         year = 2020,
        month = jul,
          eid = {182},
        pages = {182},
       adsurl = {https://ui.adsabs.harvard.edu/abs/2020sea..confE.182R}
}

@software{bradley2024,
  author       = {Larry Bradley and
                  Brigitta Sipőcz and
                  Thomas Robitaille and
                  Erik Tollerud and
                  Zé Vinícius and
                  Christoph Deil and
                  Kyle Barbary and
                  Tom J Wilson and
                  Ivo Busko and
                  Axel Donath and
                  Hans Moritz Günther and
                  Mihai Cara and
                  P. L. Lim and
                  Sebastian Meßlinger and
                  Simon Conseil and
                  Zach Burnett and
                  Azalee Bostroem and
                  Michael Droettboom and
                  E. M. Bray and
                  Lars Andersen Bratholm and
                  Adam Ginsburg and
                  William Jamieson and
                  Geert Barentsen and
                  Matt Craig and
                  Brett M. Morris and
                  Marshall Perrin and
                  Shivangee Rathi and
                  Sergio Pascual and
                  Iskren Y. Georgiev},
  title        = {astropy/photutils: 2.0.2},
  month        = oct,
  year         = 2024,
  publisher    = {Zenodo},
  version      = {2.0.2},
  doi          = {10.5281/zenodo.13989456},
  url          = {https://doi.org/10.5281/zenodo.13989456},
}

@software{synphot2018,
       author = {{STScI Development Team}},
        title = "{synphot: Synthetic photometry using Astropy}",
 howpublished = {Astrophysics Source Code Library, record ascl:1811.001},
         year = 2018,
        month = nov,
          eid = {ascl:1811.001},
archivePrefix = {ascl},
       eprint = {1811.001},
       adsurl = {https://ui.adsabs.harvard.edu/abs/2018ascl.soft11001S}
}

@ARTICLE{foreman2013,
       author = {{Foreman-Mackey}, Daniel and {Hogg}, David W. and {Lang}, Dustin and {Goodman}, Jonathan},
        title = "{emcee: The MCMC Hammer}",
      journal = {\pasp},
         year = 2013,
        month = mar,
       volume = {125},
       number = {925},
        pages = {306},
          doi = {10.1086/670067},
archivePrefix = {arXiv},
       eprint = {1202.3665},
 primaryClass = {astro-ph.IM},
       adsurl = {https://ui.adsabs.harvard.edu/abs/2013PASP..125..306F}
}

@ARTICLE{casassus2021,
       author = {{Casassus}, Simon and {Christiaens}, Valentin and {C{\'a}rcamo}, Miguel and {P{\'e}rez}, Sebasti{\'a}n and {Weber}, Philipp and {Ercolano}, Barbara and {van der Marel}, Nienke and {Pinte}, Christophe and {Dong}, Ruobing and {Baruteau}, Cl{\'e}ment and {Cieza}, Lucas and {van Dishoeck}, Ewine F. and {Jordan}, Andr{\'e}s and {Price}, Daniel J. and {Absil}, Olivier and {Arce-Tord}, Carla and {Faramaz}, Virginie and {Flores}, Christian and {Reggiani}, Maddalena},
        title = "{A dusty filament and turbulent CO spirals in HD 135344B - SAO 206462}",
      journal = {\mnras},
         year = 2021,
        month = nov,
       volume = {507},
       number = {3},
        pages = {3789-3809},
          doi = {10.1093/mnras/stab2359},
archivePrefix = {arXiv},
       eprint = {2104.08379},
 primaryClass = {astro-ph.EP},
       adsurl = {https://ui.adsabs.harvard.edu/abs/2021MNRAS.507.3789C}
}

@ARTICLE{2019A&A...631A...6L,
       author = {{Leleu}, Adrien and {Coleman}, Gavin A.~L. and {Ataiee}, Sareh},
        title = "{Stability of the co-orbital resonance under dissipation. Application to its evolution in protoplanetary discs}",
      journal = {\aap},
         year = 2019,
        month = nov,
       volume = {631},
          eid = {A6},
        pages = {A6},
          doi = {10.1051/0004-6361/201834486},
archivePrefix = {arXiv},
       eprint = {1901.07640},
 primaryClass = {astro-ph.EP},
       adsurl = {https://ui.adsabs.harvard.edu/abs/2019A&A...631A...6L}
}

@ARTICLE{dominguezjamet2025,
       author = {{Dom{\'\i}nguez-Jamett}, Oriana and {Casassus}, Simon and {Baobab Liu}, Hauyu and {Aoyama}, Yuhiko and {C{\'a}rcamo}, Miguel and {Weber}, Philipp and {Chrenko}, On{\v{d}}rej and {Marleau}, Gabriel-Dominique and {Ercolano}, Barbara and {Szul{\'a}gyi}, Judit},
        title = "{Multi-frequency observations of PDS70c: Radio emission mechanisms in the circumplanetary environment}",
      journal = {\aap},
         year = 2025,
        month = oct,
       volume = {702},
          eid = {A18},
        pages = {A18},
          doi = {10.1051/0004-6361/202554485},
archivePrefix = {arXiv},
       eprint = {2507.21970},
 primaryClass = {astro-ph.EP},
       adsurl = {https://ui.adsabs.harvard.edu/abs/2025A&A...702A..18D}
}

@ARTICLE{isella2019,
       author = {{Isella}, Andrea and {Benisty}, Myriam and {Teague}, Richard and {Bae}, Jaehan and {Keppler}, Miriam and {Facchini}, Stefano and {P{\'e}rez}, Laura},
        title = "{Detection of Continuum Submillimeter Emission Associated with Candidate Protoplanets}",
      journal = {\apjl},
         year = 2019,
        month = jul,
       volume = {879},
       number = {2},
          eid = {L25},
        pages = {L25},
          doi = {10.3847/2041-8213/ab2a12},
archivePrefix = {arXiv},
       eprint = {1906.06308},
 primaryClass = {astro-ph.EP},
       adsurl = {https://ui.adsabs.harvard.edu/abs/2019ApJ...879L..25I}
}

@ARTICLE{vip1,
       author = {{Gomez Gonzalez}, Carlos Alberto and {Wertz}, Olivier and {Absil}, Olivier and {Christiaens}, Valentin and {Defr{\`e}re}, Denis and {Mawet}, Dimitri and {Milli}, Julien and {Absil}, Pierre-Antoine and {Van Droogenbroeck}, Marc and {Cantalloube}, Faustine and {Hinz}, Philip M. and {Skemer}, Andrew J. and {Karlsson}, Mikael and {Surdej}, Jean},
        title = "{VIP: Vortex Image Processing Package for High-contrast Direct Imaging}",
      journal = {\aj},
         year = 2017,
        month = jul,
       volume = {154},
       number = {1},
          eid = {7},
        pages = {7},
          doi = {10.3847/1538-3881/aa73d7},
archivePrefix = {arXiv},
       eprint = {1705.06184},
 primaryClass = {astro-ph.IM},
       adsurl = {https://ui.adsabs.harvard.edu/abs/2017AJ....154....7G}
}

@ARTICLE{hammond2025,
       author = {{Hammond}, Iain and {Christiaens}, Valentin and {Price}, Daniel J. and {Blakely}, Dori and {Trevascus}, David and {Bonse}, Markus J. and {Cantalloube}, Faustine and {Marleau}, Gabriel-Dominique and {Pinte}, Christophe and {Juillard}, Sandrine and {Samland}, Matthias and {Thompson}, William and {Wallace}, Alex},
        title = "{Keplerian motion of a compact source orbiting the inner disc of PDS 70: a third protoplanet in resonance with b and c?}",
      journal = {\mnras},
         year = 2025,
        month = may,
       volume = {539},
       number = {2},
        pages = {1613-1627},
          doi = {10.1093/mnras/staf586},
archivePrefix = {arXiv},
       eprint = {2504.11127},
 primaryClass = {astro-ph.EP},
       adsurl = {https://ui.adsabs.harvard.edu/abs/2025MNRAS.539.1613H}
}

@ARTICLE{bonse2025,
       author = {{Bonse}, Markus J. and {Gebhard}, Timothy D. and {Dannert}, Felix A. and {Absil}, Olivier and {Cantalloube}, Faustine and {Christiaens}, Valentin and {Cugno}, Gabriele and {Garvin}, Emily O. and {Hayoz}, Jean and {Kasper}, Markus and {Matthews}, Elisabeth and {Sch{\"o}lkopf}, Bernhard and {Quanz}, Sascha P.},
        title = "{Use the 4S (Signal-Safe Speckle Subtraction): Explainable Machine Learning Reveals the Giant Exoplanet AF Lep b in High-contrast Imaging Data from 2011}",
      journal = {\aj},
         year = 2025,
        month = apr,
       volume = {169},
       number = {4},
          eid = {194},
        pages = {194},
          doi = {10.3847/1538-3881/adab79},
archivePrefix = {arXiv},
       eprint = {2406.01809},
 primaryClass = {astro-ph.EP},
       adsurl = {https://ui.adsabs.harvard.edu/abs/2025AJ....169..194B}
}

@ARTICLE{blakely2025,
       author = {{Blakely}, Dori and {Johnstone}, Doug and {Cugno}, Gabriele and {Sivaramakrishnan}, Anand and {Tuthill}, Peter and {Dong}, Ruobing and {Pope}, Benjamin J.~S. and {Albert}, Lo{\"\i}c and {Charles}, Max and {Cooper}, Rachel A. and {De Furio}, Matthew and {Desdoigts}, Louis and {Doyon}, Ren{\'e} and {Francis}, Logan and {Greenbaum}, Alexandra Z. and {Lafreni{\`e}re}, David and {Lloyd}, James P. and {Meyer}, Michael R. and {Pueyo}, Laurent and {Ray}, Shrishmoy and {S{\'a}nchez-Berm{\'u}dez}, Joel and {Soulain}, Anthony and {Thatte}, Deepashri and {Thompson}, William and {Vandal}, Thomas},
        title = "{The James Webb Interferometer: Space-based Interferometric Detections of PDS 70 b and c at 4.8 {\ensuremath{\mu}}m}",
      journal = {\aj},
         year = 2025,
        month = mar,
       volume = {169},
       number = {3},
          eid = {137},
        pages = {137},
          doi = {10.3847/1538-3881/ad9b94},
archivePrefix = {arXiv},
       eprint = {2404.13032},
 primaryClass = {astro-ph.EP},
       adsurl = {https://ui.adsabs.harvard.edu/abs/2025AJ....169..137B}
}

@ARTICLE{vip2,
       author = {{Christiaens}, Valentin and {Gonzalez}, Carlos and {Farkas}, Ralf and {Dahlqvist}, Carl-Henrik and {Nasedkin}, Evert and {Milli}, Julien and {Absil}, Olivier and {Ngo}, Henry and {Cantero}, Carles and {Rainot}, Alan and {Hammond}, Iain and {Bonse}, Markus and {Cantalloube}, Faustine and {Vigan}, Arthur and {Kompella}, Vijay and {Hancock}, Paul},
        title = "{VIP: A Python package for high-contrast imaging}",
      journal = {The Journal of Open Source Software},
         year = 2023,
        month = jan,
       volume = {8},
       number = {81},
          eid = {4774},
        pages = {4774},
          doi = {10.21105/joss.04774},
       adsurl = {https://ui.adsabs.harvard.edu/abs/2023JOSS....8.4774C}
}

@ARTICLE{chabrier2000,
       author = {{Chabrier}, G. and {Baraffe}, I. and {Allard}, F. and {Hauschildt}, P.},
        title = "{Evolutionary Models for Very Low-Mass Stars and Brown Dwarfs with Dusty Atmospheres}",
      journal = {\apj},
         year = 2000,
        month = oct,
       volume = {542},
       number = {1},
        pages = {464-472},
          doi = {10.1086/309513},
archivePrefix = {arXiv},
       eprint = {astro-ph/0005557},
 primaryClass = {astro-ph},
       adsurl = {https://ui.adsabs.harvard.edu/abs/2000ApJ...542..464C}
}

@ARTICLE{montesinos2021,
       author = {{Montesinos}, Mat{\'\i}as and {Cuello}, Nicol{\'a}s and {Olofsson}, Johan and {Cuadra}, Jorge and {Bayo}, Amelia and {Bertrang}, Gesa H.-M. and {Perrot}, Cl{\'e}ment},
        title = "{Radiative Scale Height and Shadows in Protoplanetary Disks}",
      journal = {\apj},
         year = 2021,
        month = mar,
       volume = {910},
       number = {1},
          eid = {31},
        pages = {31},
          doi = {10.3847/1538-4357/abe3fc},
archivePrefix = {arXiv},
       eprint = {2102.02874},
 primaryClass = {astro-ph.EP},
       adsurl = {https://ui.adsabs.harvard.edu/abs/2021ApJ...910...31M}
}

@ARTICLE{akansoy2025,
       author = {{Akansoy}, Deniz and {Petrou}, Helen and {Ballabio}, Giulia and {Penzlin}, Anna},
        title = "{Modelling shadows in scattered light observations as signals from companions in protoplanetary discs}",
      journal = {\mnras},
         year = 2025,
        month = jul,
       volume = {540},
       number = {4},
        pages = {3186-3203},
          doi = {10.1093/mnras/staf925},
archivePrefix = {arXiv},
       eprint = {2506.04415},
 primaryClass = {astro-ph.EP},
       adsurl = {https://ui.adsabs.harvard.edu/abs/2025MNRAS.540.3186A}
}

@ARTICLE{stolker2016,
       author = {{Stolker}, T. and {Dominik}, C. and {Avenhaus}, H. and {Min}, M. and {de Boer}, J. and {Ginski}, C. and {Schmid}, H.~M. and {Juhasz}, A. and {Bazzon}, A. and {Waters}, L.~B.~F.~M. and {Garufi}, A. and {Augereau}, J.-C. and {Benisty}, M. and {Boccaletti}, A. and {Henning}, Th. and {Langlois}, M. and {Maire}, A.-L. and {M{\'e}nard}, F. and {Meyer}, M.~R. and {Pinte}, C. and {Quanz}, S.~P. and {Thalmann}, C. and {Beuzit}, J.-L. and {Carbillet}, M. and {Costille}, A. and {Dohlen}, K. and {Feldt}, M. and {Gisler}, D. and {Mouillet}, D. and {Pavlov}, A. and {Perret}, D. and {Petit}, C. and {Pragt}, J. and {Rochat}, S. and {Roelfsema}, R. and {Salasnich}, B. and {Soenke}, C. and {Wildi}, F.},
        title = "{Shadows cast on the transition disk of HD 135344B. Multiwavelength VLT/SPHERE polarimetric differential imaging}",
      journal = {\aap},
         year = 2016,
        month = nov,
       volume = {595},
          eid = {A113},
        pages = {A113},
          doi = {10.1051/0004-6361/201528039},
archivePrefix = {arXiv},
       eprint = {1603.00481},
 primaryClass = {astro-ph.EP},
       adsurl = {https://ui.adsabs.harvard.edu/abs/2016A&A...595A.113S}
}

@ARTICLE{marino2015,
       author = {{Marino}, S. and {Perez}, S. and {Casassus}, S.},
        title = "{Shadows Cast by a Warp in the HD 142527 Protoplanetary Disk}",
      journal = {\apjl},
         year = 2015,
        month = jan,
       volume = {798},
       number = {2},
          eid = {L44},
        pages = {L44},
          doi = {10.1088/2041-8205/798/2/L44},
archivePrefix = {arXiv},
       eprint = {1412.4632},
 primaryClass = {astro-ph.EP},
       adsurl = {https://ui.adsabs.harvard.edu/abs/2015ApJ...798L..44M}
}

@ARTICLE{canovas2013,
       author = {{Canovas}, H. and {M{\'e}nard}, F. and {Hales}, A. and {Jord{\'a}n}, A. and {Schreiber}, M.~R. and {Casassus}, S. and {Gledhill}, T.~M. and {Pinte}, C.},
        title = "{Near-infrared imaging polarimetry of HD 142527}",
      journal = {\aap},
         year = 2013,
        month = aug,
       volume = {556},
          eid = {A123},
        pages = {A123},
          doi = {10.1051/0004-6361/201321924},
archivePrefix = {arXiv},
       eprint = {1306.6379},
 primaryClass = {astro-ph.SR},
       adsurl = {https://ui.adsabs.harvard.edu/abs/2013A&A...556A.123C}
}

@ARTICLE{benisty17,
       author = {{Benisty}, M. and {Stolker}, T. and {Pohl}, A. and {de Boer}, J. and {Lesur}, G. and {Dominik}, C. and {Dullemond}, C.~P. and {Langlois}, M. and {Min}, M. and {Wagner}, K. and {Henning}, T. and {Juhasz}, A. and {Pinilla}, P. and {Facchini}, S. and {Apai}, D. and {van Boekel}, R. and {Garufi}, A. and {Ginski}, C. and {M{\'e}nard}, F. and {Pinte}, C. and {Quanz}, S.~P. and {Zurlo}, A. and {Boccaletti}, A. and {Bonnefoy}, M. and {Beuzit}, J.~L. and {Chauvin}, G. and {Cudel}, M. and {Desidera}, S. and {Feldt}, M. and {Fontanive}, C. and {Gratton}, R. and {Kasper}, M. and {Lagrange}, A.-M. and {LeCoroller}, H. and {Mouillet}, D. and {Mesa}, D. and {Sissa}, E. and {Vigan}, A. and {Antichi}, J. and {Buey}, T. and {Fusco}, T. and {Gisler}, D. and {Llored}, M. and {Magnard}, Y. and {Moeller-Nilsson}, O. and {Pragt}, J. and {Roelfsema}, R. and {Sauvage}, J.-F. and {Wildi}, F.},
        title = "{Shadows and spirals in the protoplanetary disk HD 100453}",
      journal = {\aap},
         year = 2017,
        month = jan,
       volume = {597},
          eid = {A42},
        pages = {A42},
          doi = {10.1051/0004-6361/201629798},
archivePrefix = {arXiv},
       eprint = {1610.10089},
 primaryClass = {astro-ph.EP},
       adsurl = {https://ui.adsabs.harvard.edu/abs/2017A&A...597A..42B}
}

@ARTICLE{lagrange2025,
       author = {{Lagrange}, A. -M. and {Wilkinson}, C. and {M{\^a}lin}, M. and {Boccaletti}, A. and {Perrot}, C. and {Matr{\`a}}, L. and {Combes}, F. and {Beust}, H. and {Rouan}, D. and {Chomez}, A. and {Milli}, J. and {Charnay}, B. and {Mazevet}, S. and {Flasseur}, O. and {Olofsson}, J. and {Bayo}, A. and {Kral}, Q. and {Carter}, A. and {Crotts}, K.~A. and {Delorme}, P. and {Chauvin}, G. and {Thebault}, P. and {Rubini}, P. and {Kiefer}, F. and {Radcliffe}, A. and {Mazoyer}, J. and {Bodrito}, T. and {Stasevic}, S. and {Langlois}, M.},
        title = "{Evidence for a sub-Jovian planet in the young TWA 7 disk}",
      journal = {\nat},
         year = 2025,
        month = jun,
       volume = {642},
       number = {8069},
        pages = {905-908},
          doi = {10.1038/s41586-025-09150-4},
archivePrefix = {arXiv},
       eprint = {2502.15081},
 primaryClass = {astro-ph.EP},
       adsurl = {https://ui.adsabs.harvard.edu/abs/2025Natur.642..905L}
}

@ARTICLE{pairet2019,
       author = {{Pairet}, Beno{\^\i}t and {Cantalloube}, Faustine and {Gomez Gonzalez}, Carlos A. and {Absil}, Olivier and {Jacques}, Laurent},
        title = "{STIM map: detection map for exoplanets imaging beyond asymptotic Gaussian residual speckle noise}",
      journal = {\mnras},
         year = 2019,
        month = aug,
       volume = {487},
       number = {2},
        pages = {2262-2277},
          doi = {10.1093/mnras/stz1350},
archivePrefix = {arXiv},
       eprint = {1810.06895},
 primaryClass = {astro-ph.IM},
       adsurl = {https://ui.adsabs.harvard.edu/abs/2019MNRAS.487.2262P}
}

@ARTICLE{tremblin2015,
       author = {{Tremblin}, P. and {Amundsen}, D.~S. and {Mourier}, P. and {Baraffe}, I. and {Chabrier}, G. and {Drummond}, B. and {Homeier}, D. and {Venot}, O.},
        title = "{Fingering Convection and Cloudless Models for Cool Brown Dwarf Atmospheres}",
      journal = {\apjl},
         year = 2015,
        month = may,
       volume = {804},
       number = {1},
          eid = {L17},
        pages = {L17},
          doi = {10.1088/2041-8205/804/1/L17},
archivePrefix = {arXiv},
       eprint = {1504.03334},
 primaryClass = {astro-ph.SR},
       adsurl = {https://ui.adsabs.harvard.edu/abs/2015ApJ...804L..17T}
}

@ARTICLE{stolker2020,
       author = {{Stolker}, T. and {Quanz}, S.~P. and {Todorov}, K.~O. and {K{\"u}hn}, J. and {Molli{\`e}re}, P. and {Meyer}, M.~R. and {Currie}, T. and {Daemgen}, S. and {Lavie}, B.},
        title = "{MIRACLES: atmospheric characterization of directly imaged planets and substellar companions at 4-5 {\ensuremath{\mu}}m. I. Photometric analysis of {\ensuremath{\beta}} Pic b, HIP 65426 b, PZ Tel B, and HD 206893 B}",
      journal = {\aap},
         year = 2020,
        month = mar,
       volume = {635},
          eid = {A182},
        pages = {A182},
          doi = {10.1051/0004-6361/201937159},
archivePrefix = {arXiv},
       eprint = {1912.13316},
 primaryClass = {astro-ph.EP},
       adsurl = {https://ui.adsabs.harvard.edu/abs/2020A&A...635A.182S}
}

@ARTICLE{hammerz2019,
       author = {{Hammer}, Michael and {Pinilla}, Paola and {Kratter}, Kaitlin M. and {Lin}, Min-Kai},
        title = "{Observational diagnostics of elongated planet-induced vortices with realistic planet formation time-scales}",
      journal = {\mnras},
         year = 2019,
        month = jan,
       volume = {482},
       number = {3},
        pages = {3609-3621},
          doi = {10.1093/mnras/sty2946},
archivePrefix = {arXiv},
       eprint = {1809.07391},
 primaryClass = {astro-ph.EP},
       adsurl = {https://ui.adsabs.harvard.edu/abs/2019MNRAS.482.3609H}
}

@ARTICLE{esposito2013,
       author = {{Esposito}, S. and {Mesa}, D. and {Skemer}, A. and {Arcidiacono}, C. and {Claudi}, R.~U. and {Desidera}, S. and {Gratton}, R. and {Mannucci}, F. and {Marzari}, F. and {Masciadri}, E. and {Close}, L. and {Hinz}, P. and {Kulesa}, C. and {McCarthy}, D. and {Males}, J. and {Agapito}, G. and {Argomedo}, J. and {Boutsia}, K. and {Briguglio}, R. and {Brusa}, G. and {Busoni}, L. and {Cresci}, G. and {Fini}, L. and {Fontana}, A. and {Guerra}, J.~C. and {Hill}, J.~M. and {Miller}, D. and {Paris}, D. and {Pinna}, E. and {Puglisi}, A. and {Quiros-Pacheco}, F. and {Riccardi}, A. and {Stefanini}, P. and {Testa}, V. and {Xompero}, M. and {Woodward}, C.},
        title = "{LBT observations of the HR 8799 planetary system. First detection of HR 8799e in H band}",
      journal = {\aap},
         year = 2013,
        month = jan,
       volume = {549},
          eid = {A52},
        pages = {A52},
          doi = {10.1051/0004-6361/201219212},
archivePrefix = {arXiv},
       eprint = {1203.2735},
 primaryClass = {astro-ph.EP},
       adsurl = {https://ui.adsabs.harvard.edu/abs/2013A&A...549A..52E}
}

@ARTICLE{zurlo2022,
       author = {{Zurlo}, A. and {Go{\'z}dziewski}, K. and {Lazzoni}, C. and {Mesa}, D. and {Nogueira}, P. and {Desidera}, S. and {Gratton}, R. and {Marzari}, F. and {Langlois}, M. and {Pinna}, E. and {Chauvin}, G. and {Delorme}, P. and {Girard}, J.~H. and {Hagelberg}, J. and {Henning}, Th. and {Janson}, M. and {Rickman}, E. and {Kervella}, P. and {Avenhaus}, H. and {Bhowmik}, T. and {Biller}, B. and {Boccaletti}, A. and {Bonaglia}, M. and {Bonavita}, M. and {Bonnefoy}, M. and {Cantalloube}, F. and {Cheetham}, A. and {Claudi}, R. and {D'Orazi}, V. and {Feldt}, M. and {Galicher}, R. and {Ghose}, E. and {Lagrange}, A.-M. and {le Coroller}, H. and {Ligi}, R. and {Kasper}, M. and {Maire}, A.-L. and {Medard}, F. and {Meyer}, M. and {Peretti}, S. and {Perrot}, C. and {Puglisi}, A.~T. and {Rossi}, F. and {Rothberg}, B. and {Schmidt}, T. and {Sissa}, E. and {Vigan}, A. and {Wahhaj}, Z.},
        title = "{Orbital and dynamical analysis of the system around HR 8799. New astrometric epochs from VLT/SPHERE and LBT/LUCI}",
      journal = {\aap},
         year = 2022,
        month = oct,
       volume = {666},
          eid = {A133},
        pages = {A133},
          doi = {10.1051/0004-6361/202243862},
archivePrefix = {arXiv},
       eprint = {2207.10684},
 primaryClass = {astro-ph.EP},
       adsurl = {https://ui.adsabs.harvard.edu/abs/2022A&A...666A.133Z}
}

@ARTICLE{pourre2024,
       author = {{Pourr{\'e}}, N. and {Winterhalder}, T.~O. and {Le Bouquin}, J.-B. and {Lacour}, S. and {Bidot}, A. and {Nowak}, M. and {Maire}, A.-L. and {Mouillet}, D. and {Babusiaux}, C. and {Woillez}, J. and {Abuter}, R. and {Amorim}, A. and {Asensio-Torres}, R. and {Balmer}, W.~O. and {Benisty}, M. and {Berger}, J.-P. and {Beust}, H. and {Blunt}, S. and {Boccaletti}, A. and {Bonnefoy}, M. and {Bonnet}, H. and {Bordoni}, M.~S. and {Bourdarot}, G. and {Brandner}, W. and {Cantalloube}, F. and {Caselli}, P. and {Charnay}, B. and {Chauvin}, G. and {Chavez}, A. and {Choquet}, E. and {Christiaens}, V. and {Cl{\'e}net}, Y. and {du Foresto}, V. Coud{\'e} and {Cridland}, A. and {Davies}, R. and {Defr{\`e}re}, D. and {Dembet}, R. and {Dexter}, J. and {Drescher}, A. and {Duvert}, G. and {Eckart}, A. and {Eisenhauer}, F. and {Schreiber}, N.~M. F{\"o}rster and {Garcia}, P. and {Lopez}, R. Garcia and {Gendron}, E. and {Genzel}, R. and {Gillessen}, S. and {Girard}, J.~H. and {Gonte}, F. and {Grant}, S. and {Haubois}, X. and {Hei{\ss}el}, G. and {Henning}, Th. and {Hinkley}, S. and {Hippler}, S. and {H{\"o}nig}, S.~F. and {Houll{\'e}}, M. and {Hubert}, Z. and {Jocou}, L. and {Kammerer}, J. and {Kenworthy}, M. and {Keppler}, M. and {Kervella}, P. and {Kreidberg}, L. and {Kurtovic}, N.~T. and {Lagrange}, A.-M. and {Lapeyr{\`e}re}, V. and {Lutz}, D. and {Mang}, F. and {Marleau}, G.-D. and {M{\'e}rand}, A. and {Millour}, F. and {Molli{\`e}re}, P. and {Monnier}, J.~D. and {Mordasini}, C. and {Nasedkin}, E. and {Oberti}, S. and {Ott}, T. and {Otten}, G.~P.~P.~L. and {Paladini}, C. and {Paumard}, T. and {Perraut}, K. and {Perrin}, G. and {Pfuhl}, O. and {Pueyo}, L. and {Ribeiro}, D.~C. and {Rickman}, E. and {Rustamkulov}, Z. and {Shangguan}, J. and {Shimizu}, T. and {Sing}, D. and {Soulez}, F. and {Stadler}, J. and {Stolker}, T. and {Straub}, O. and {Straubmeier}, C. and {Sturm}, E. and {Sykes}, C. and {Tacconi}, L.~J. and {van Dishoeck}, E.~F. and {Vigan}, A. and {Vincent}, F. and {von Fellenberg}, S.~D. and {Wang}, J.~J. and {Widmann}, F. and {Yazici}, S. and {Abad}, J.~A. and {Carpentier}, E. Aller and {Alonso}, J. and {Andolfato}, L. and {Barriga}, P. and {Beuzit}, J.-L. and {Bourget}, P. and {Brast}, R. and {Caniguante}, L. and {Cottalorda}, E. and {Darr{\'e}}, P. and {Delabre}, B. and {Delboulb{\'e}}, A. and {Delplancke-Str{\"o}bele}, F. and {Donaldson}, R. and {Dorn}, R. and {Dupuy}, C. and {Egner}, S. and {Fischer}, G. and {Frank}, C. and {Fuenteseca}, E. and {Gitton}, P. and {Guerlet}, T. and {Guieu}, S. and {Gutierrez}, P. and {Haguenauer}, P. and {Haimerl}, A. and {Heritier}, C.~T. and {Huber}, S. and {Hubin}, N. and {Jolley}, P. and {Kirchbauer}, J.-P. and {Kolb}, J. and {Kosmalski}, J. and {Krempl}, P. and {Le Louarn}, M. and {Lilley}, P. and {Lopez}, B. and {Magnard}, Y. and {Mclay}, S. and {Meilland}, A. and {Meister}, A. and {Moulin}, T. and {Pasquini}, L. and {Paufique}, J. and {Percheron}, I. and {Pettazzi}, L. and {Phan}, D. and {Pirani}, W. and {Quentin}, J. and {Rakich}, A. and {Ridings}, R. and {Reyes}, J. and {Rochat}, S. and {Schmid}, C. and {Schuhler}, N. and {Shchekaturov}, P. and {Seidel}, M. and {Soenke}, C. and {Stadler}, E. and {Stephan}, C. and {Su{\'a}rez}, M. and {Todorovic}, M. and {Valdes}, G. and {Verinaud}, C. and {Zins}, G. and {Z{\'u}{\~n}iga-Fern{\'a}ndez}, S.},
        title = "{High contrast at short separation with VLTI/GRAVITY: Bringing Gaia companions to light}",
      journal = {\aap},
         year = 2024,
        month = jun,
       volume = {686},
          eid = {A258},
        pages = {A258},
          doi = {10.1051/0004-6361/202449507},
archivePrefix = {arXiv},
       eprint = {2406.04003},
 primaryClass = {astro-ph.IM},
       adsurl = {https://ui.adsabs.harvard.edu/abs/2024A&A...686A.258P}
}

@ARTICLE{casassus2022,
       author = {{Casassus}, Simon and {C{\'a}rcamo}, Miguel},
        title = "{Variable structure in the PDS 70 disc and uncertainties in radio-interferometric image restoration}",
      journal = {\mnras},
         year = 2022,
        month = jul,
       volume = {513},
       number = {4},
        pages = {5790-5798},
          doi = {10.1093/mnras/stac1285},
archivePrefix = {arXiv},
       eprint = {2204.08589},
 primaryClass = {astro-ph.EP},
       adsurl = {https://ui.adsabs.harvard.edu/abs/2022MNRAS.513.5790C}
}

@ARTICLE{benisty2021,
       author = {{Benisty}, Myriam and {Bae}, Jaehan and {Facchini}, Stefano and {Keppler}, Miriam and {Teague}, Richard and {Isella}, Andrea and {Kurtovic}, Nicolas T. and {P{\'e}rez}, Laura M. and {Sierra}, Anibal and {Andrews}, Sean M. and {Carpenter}, John and {Czekala}, Ian and {Dominik}, Carsten and {Henning}, Thomas and {Menard}, Francois and {Pinilla}, Paola and {Zurlo}, Alice},
        title = "{A Circumplanetary Disk around PDS70c}",
      journal = {\apjl},
         year = 2021,
        month = jul,
       volume = {916},
       number = {1},
          eid = {L2},
        pages = {L2},
          doi = {10.3847/2041-8213/ac0f83},
archivePrefix = {arXiv},
       eprint = {2108.07123},
 primaryClass = {astro-ph.EP},
       adsurl = {https://ui.adsabs.harvard.edu/abs/2021ApJ...916L...2B}
}

@ARTICLE{haffert2019,
       author = {{Haffert}, S.~Y. and {Bohn}, A.~J. and {de Boer}, J. and {Snellen}, I.~A.~G. and {Brinchmann}, J. and {Girard}, J.~H. and {Keller}, C.~U. and {Bacon}, R.},
        title = "{Two accreting protoplanets around the young star PDS 70}",
      journal = {Nature Astronomy},
         year = 2019,
        month = jun,
       volume = {3},
        pages = {749-754},
          doi = {10.1038/s41550-019-0780-5},
archivePrefix = {arXiv},
       eprint = {1906.01486},
 primaryClass = {astro-ph.EP},
       adsurl = {https://ui.adsabs.harvard.edu/abs/2019NatAs...3..749H}
}

@ARTICLE{irdap,
       author = {{van Holstein}, R.~G. and {Girard}, J.~H. and {de Boer}, J. and {Snik}, F. and {Milli}, J. and {Stam}, D.~M. and {Ginski}, C. and {Mouillet}, D. and {Wahhaj}, Z. and {Schmid}, H.~M. and {Keller}, C.~U. and {Langlois}, M. and {Dohlen}, K. and {Vigan}, A. and {Pohl}, A. and {Carbillet}, M. and {Fantinel}, D. and {Maurel}, D. and {Orign{\'e}}, A. and {Petit}, C. and {Ramos}, J. and {Rigal}, F. and {Sevin}, A. and {Boccaletti}, A. and {Le Coroller}, H. and {Dominik}, C. and {Henning}, T. and {Lagadec}, E. and {M{\'e}nard}, F. and {Turatto}, M. and {Udry}, S. and {Chauvin}, G. and {Feldt}, M. and {Beuzit}, J. -L.},
        title = "{Polarimetric imaging mode of VLT/SPHERE/IRDIS. II. Characterization and correction of instrumental polarization effects}",
      journal = {\aap},
         year = 2020,
        month = jan,
       volume = {633},
          eid = {A64},
        pages = {A64},
          doi = {10.1051/0004-6361/201834996},
archivePrefix = {arXiv},
       eprint = {1909.13108},
 primaryClass = {astro-ph.IM},
       adsurl = {https://ui.adsabs.harvard.edu/abs/2020A&A...633A..64V}
}

@ARTICLE{wagner2018,
       author = {{Wagner}, Kevin and {Follete}, Katherine B. and {Close}, Laird M. and {Apai}, D{\'a}niel and {Gibbs}, Aidan and {Keppler}, Miriam and {M{\"u}ller}, Andr{\'e} and {Henning}, Thomas and {Kasper}, Markus and {Wu}, Ya-Lin and {Long}, Joseph and {Males}, Jared and {Morzinski}, Katie and {McClure}, Melissa},
        title = "{Magellan Adaptive Optics Imaging of PDS 70: Measuring the Mass Accretion Rate of a Young Giant Planet within a Gapped Disk}",
      journal = {\apjl},
         year = 2018,
        month = aug,
       volume = {863},
       number = {1},
          eid = {L8},
        pages = {L8},
          doi = {10.3847/2041-8213/aad695},
archivePrefix = {arXiv},
       eprint = {1807.10766},
 primaryClass = {astro-ph.EP},
       adsurl = {https://ui.adsabs.harvard.edu/abs/2018ApJ...863L...8W}
}

@ARTICLE{riaud2006,
       author = {{Riaud}, P. and {Mawet}, D. and {Absil}, O. and {Boccaletti}, A. and {Baudoz}, P. and {Herwats}, E. and {Surdej}, J.},
        title = "{Coronagraphic imaging of three weak-line T Tauri stars: evidence of planetary formation around PDS 70}",
      journal = {\aap},
         year = 2006,
        month = oct,
       volume = {458},
       number = {1},
        pages = {317-325},
          doi = {10.1051/0004-6361:20065232},
       adsurl = {https://ui.adsabs.harvard.edu/abs/2006A&A...458..317R}
}

@ARTICLE{dong2012,
       author = {{Dong}, Ruobing and {Hashimoto}, Jun and {Rafikov}, Roman and {Zhu}, Zhaohuan and {Whitney}, Barbara and {Kudo}, Tomoyuki and {Muto}, Takayuki and {Brandt}, Timothy and {McClure}, Melissa K. and {Wisniewski}, John and {Abe}, L. and {Brandner}, W. and {Carson}, J. and {Egner}, S. and {Feldt}, M. and {Goto}, M. and {Grady}, C. and {Guyon}, O. and {Hayano}, Y. and {Hayashi}, M. and {Hayashi}, S. and {Henning}, T. and {Hodapp}, K.~W. and {Ishii}, M. and {Iye}, M. and {Janson}, M. and {Kandori}, R. and {Knapp}, G.~R. and {Kusakabe}, N. and {Kuzuhara}, M. and {Kwon}, J. and {Matsuo}, T. and {McElwain}, M. and {Miyama}, S. and {Morino}, J.-I. and {Moro-Martin}, A. and {Nishimura}, T. and {Pyo}, T.-S. and {Serabyn}, E. and {Suto}, H. and {Suzuki}, R. and {Takami}, M. and {Takato}, N. and {Terada}, H. and {Thalmann}, C. and {Tomono}, D. and {Turner}, E. and {Watanabe}, M. and {Yamada}, T. and {Takami}, H. and {Usuda}, T. and {Tamura}, M.},
        title = "{The Structure of Pre-transitional Protoplanetary Disks. I. Radiative Transfer Modeling of the Disk+Cavity in the PDS 70 System}",
      journal = {\apj},
         year = 2012,
        month = dec,
       volume = {760},
       number = {2},
          eid = {111},
        pages = {111},
          doi = {10.1088/0004-637X/760/2/111},
archivePrefix = {arXiv},
       eprint = {1209.3772},
 primaryClass = {astro-ph.EP},
       adsurl = {https://ui.adsabs.harvard.edu/abs/2012ApJ...760..111D}
}

@ARTICLE{hashimoto2012,
       author = {{Hashimoto}, J. and {Dong}, R. and {Kudo}, T. and {Honda}, M. and {McClure}, M.~K. and {Zhu}, Z. and {Muto}, T. and {Wisniewski}, J. and {Abe}, L. and {Brandner}, W. and {Brandt}, T. and {Carson}, J. and {Egner}, S. and {Feldt}, M. and {Fukagawa}, M. and {Goto}, M. and {Grady}, C.~A. and {Guyon}, O. and {Hayano}, Y. and {Hayashi}, M. and {Hayashi}, S. and {Henning}, T. and {Hodapp}, K. and {Ishii}, M. and {Iye}, M. and {Janson}, M. and {Kandori}, R. and {Knapp}, G. and {Kusakabe}, N. and {Kuzuhara}, M. and {Kwon}, J. and {Matsuo}, T. and {Mayama}, S. and {McElwain}, M.~W. and {Miyama}, S. and {Morino}, J.-I. and {Moro-Martin}, A. and {Nishimura}, T. and {Pyo}, T.-S. and {Serabyn}, G. and {Suenaga}, T. and {Suto}, H. and {Suzuki}, R. and {Takahashi}, Y. and {Takami}, M. and {Takato}, N. and {Terada}, H. and {Thalmann}, C. and {Tomono}, D. and {Turner}, E.~L. and {Watanabe}, M. and {Yamada}, T. and {Takami}, H. and {Usuda}, T. and {Tamura}, M.},
        title = "{Polarimetric Imaging of Large Cavity Structures in the Pre-transitional Protoplanetary Disk around PDS 70: Observations of the Disk}",
      journal = {\apjl},
         year = 2012,
        month = oct,
       volume = {758},
       number = {1},
          eid = {L19},
        pages = {L19},
          doi = {10.1088/2041-8205/758/1/L19},
archivePrefix = {arXiv},
       eprint = {1208.2075},
 primaryClass = {astro-ph.SR},
       adsurl = {https://ui.adsabs.harvard.edu/abs/2012ApJ...758L..19H}
}

@ARTICLE{bohn2022,
       author = {{Bohn}, A.~J. and {Benisty}, M. and {Perraut}, K. and {van der Marel}, N. and {W{\"o}lfer}, L. and {van Dishoeck}, E.~F. and {Facchini}, S. and {Manara}, C.~F. and {Teague}, R. and {Francis}, L. and {Berger}, J.-P. and {Garcia-Lopez}, R. and {Ginski}, C. and {Henning}, T. and {Kenworthy}, M. and {Kraus}, S. and {M{\'e}nard}, F. and {M{\'e}rand}, A. and {P{\'e}rez}, L.~M.},
        title = "{Probing inner and outer disk misalignments in transition disks. Constraints from VLTI/GRAVITY and ALMA observations}",
      journal = {\aap},
         year = 2022,
        month = feb,
       volume = {658},
          eid = {A183},
        pages = {A183},
          doi = {10.1051/0004-6361/202142070},
archivePrefix = {arXiv},
       eprint = {2112.00123},
 primaryClass = {astro-ph.EP},
       adsurl = {https://ui.adsabs.harvard.edu/abs/2022A&A...658A.183B}
}

@ARTICLE{ma2024,
       author = {{Ma}, J. and {Ginski}, C. and {Tazaki}, R. and {Dominik}, C. and {Schmid}, H.~M. and {M{\'e}nard}, F.},
        title = "{Temporal and chromatic variation of polarized scattered light in the outer disk of PDS 70}",
      journal = {\aap},
         year = 2024,
        month = nov,
       volume = {691},
          eid = {L16},
        pages = {L16},
          doi = {10.1051/0004-6361/202451614},
archivePrefix = {arXiv},
       eprint = {2411.04091},
 primaryClass = {astro-ph.EP},
       adsurl = {https://ui.adsabs.harvard.edu/abs/2024A&A...691L..16M}
}

@ARTICLE{wang2021,
       author = {{Wang}, J.~J. and {Vigan}, A. and {Lacour}, S. and {Nowak}, M. and {Stolker}, T. and {De Rosa}, R.~J. and {Ginzburg}, S. and {Gao}, P. and {Abuter}, R. and {Amorim}, A. and {Asensio-Torres}, R. and {Baub{\"o}ck}, M. and {Benisty}, M. and {Berger}, J.~P. and {Beust}, H. and {Beuzit}, J. -L. and {Blunt}, S. and {Boccaletti}, A. and {Bohn}, A. and {Bonnefoy}, M. and {Bonnet}, H. and {Brandner}, W. and {Cantalloube}, F. and {Caselli}, P. and {Charnay}, B. and {Chauvin}, G. and {Choquet}, E. and {Christiaens}, V. and {Cl{\'e}net}, Y. and {Coud{\'e} Du Foresto}, V. and {Cridland}, A. and {de Zeeuw}, P.~T. and {Dembet}, R. and {Dexter}, J. and {Drescher}, A. and {Duvert}, G. and {Eckart}, A. and {Eisenhauer}, F. and {Facchini}, S. and {Gao}, F. and {Garcia}, P. and {Garcia Lopez}, R. and {Gardner}, T. and {Gendron}, E. and {Genzel}, R. and {Gillessen}, S. and {Girard}, J. and {Haubois}, X. and {Hei{\ss}el}, G. and {Henning}, T. and {Hinkley}, S. and {Hippler}, S. and {Horrobin}, M. and {Houll{\'e}}, M. and {Hubert}, Z. and {Jim{\'e}nez-Rosales}, A. and {Jocou}, L. and {Kammerer}, J. and {Keppler}, M. and {Kervella}, P. and {Meyer}, M. and {Kreidberg}, L. and {Lagrange}, A. -M. and {Lapeyr{\`e}re}, V. and {Le Bouquin}, J. -B. and {L{\'e}na}, P. and {Lutz}, D. and {Maire}, A. -L. and {M{\'e}nard}, F. and {M{\'e}rand}, A. and {Molli{\`e}re}, P. and {Monnier}, J.~D. and {Mouillet}, D. and {M{\"u}ller}, A. and {Nasedkin}, E. and {Ott}, T. and {Otten}, G.~P.~P.~L. and {Paladini}, C. and {Paumard}, T. and {Perraut}, K. and {Perrin}, G. and {Pfuhl}, O. and {Pueyo}, L. and {Rameau}, J. and {Rodet}, L. and {Rodr{\'\i}guez-Coira}, G. and {Rousset}, G. and {Scheithauer}, S. and {Shangguan}, J. and {Shimizu}, T. and {Stadler}, J. and {Straub}, O. and {Straubmeier}, C. and {Sturm}, E. and {Tacconi}, L.~J. and {van Dishoeck}, E.~F. and {Vincent}, F. and {von Fellenberg}, S.~D. and {Ward-Duong}, K. and {Widmann}, F. and {Wieprecht}, E. and {Wiezorrek}, E. and {Woillez}, J. and {Gravity Collaboration}},
        title = "{Constraining the Nature of the PDS 70 Protoplanets with VLTI/GRAVITY}",
      journal = {\aj},
         year = 2021,
        month = mar,
       volume = {161},
       number = {3},
          eid = {148},
        pages = {148},
          doi = {10.3847/1538-3881/abdb2d},
archivePrefix = {arXiv},
       eprint = {2101.04187},
 primaryClass = {astro-ph.EP},
       adsurl = {https://ui.adsabs.harvard.edu/abs/2021AJ....161..148W}
}

@ARTICLE{lagrange2010,
       author = {{Lagrange}, A.-M. and {Bonnefoy}, M. and {Chauvin}, G. and {Apai}, D. and {Ehrenreich}, D. and {Boccaletti}, A. and {Gratadour}, D. and {Rouan}, D. and {Mouillet}, D. and {Lacour}, S. and {Kasper}, M.},
        title = "{A Giant Planet Imaged in the Disk of the Young Star {\ensuremath{\beta}} Pictoris}",
      journal = {Science},
         year = 2010,
        month = jul,
       volume = {329},
       number = {5987},
        pages = {57},
          doi = {10.1126/science.1187187},
archivePrefix = {arXiv},
       eprint = {1006.3314},
 primaryClass = {astro-ph.EP},
       adsurl = {https://ui.adsabs.harvard.edu/abs/2010Sci...329...57L}
}

@ARTICLE{fasano2025,
       author = {{Fasano}, D. and {Benisty}, M. and {Curone}, P. and {Facchini}, S. and {Zagaria}, F. and {Yoshida}, T.~C. and {Doi}, K. and {Sierra}, A. and {Andrews}, S. and {Bae}, J. and {Isella}, A. and {Kurtovic}, N. and {P{\'e}rez}, L.~M. and {Pinilla}, P. and {Rampinelli}, L. and {Teague}, R.},
        title = "{Inner disc and circumplanetary material in the PDS 70 system: Insights from multi-epoch, multi-frequency ALMA observations}",
      journal = {\aap},
         year = 2025,
        month = jul,
       volume = {699},
          eid = {A373},
        pages = {A373},
          doi = {10.1051/0004-6361/202554959},
archivePrefix = {arXiv},
       eprint = {2506.11709},
 primaryClass = {astro-ph.EP},
       adsurl = {https://ui.adsabs.harvard.edu/abs/2025A&A...699A.373F}
}

@ARTICLE{muller2018,
       author = {{M{\"u}ller}, A. and {Keppler}, M. and {Henning}, Th. and {Samland}, M. and {Chauvin}, G. and {Beust}, H. and {Maire}, A. -L. and {Molaverdikhani}, K. and {van Boekel}, R. and {Benisty}, M. and {Boccaletti}, A. and {Bonnefoy}, M. and {Cantalloube}, F. and {Charnay}, B. and {Baudino}, J. -L. and {Gennaro}, M. and {Long}, Z.~C. and {Cheetham}, A. and {Desidera}, S. and {Feldt}, M. and {Fusco}, T. and {Girard}, J. and {Gratton}, R. and {Hagelberg}, J. and {Janson}, M. and {Lagrange}, A. -M. and {Langlois}, M. and {Lazzoni}, C. and {Ligi}, R. and {M{\'e}nard}, F. and {Mesa}, D. and {Meyer}, M. and {Molli{\`e}re}, P. and {Mordasini}, C. and {Moulin}, T. and {Pavlov}, A. and {Pawellek}, N. and {Quanz}, S.~P. and {Ramos}, J. and {Rouan}, D. and {Sissa}, E. and {Stadler}, E. and {Vigan}, A. and {Wahhaj}, Z. and {Weber}, L. and {Zurlo}, A.},
        title = "{Orbital and atmospheric characterization of the planet within the gap of the PDS 70 transition disk}",
      journal = {\aap},
         year = 2018,
        month = sep,
       volume = {617},
          eid = {L2},
        pages = {L2},
          doi = {10.1051/0004-6361/201833584},
archivePrefix = {arXiv},
       eprint = {1806.11567},
 primaryClass = {astro-ph.EP},
       adsurl = {https://ui.adsabs.harvard.edu/abs/2018A&A...617L...2M}
}

@ARTICLE{keppler2018,
       author = {{Keppler}, M. and {Benisty}, M. and {M{\"u}ller}, A. and {Henning}, Th. and {van Boekel}, R. and {Cantalloube}, F. and {Ginski}, C. and {van Holstein}, R.~G. and {Maire}, A. -L. and {Pohl}, A. and {Samland}, M. and {Avenhaus}, H. and {Baudino}, J. -L. and {Boccaletti}, A. and {de Boer}, J. and {Bonnefoy}, M. and {Chauvin}, G. and {Desidera}, S. and {Langlois}, M. and {Lazzoni}, C. and {Marleau}, G. -D. and {Mordasini}, C. and {Pawellek}, N. and {Stolker}, T. and {Vigan}, A. and {Zurlo}, A. and {Birnstiel}, T. and {Brandner}, W. and {Feldt}, M. and {Flock}, M. and {Girard}, J. and {Gratton}, R. and {Hagelberg}, J. and {Isella}, A. and {Janson}, M. and {Juhasz}, A. and {Kemmer}, J. and {Kral}, Q. and {Lagrange}, A. -M. and {Launhardt}, R. and {Matter}, A. and {M{\'e}nard}, F. and {Milli}, J. and {Molli{\`e}re}, P. and {Olofsson}, J. and {P{\'e}rez}, L. and {Pinilla}, P. and {Pinte}, C. and {Quanz}, S.~P. and {Schmidt}, T. and {Udry}, S. and {Wahhaj}, Z. and {Williams}, J.~P. and {Buenzli}, E. and {Cudel}, M. and {Dominik}, C. and {Galicher}, R. and {Kasper}, M. and {Lannier}, J. and {Mesa}, D. and {Mouillet}, D. and {Peretti}, S. and {Perrot}, C. and {Salter}, G. and {Sissa}, E. and {Wildi}, F. and {Abe}, L. and {Antichi}, J. and {Augereau}, J. -C. and {Baruffolo}, A. and {Baudoz}, P. and {Bazzon}, A. and {Beuzit}, J. -L. and {Blanchard}, P. and {Brems}, S.~S. and {Buey}, T. and {De Caprio}, V. and {Carbillet}, M. and {Carle}, M. and {Cascone}, E. and {Cheetham}, A. and {Claudi}, R. and {Costille}, A. and {Delboulb{\'e}}, A. and {Dohlen}, K. and {Fantinel}, D. and {Feautrier}, P. and {Fusco}, T. and {Giro}, E. and {Gluck}, L. and {Gry}, C. and {Hubin}, N. and {Hugot}, E. and {Jaquet}, M. and {Le Mignant}, D. and {Llored}, M. and {Madec}, F. and {Magnard}, Y. and {Martinez}, P. and {Maurel}, D. and {Meyer}, M. and {M{\"o}ller-Nilsson}, O. and {Moulin}, T. and {Mugnier}, L. and {Orign{\'e}}, A. and {Pavlov}, A. and {Perret}, D. and {Petit}, C. and {Pragt}, J. and {Puget}, P. and {Rabou}, P. and {Ramos}, J. and {Rigal}, F. and {Rochat}, S. and {Roelfsema}, R. and {Rousset}, G. and {Roux}, A. and {Salasnich}, B. and {Sauvage}, J. -F. and {Sevin}, A. and {Soenke}, C. and {Stadler}, E. and {Suarez}, M. and {Turatto}, M. and {Weber}, L.},
        title = "{Discovery of a planetary-mass companion within the gap of the transition disk around PDS 70}",
      journal = {\aap},
         year = 2018,
        month = sep,
       volume = {617},
          eid = {A44},
        pages = {A44},
          doi = {10.1051/0004-6361/201832957},
archivePrefix = {arXiv},
       eprint = {1806.11568},
 primaryClass = {astro-ph.EP},
       adsurl = {https://ui.adsabs.harvard.edu/abs/keppler2018}
}

@ARTICLE{nowak20,
       author = {{Nowak}, M. and {Lacour}, S. and {Lagrange}, A. -M. and {Rubini}, P. and {Wang}, J. and {Stolker}, T. and {Abuter}, R. and {Amorim}, A. and {Asensio-Torres}, R. and {Baub{\"o}ck}, M. and {Benisty}, M. and {Berger}, J.~P. and {Beust}, H. and {Blunt}, S. and {Boccaletti}, A. and {Bonnefoy}, M. and {Bonnet}, H. and {Brandner}, W. and {Cantalloube}, F. and {Charnay}, B. and {Choquet}, E. and {Christiaens}, V. and {Cl{\'e}net}, Y. and {Coud{\'e} Du Foresto}, V. and {Cridland}, A. and {de Zeeuw}, P.~T. and {Dembet}, R. and {Dexter}, J. and {Drescher}, A. and {Duvert}, G. and {Eckart}, A. and {Eisenhauer}, F. and {Gao}, F. and {Garcia}, P. and {Garcia Lopez}, R. and {Gardner}, T. and {Gendron}, E. and {Genzel}, R. and {Gillessen}, S. and {Girard}, J. and {Grandjean}, A. and {Haubois}, X. and {Hei{\ss}el}, G. and {Henning}, T. and {Hinkley}, S. and {Hippler}, S. and {Horrobin}, M. and {Houll{\'e}}, M. and {Hubert}, Z. and {Jim{\'e}nez-Rosales}, A. and {Jocou}, L. and {Kammerer}, J. and {Kervella}, P. and {Keppler}, M. and {Kreidberg}, L. and {Kulikauskas}, M. and {Lapeyr{\`e}re}, V. and {Le Bouquin}, J. -B. and {L{\'e}na}, P. and {M{\'e}rand}, A. and {Maire}, A. -L. and {Molli{\`e}re}, P. and {Monnier}, J.~D. and {Mouillet}, D. and {M{\"u}ller}, A. and {Nasedkin}, E. and {Ott}, T. and {Otten}, G. and {Paumard}, T. and {Paladini}, C. and {Perraut}, K. and {Perrin}, G. and {Pueyo}, L. and {Pfuhl}, O. and {Rameau}, J. and {Rodet}, L. and {Rodr{\'\i}guez-Coira}, G. and {Rousset}, G. and {Scheithauer}, S. and {Shangguan}, J. and {Stadler}, J. and {Straub}, O. and {Straubmeier}, C. and {Sturm}, E. and {Tacconi}, L.~J. and {van Dishoeck}, E.~F. and {Vigan}, A. and {Vincent}, F. and {von Fellenberg}, S.~D. and {Ward-Duong}, K. and {Widmann}, F. and {Wieprecht}, E. and {Wiezorrek}, E. and {Woillez}, J. and {GRAVITY Collaboration}},
        title = "{Direct confirmation of the radial-velocity planet {\ensuremath{\beta}} Pictoris c}",
      journal = {\aap},
         year = 2020,
        month = oct,
       volume = {642},
          eid = {L2},
        pages = {L2},
          doi = {10.1051/0004-6361/202039039},
archivePrefix = {arXiv},
       eprint = {2010.04442},
 primaryClass = {astro-ph.EP},
       adsurl = {https://ui.adsabs.harvard.edu/abs/2020A&A...642L...2N}
}

@ARTICLE{trevascus26,
       author = {{Trevascus}, David and {Brandner}, Wolfgang and {Balsalobre-Ruza}, Olga and {Lacour}, Sylvestre and {El Dayem}, Karim Abd and {Aimar}, Nicolas and {Berdeu}, Anthony and {Berger}, Jean-Philppe and {Bourdarot}, Guillaume and {Christiaens}, Valentin and {Correia}, Carlos and {Davies}, Richard and {Defr{\`e}re}, Denis and {Drescher}, Antonia and {Eckart}, Andreas and {Eisenhauer}, Frank and {Fabricius}, Maximilian and {Feuchtgruber}, Helmut and {Flesch}, Simon and {F{\"o}rster Schreiber}, Natascha M. and {Foschi}, Arianna and {Fournier}, Quentin and {Garcia}, Paulo and {Garcia Lopez}, Rebeca and {Genzel}, Reinhard and {Gillessen}, Stefan and {Hammond}, Iain and {H{\"o}nig}, Sebastian F. and {Houll{\'e}}, Mathias and {Joharle}, Simran and {Kervella}, Pierre and {Kreidberg}, Laura and {Labadie}, Lucas and {Lai}, Olivier and {Laugier}, Romain and {Le Bouquin}, Jean-Baptiste and {Leftley}, James and {Li}, Ruancun and {Lopez}, Bruno and {Lutz}, Dieter and {Marleau}, Gabriel-Dominique and {Mang}, Felix and {M{\'e}rand}, Antoine and {Millour}, Florentin and {Montarg{\`e}s}, Miguel and {Moruj{\~a}o}, Nuno and {Nowacki}, Hugo and {Nowak}, Mathias and {Osorno}, Juan and {Ott}, Thomas and {Pappert}, Sarah and {Paumard}, Thibaut and {Perraut}, Karine and {Perrin}, Guy and {Petrov}, Romain and {Petrucci}, Pierre-Olivier and {Pourr{\'e}}, Nicolas and {Rabien}, Sebastian and {Ribeiro}, Diogo C. and {Robbe-Dubois}, Sylvie and {Sadun Bordoni}, Matteo and {S{\'a}nchez Berm{\'u}dez}, Joel and {Santos}, Daryl and {Sauter}, Jonas and {Scigliuto}, Jules and {Shangguan}, Jinyi and {Shimizu}, Taro T. and {Soulez}, Ferr{\'e}ol and {Straubmeier}, Christian and {Sturm}, Eckhard and {Subroweit}, Matthias and {Sykes}, Calvin and {Tacconi}, Linda and {Th{\'e}venet}, Paloma and {Urso}, Irene and {Vincent}, Fr{\'e}d{\'e}ric and {Woillez}, Julien and {the GRAVITY+ Collaboration}},
        title = "{Using VLTI/GRAVITY+ to determine the identity of a third planet candidate in the PDS 70 system}",
      journal = {arXiv e-prints},
         year = 2026,
        month = jun,
          eid = {arXiv:2606.26249},
        pages = {arXiv:2606.26249},
          doi = {10.48550/arXiv.2606.26249},
archivePrefix = {arXiv},
       eprint = {2606.26249},
 primaryClass = {astro-ph.EP},
       adsurl = {https://ui.adsabs.harvard.edu/abs/2026arXiv260626249T}
}

@ARTICLE{trevascus2025,
       author = {{Trevascus}, David and {Blunt}, Sarah and {Christiaens}, Valentin and {Matthews}, Elisabeth and {Hammond}, Iain and {Brandner}, Wolfgang and {Wang}, Jason and {Lacour}, Sylvestre and {Vigan}, Arthur and {Balmer}, William O. and {Bonnefoy}, Mickael and {Burn}, Remo and {Chauvin}, Ga{\"e}l and {Gratton}, Raffaele and {Houll{\'e}}, Mathis and {Hinkley}, Sasha and {Kammerer}, Jens and {Kreidberg}, Laura and {Marleau}, Gabriel-Dominique and {Mesa}, Dino and {Otten}, Gilles and {Nowak}, Mathias and {Rickman}, Emily and {Sanchez-Bermudez}, Joel and {Sauter}, Jonas},
        title = "{Differentiating formation models with new dynamical masses for the PDS 70 protoplanets}",
      journal = {\aap},
         year = 2025,
        month = jun,
       volume = {698},
          eid = {A19},
        pages = {A19},
          doi = {10.1051/0004-6361/202553936},
archivePrefix = {arXiv},
       eprint = {2504.11210},
 primaryClass = {astro-ph.EP},
       adsurl = {https://ui.adsabs.harvard.edu/abs/2025A&A...698A..19T}
}

@ARTICLE{zhang2018,
       author = {{Zhang}, Shangjia and {Zhu}, Zhaohuan and {Huang}, Jane and {Guzm{\'a}n}, Viviana V. and {Andrews}, Sean M. and {Birnstiel}, Tilman and {Dullemond}, Cornelis P. and {Carpenter}, John M. and {Isella}, Andrea and {P{\'e}rez}, Laura M. and {Benisty}, Myriam and {Wilner}, David J. and {Baruteau}, Cl{\'e}ment and {Bai}, Xue-Ning and {Ricci}, Luca},
        title = "{The Disk Substructures at High Angular Resolution Project (DSHARP). VII. The Planet-Disk Interactions Interpretation}",
      journal = {\apjl},
         year = 2018,
        month = dec,
       volume = {869},
       number = {2},
          eid = {L47},
        pages = {L47},
          doi = {10.3847/2041-8213/aaf744},
archivePrefix = {arXiv},
       eprint = {1812.04045},
 primaryClass = {astro-ph.EP},
       adsurl = {https://ui.adsabs.harvard.edu/abs/2018ApJ...869L..47Z}
}

@ARTICLE{heron2026,
       author = {{Heron}, Agustin and {Petrovich}, Cristobal and {Ben{\'\i}tez-Llambay}, Pablo and {Garrido-Deutelmoser}, Juan},
        title = "{Origin of the Asymmetric Gas Distribution near the Co-orbital Lagrange Points of an Embedded Planet}",
      journal = {\apj},
         year = 2026,
        month = feb,
       volume = {998},
       number = {1},
          eid = {169},
        pages = {169},
          doi = {10.3847/1538-4357/ae3a89},
archivePrefix = {arXiv},
       eprint = {2505.07937},
 primaryClass = {astro-ph.EP},
       adsurl = {https://ui.adsabs.harvard.edu/abs/2026ApJ...998..169H}
}

@ARTICLE{long2022,
       author = {{Long}, Feng and {Andrews}, Sean M. and {Zhang}, Shangjia and {Qi}, Chunhua and {Benisty}, Myriam and {Facchini}, Stefano and {Isella}, Andrea and {Wilner}, David J. and {Bae}, Jaehan and {Huang}, Jane and {Loomis}, Ryan A. and {{\"O}berg}, Karin I. and {Zhu}, Zhaohuan},
        title = "{ALMA Detection of Dust Trapping around Lagrangian Points in the LkCa 15 Disk}",
      journal = {\apjl},
         year = 2022,
        month = sep,
       volume = {937},
       number = {1},
          eid = {L1},
        pages = {L1},
          doi = {10.3847/2041-8213/ac8b10},
archivePrefix = {arXiv},
       eprint = {2209.05535},
 primaryClass = {astro-ph.EP},
       adsurl = {https://ui.adsabs.harvard.edu/abs/2022ApJ...937L...1L}
}

@ARTICLE{rodenkirch2021,
       author = {{Rodenkirch}, P.~J. and {Rometsch}, T. and {Dullemond}, C.~P. and {Weber}, P. and {Kley}, W.},
        title = "{Modeling the nonaxisymmetric structure in the HD 163296 disk with planet-disk interaction}",
      journal = {\aap},
         year = 2021,
        month = mar,
       volume = {647},
          eid = {A174},
        pages = {A174},
          doi = {10.1051/0004-6361/202038484},
archivePrefix = {arXiv},
       eprint = {2012.09217},
 primaryClass = {astro-ph.EP},
       adsurl = {https://ui.adsabs.harvard.edu/abs/2021A&A...647A.174R}
}

@ARTICLE{isella2018,
       author = {{Isella}, Andrea and {Huang}, Jane and {Andrews}, Sean M. and {Dullemond}, Cornelis P. and {Birnstiel}, Tilman and {Zhang}, Shangjia and {Zhu}, Zhaohuan and {Guzm{\'a}n}, Viviana V. and {P{\'e}rez}, Laura M. and {Bai}, Xue-Ning and {Benisty}, Myriam and {Carpenter}, John M. and {Ricci}, Luca and {Wilner}, David J.},
        title = "{The Disk Substructures at High Angular Resolution Project (DSHARP). IX. A High-definition Study of the HD 163296 Planet-forming Disk}",
      journal = {\apjl},
         year = 2018,
        month = dec,
       volume = {869},
       number = {2},
          eid = {L49},
        pages = {L49},
          doi = {10.3847/2041-8213/aaf747},
archivePrefix = {arXiv},
       eprint = {1812.04047},
 primaryClass = {astro-ph.SR},
       adsurl = {https://ui.adsabs.harvard.edu/abs/2018ApJ...869L..49I}
}

@ARTICLE{andrews18,
       author = {{Andrews}, Sean M. and {Huang}, Jane and {P{\'e}rez}, Laura M. and {Isella}, Andrea and {Dullemond}, Cornelis P. and {Kurtovic}, Nicol{\'a}s T. and {Guzm{\'a}n}, Viviana V. and {Carpenter}, John M. and {Wilner}, David J. and {Zhang}, Shangjia and {Zhu}, Zhaohuan and {Birnstiel}, Tilman and {Bai}, Xue-Ning and {Benisty}, Myriam and {Hughes}, A. Meredith and {{\"O}berg}, Karin I. and {Ricci}, Luca},
        title = "{The Disk Substructures at High Angular Resolution Project (DSHARP). I. Motivation, Sample, Calibration, and Overview}",
      journal = {\apjl},
         year = 2018,
        month = dec,
       volume = {869},
       number = {2},
          eid = {L41},
        pages = {L41},
          doi = {10.3847/2041-8213/aaf741},
archivePrefix = {arXiv},
       eprint = {1812.04040},
 primaryClass = {astro-ph.SR},
       adsurl = {https://ui.adsabs.harvard.edu/abs/2018ApJ...869L..41A}
}

@INPROCEEDINGS{benisty2023,
       author = {{Benisty}, M. and {Dominik}, C. and {Follette}, K. and {Garufi}, A. and {Ginski}, C. and {Hashimoto}, J. and {Keppler}, M. and {Kley}, W. and {Monnier}, J.},
        title = "{Optical and Near-infrared View of Planet-forming Disks and Protoplanets}",
    booktitle = {Protostars and Planets VII},
         year = 2023,
       editor = {{Inutsuka}, S. and {Aikawa}, Y. and {Muto}, T. and {Tomida}, K. and {Tamura}, M.},
       series = {Astronomical Society of the Pacific Conference Series},
       volume = {534},
        month = jul,
        pages = {605},
          doi = {10.48550/arXiv.2203.09991},
archivePrefix = {arXiv},
       eprint = {2203.09991},
 primaryClass = {astro-ph.EP},
       adsurl = {https://ui.adsabs.harvard.edu/abs/2023ASPC..534..605B}
}

@ARTICLE{sierra2021,
       author = {{Sierra}, Anibal and {P{\'e}rez}, Laura M. and {Zhang}, Ke and {Law}, Charles J. and {Guzm{\'a}n}, Viviana V. and {Qi}, Chunhua and {Bosman}, Arthur D. and {{\"O}berg}, Karin I. and {Andrews}, Sean M. and {Long}, Feng and {Teague}, Richard and {Booth}, Alice S. and {Walsh}, Catherine and {Wilner}, David J. and {M{\'e}nard}, Fran{\c{c}}ois and {Cataldi}, Gianni and {Czekala}, Ian and {Bae}, Jaehan and {Huang}, Jane and {Bergner}, Jennifer B. and {Ilee}, John D. and {Benisty}, Myriam and {Le Gal}, Romane and {Loomis}, Ryan A. and {Tsukagoshi}, Takashi and {Liu}, Yao and {Yamato}, Yoshihide and {Aikawa}, Yuri},
        title = "{Molecules with ALMA at Planet-forming Scales (MAPS). XIV. Revealing Disk Substructures in Multiwavelength Continuum Emission}",
      journal = {\apjs},
         year = 2021,
        month = nov,
       volume = {257},
       number = {1},
          eid = {14},
        pages = {14},
          doi = {10.3847/1538-4365/ac1431},
archivePrefix = {arXiv},
       eprint = {2109.06433},
 primaryClass = {astro-ph.EP},
       adsurl = {https://ui.adsabs.harvard.edu/abs/2021ApJS..257...14S}
}

@article{bayo2008,
	adsurl = {http://adsabs.harvard.edu/abs/2008A%26A...492..277B},
	archiveprefix = {arXiv},
	author = {{Bayo}, A. and {Rodrigo}, C. and {Barrado y Navascu{\'e}s}, D. and {Solano}, E. and {Guti{\'e}rrez}, R. and {Morales-Calder{\'o}n}, M. and others},
	doi = {10.1051/0004-6361:200810395},
	eprint = {0808.0270},
	journal = {\aap},
	month = dec,
	pages = {277--287},
	title = {{VOSA: virtual observatory SED analyzer. An application to the Collinder 69 open cluster}},
	volume = 492,
	year = 2008}

@INPROCEEDINGS{castelli2003,
       author = {{Castelli}, F. and {Kurucz}, R.~L.},
        title = "{New Grids of ATLAS9 Model Atmospheres}",
    booktitle = {Modelling of Stellar Atmospheres},
         year = 2003,
       editor = {{Piskunov}, N. and {Weiss}, W.~W. and {Gray}, D.~F.},
       series = {IAU Symposium},
       volume = {210},
        month = jan,
        pages = {A20},
          doi = {10.48550/arXiv.astro-ph/0405087},
archivePrefix = {arXiv},
       eprint = {astro-ph/0405087},
 primaryClass = {astro-ph},
       adsurl = {https://ui.adsabs.harvard.edu/abs/2003IAUS..210P.A20C}
}

@ARTICLE{garufi2014,
       author = {{Garufi}, A. and {Quanz}, S.~P. and {Schmid}, H.~M. and {Avenhaus}, H. and {Buenzli}, E. and {Wolf}, S.},
        title = "{Shadows and cavities in protoplanetary disks: HD 163296, HD 141569A, and HD 150193A in polarized light}",
      journal = {\aap},
         year = 2014,
        month = aug,
       volume = {568},
          eid = {A40},
        pages = {A40},
          doi = {10.1051/0004-6361/201424262},
archivePrefix = {arXiv},
       eprint = {1406.7387},
 primaryClass = {astro-ph.SR},
       adsurl = {https://ui.adsabs.harvard.edu/abs/2014A&A...568A..40G}
}

@ARTICLE{montesinos2020,
       author = {{Montesinos}, Mat{\'\i}as and {Garrido-Deutelmoser}, Juan and {Olofsson}, Johan and {Giuppone}, Cristian A. and {Cuadra}, Jorge and {Bayo}, Amelia and {Sucerquia}, Mario and {Cuello}, Nicol{\'a}s},
        title = "{Dust trapping around Lagrangian points in protoplanetary disks}",
      journal = {\aap},
         year = 2020,
        month = oct,
       volume = {642},
          eid = {A224},
        pages = {A224},
          doi = {10.1051/0004-6361/202038758},
archivePrefix = {arXiv},
       eprint = {2009.10768},
 primaryClass = {astro-ph.EP},
       adsurl = {https://ui.adsabs.harvard.edu/abs/2020A&A...642A.224M}
}

@ARTICLE{2018A&A...609A..96L,
       author = {{Lillo-Box}, J. and {Barrado}, D. and {Figueira}, P. and {Leleu}, A. and {Santos}, N.~C. and {Correia}, A.~C.~M. and {Robutel}, P. and {Faria}, J.~P.},
        title = "{The TROY project: Searching for co-orbital bodies to known planets. I. Project goals and first results from archival radial velocity}",
      journal = {\aap},
         year = 2018,
        month = jan,
       volume = {609},
          eid = {A96},
        pages = {A96},
          doi = {10.1051/0004-6361/201730652},
archivePrefix = {arXiv},
       eprint = {1710.01138},
 primaryClass = {astro-ph.EP},
       adsurl = {https://ui.adsabs.harvard.edu/abs/2018A&A...609A..96L}
}

@ARTICLE{portillarevelo2023,
       author = {{Portilla-Revelo}, B. and {Kamp}, I. and {Facchini}, S. and {van Dishoeck}, E.~F. and {Law}, C. and {Rab}, Ch. and {Bae}, J. and {Benisty}, M. and {{\"O}berg}, K. and {Teague}, R.},
        title = "{Constraining the gas distribution in the PDS 70 disc as a method to assess the effect of planet-disc interactions}",
      journal = {\aap},
         year = 2023,
        month = sep,
       volume = {677},
          eid = {A76},
        pages = {A76},
          doi = {10.1051/0004-6361/202346607},
archivePrefix = {arXiv},
       eprint = {2306.16850},
 primaryClass = {astro-ph.EP},
       adsurl = {https://ui.adsabs.harvard.edu/abs/2023A&A...677A..76P}
}

@ARTICLE{ginski2025,
       author = {{Ginski}, C. and {Pinilla}, P. and {Benisty}, M. and {Pinte}, C. and {Claes}, R. and {Mamajek}, E. and {Kenworthy}, M. and {Murphy}, M. and {Manara}, C. and {Bae}, J. and {Birnstiel}, T. and {Byrne}, J. and {Dominik}, C. and {Facchini}, S. and {Garufi}, A. and {Gratton}, R. and {Hogerheijde}, M. and {van Holstein}, R. and {Huang}, J. and {Langlois}, M. and {Lawlor}, C. and {Ma}, J. and {McLachlan}, D. and {Menard}, F. and {Rigliaco}, R. and {Ribas}, A. and {Schmidt}, T. and {Sierra}, A. and {Tazaki}, R. and {Williams}, J. and {Zurlo}, A.},
        title = "{Disk Evolution Study Through Imaging of Nearby Young Stars (DESTINYS): Evidence of planet─disk interaction in the 2MASSJ16120668-3010270 system}",
      journal = {\aap},
         year = 2025,
        month = jul,
       volume = {699},
          eid = {A237},
        pages = {A237},
          doi = {10.1051/0004-6361/202451647},
archivePrefix = {arXiv},
       eprint = {2506.05892},
 primaryClass = {astro-ph.EP},
       adsurl = {https://ui.adsabs.harvard.edu/abs/2025A&A...699A.237G}
}

@ARTICLE{deboer2021,
       author = {{de Boer}, J. and {Ginski}, C. and {Chauvin}, G. and {M{\'e}nard}, F. and {Benisty}, M. and {Dominik}, C. and {Maaskant}, K. and {Girard}, J.~H. and {van der Plas}, G. and {Garufi}, A. and {Perrot}, C. and {Stolker}, T. and {Avenhaus}, H. and {Bohn}, A. and {Delboulb{\'e}}, A. and {Jaquet}, M. and {Buey}, T. and {M{\"o}ller-Nilsson}, O. and {Pragt}, J. and {Fusco}, T.},
        title = "{Possible single-armed spiral in the protoplanetary disk around HD 34282}",
      journal = {\aap},
         year = 2021,
        month = may,
       volume = {649},
          eid = {A25},
        pages = {A25},
          doi = {10.1051/0004-6361/201936787},
archivePrefix = {arXiv},
       eprint = {2010.12202},
 primaryClass = {astro-ph.EP},
       adsurl = {https://ui.adsabs.harvard.edu/abs/2021A&A...649A..25D}
}

@ARTICLE{dong2017,
       author = {{Dong}, Ruobing and {Fung}, Jeffrey},
        title = "{How Bright are Planet-induced Spiral Arms in Scattered Light?}",
      journal = {\apj},
         year = 2017,
        month = jan,
       volume = {835},
       number = {1},
          eid = {38},
        pages = {38},
          doi = {10.3847/1538-4357/835/1/38},
archivePrefix = {arXiv},
       eprint = {1612.00446},
 primaryClass = {astro-ph.EP},
       adsurl = {https://ui.adsabs.harvard.edu/abs/2017ApJ...835...38D}
}

@ARTICLE{beauge2007,
       author = {{Beaug{\'e}}, C. and {S{\'a}ndor}, Zs. and {{\'E}rdi}, B. and {S{\"u}li}, {\'A}.},
        title = "{Co-orbital terrestrial planets in exoplanetary systems: a formation scenario}",
      journal = {\aap},
         year = 2007,
        month = feb,
       volume = {463},
       number = {1},
        pages = {359-367},
          doi = {10.1051/0004-6361:20066582},
       adsurl = {https://ui.adsabs.harvard.edu/abs/2007A&A...463..359B}
}

@ARTICLE{gravity+2026,
       author = {{Gravity+ Collaboration} and {Abuter}, R. and {Allouche}, F. and {Bailet}, C. and {Benisty}, M. and {Berdeu}, A. and {Berger}, J.-P. and {Berio}, P. and {Bigioli}, A. and {Blanchard}, C. and {Boebion}, O. and {Bonnet}, H. and {Bourdarot}, G. and {Bourget}, P. and {Brandner}, W. and {Brul{\'e}}, J. and {Burgos}, P. and {Carbillet}, M. and {Correia}, C. and {Courtney-Barrer}, B. and {Curaba}, S. and {Davies}, R. and {Defr{\`e}re}, D. and {Delboulb{\'e}}, A. and {Delplancke}, F. and {Dembet}, R. and {Drescher}, A. and {Dubost}, N. and {Eckart}, A. and {{\'E}douard}, C. and {Eisenhauer}, F. and {Esteras Otal}, L. and {Fabricius}, M. and {Feuchtgruber}, H. and {F{\'e}dou}, P. and {Finger}, G. and {Schreiber}, N.~M. F{\"o}rster and {Frahm}, R. and {Garcia}, E. and {Garcia}, P. and {Lopez}, R. Garcia and {Genzel}, R. and {Gil}, J.~P. and {Gillessen}, S. and {Gomes}, T. and {Gont{\'e}}, F. and {Gopinath}, V. and {Gouvret}, C. and {Graf}, J. and {Guajardo}, P. and {Guieu}, S. and {Hackenberg}, W. and {Hartl}, M. and {Haubois}, X. and {Hau{\ss}mann}, F. and {Henning}, T. and {Hibon}, P. and {H{\"o}nig}, S. and {Horrobin}, M. and {Houll{\'e}}, M. and {Hubin}, N. and {Taieb}, I. Ibn and {Jochum}, L. and {Jocou}, L. and {Jost}, A. and {Kammerer}, J. and {Karl}, L. and {Kaufer}, A. and {Kern}, P. and {Kervella}, P. and {Kolb}, J. and {Korhonen}, H. and {Kreidberg}, L. and {Krempl}, P. and {Lacour}, S. and {Lagarde}, S. and {Lai}, O. and {Lapeyr{\`e}re}, V. and {Laugier}, R. and {Leal}, V. and {Le Bouquin}, J.-B. and {Leftley}, J. and {L{\'e}na}, P. and {Lopez}, B. and {Lutz}, D. and {Magnard}, Y. and {Mang}, F. and {Marcotto}, A. and {Maurel}, D. and {M{\'e}rand}, A. and {Millour}, F. and {Montarges}, M. and {More}, N. and {Moruj{\~a}o}, N. and {Moulin}, T. and {Nowacki}, H. and {Nowak}, M. and {Oberti}, S. and {Ott}, T. and {Pallanca}, L. and {Patru}, F. and {Paumard}, T. and {Perraut}, K. and {Perrin}, G. and {Petrucci}, P.~O. and {Petrov}, R. and {Pfuhl}, O. and {Pourr{\'e}}, N. and {Rabien}, S. and {Rau}, C. and {Riquelme}, M. and {Robbe-Dubois}, S. and {Rochat}, S. and {Salman}, M. and {S{\'a}nchez-Berm{\'u}dez}, J. and {Schubert}, J. and {Scigliuto}, J. and {Shchekaturov}, P. and {Schuhler}, N. and {Shangguan}, J. and {Shimizu}, T. and {Scheithauer}, S. and {Soenke}, C. and {Soulez}, F. and {Stadler}, E. and {Stadler}, J. and {Straubmeier}, C. and {Sturm}, E. and {Subroweit}, M. and {Sykes}, C. and {Tacconi}, L.~J. and {Tristram}, K.~R.~W. and {Uysal}, S. and {von Fellenberg}, S. and {Widmann}, F. and {Wieprecht}, E. and {Wiezorrek}, E. and {Woillez}, J. and {Yazici}, S. and {Zins}, G.},
        title = "{First light for the GRAVITY+ Adaptive Optics: Extreme adaptive optics for the Very Large Telescope Interferometer}",
      journal = {\aap},
         year = 2026,
        month = mar,
       volume = {707},
          eid = {A115},
        pages = {A115},
          doi = {10.1051/0004-6361/202555666},
archivePrefix = {arXiv},
       eprint = {2509.21431},
 primaryClass = {astro-ph.IM},
       adsurl = {https://ui.adsabs.harvard.edu/abs/2026A&A...707A.115G}
}

@ARTICLE{metis,
       author = {{Brandl}, B. and {Bettonvil}, F. and {van Boekel}, R. and {Glauser}, A. and {Quanz}, S. and {Absil}, O. and {Amorim}, A. and {Feldt}, M. and {Glasse}, A. and {G{\"u}del}, M. and {Ho}, P. and {Labadie}, L. and {Meyer}, M. and {Pantin}, E. and {van Winckel}, H. and {METIS Consortium}},
        title = "{METIS: The Mid-infrared ELT Imager and Spectrograph}",
      journal = {The Messenger},
         year = 2021,
        month = mar,
       volume = {182},
        pages = {22-26},
          doi = {10.18727/0722-6691/5218},
archivePrefix = {arXiv},
       eprint = {2103.11208},
 primaryClass = {astro-ph.IM},
       adsurl = {https://ui.adsabs.harvard.edu/abs/2021Msngr.182...22B}
}

@ARTICLE{garrido2022,
       author = {{Garrido-Deutelmoser}, Juan and {Petrovich}, Cristobal and {Krapp}, Leonardo and {Kratter}, Kaitlin M. and {Dong}, Ruobing},
        title = "{Substructures in Protoplanetary Disks Imprinted by Compact Planetary Systems}",
      journal = {\apj},
         year = 2022,
        month = jun,
       volume = {932},
       number = {1},
          eid = {41},
        pages = {41},
          doi = {10.3847/1538-4357/ac6bfd},
archivePrefix = {arXiv},
       eprint = {2204.09074},
 primaryClass = {astro-ph.EP},
       adsurl = {https://ui.adsabs.harvard.edu/abs/2022ApJ...932...41G}
}

@ARTICLE{garufi2026,
       author = {{Garufi}, A. and {Ginski}, C. and {Benisty}, M. and {Vioque}, M. and {Winter}, A. and {Huang}, J. and {Manara}, C.~F. and {Dominik}, C.},
        title = "{Planet-forming disks and their environment across regions and time from the full NIR census}",
      journal = {\aap},
         year = 2026,
        month = may,
       volume = {709},
          eid = {A269},
        pages = {A269},
          doi = {10.1051/0004-6361/202558522},
archivePrefix = {arXiv},
       eprint = {2603.01703},
 primaryClass = {astro-ph.SR},
       adsurl = {https://ui.adsabs.harvard.edu/abs/2026A&A...709A.269G}
}

@ARTICLE{bottke2023,
       author = {{Bottke}, William F. and {Marschall}, Raphael and {Nesvorn{\'y}}, David and {Vokrouhlick{\'y}}, David},
        title = "{Origin and Evolution of Jupiter's Trojan Asteroids}",
      journal = {\ssr},
         year = 2023,
        month = dec,
       volume = {219},
       number = {8},
          eid = {83},
        pages = {83},
          doi = {10.1007/s11214-023-01031-4},
archivePrefix = {arXiv},
       eprint = {2312.02864},
 primaryClass = {astro-ph.EP},
       adsurl = {https://ui.adsabs.harvard.edu/abs/2023SSRv..219...83B}
}

@ARTICLE{2018A&A...618A..42L,
       author = {{Lillo-Box}, J. and {Leleu}, A. and {Parviainen}, H. and {Figueira}, P. and {Mallonn}, M. and {Correia}, A.~C.~M. and {Santos}, N.~C. and {Robutel}, P. and {Lendl}, M. and {Boffin}, H.~M.~J. and {Faria}, J.~P. and {Barrado}, D. and {Neal}, J.},
        title = "{The TROY project. II. Multi-technique constraints on exotrojans in nine planetary systems}",
      journal = {\aap},
         year = 2018,
        month = oct,
       volume = {618},
          eid = {A42},
        pages = {A42},
          doi = {10.1051/0004-6361/201833312},
archivePrefix = {arXiv},
       eprint = {1807.00773},
 primaryClass = {astro-ph.EP},
       adsurl = {https://ui.adsabs.harvard.edu/abs/2018A&A...618A..42L}
}

@ARTICLE{2017A&A...599L...7L,
       author = {{Leleu}, A. and {Robutel}, P. and {Correia}, A.~C.~M. and {Lillo-Box}, J.},
        title = "{Detection of co-orbital planets by combining transit and radial-velocity measurements}",
      journal = {\aap},
         year = 2017,
        month = mar,
       volume = {599},
          eid = {L7},
        pages = {L7},
          doi = {10.1051/0004-6361/201630073},
archivePrefix = {arXiv},
       eprint = {1702.08775},
 primaryClass = {astro-ph.EP},
       adsurl = {https://ui.adsabs.harvard.edu/abs/2017A&A...599L...7L}
}

@BBOK{murray1999,
      author={{Murray} and {Dermott}},
      year=1999,
      title="{Solar System Dynamics}",
      doi={10.0117/CBO9781139174817}
        }

@ARTICLE{christiaens2024,
       author = {{Christiaens}, V. and {Samland}, M. and {Henning}, Th. and {Portilla-Revelo}, B. and {Perotti}, G. and {Matthews}, E. and {Absil}, O. and {Decin}, L. and {Kamp}, I. and {Boccaletti}, A. and {Tabone}, B. and {Marleau}, G. -D. and {van Dishoeck}, E.~F. and {G{\"u}del}, M. and {Lagage}, P. -O. and {Barrado}, D. and {Caratti o Garatti}, A. and {Glauser}, A.~M. and {Olofsson}, G. and {Ray}, T.~P. and {Scheithauer}, S. and {Vandenbussche}, B. and {Waters}, L.~B.~F.~M. and {Arabhavi}, A.~M. and {Grant}, S.~L. and {Jang}, H. and {Kanwar}, J. and {Schreiber}, J. and {Schwarz}, K. and {Temmink}, M. and {{\"O}stlin}, G.},
        title = "{MINDS: JWST/NIRCam imaging of the protoplanetary disk PDS 70. A spiral accretion stream and a potential third protoplanet}",
      journal = {\aap},
         year = 2024,
        month = may,
       volume = {685},
          eid = {L1},
        pages = {L1},
          doi = {10.1051/0004-6361/202349089},
archivePrefix = {arXiv},
       eprint = {2403.04855},
 primaryClass = {astro-ph.EP},
       adsurl = {https://ui.adsabs.harvard.edu/abs/2024A&A...685L...1C}
}

@ARTICLE{gregorio1992,
       author = {{Gregorio-Hetem}, J. and {Lepine}, J.~R.~D. and {Quast}, G.~R. and {Torres}, C.~A.~O. and {de La Reza}, R.},
        title = "{A Search for T Tauri Stars Based on the IRAS Point Source Catalog. I.}",
      journal = {\aj},
         year = 1992,
        month = feb,
       volume = {103},
        pages = {549},
          doi = {10.1086/116082},
       adsurl = {https://ui.adsabs.harvard.edu/abs/1992AJ....103..549G}
}

@ARTICLE{2002AJ....124..592L,
       author = {{Laughlin}, Gregory and {Chambers}, John E.},
        title = "{Extrasolar Trojans: The Viability and Detectability of Planets in the 1:1 Resonance}",
      journal = {\aj},
         year = 2002,
        month = jul,
       volume = {124},
       number = {1},
        pages = {592-600},
          doi = {10.1086/341173},
archivePrefix = {arXiv},
       eprint = {astro-ph/0204091},
 primaryClass = {astro-ph},
       adsurl = {https://ui.adsabs.harvard.edu/abs/2002AJ....124..592L}
}

@dataset{2003yCat.2246....0C,
       author = {{Cutri}, R.~M. and {Skrutskie}, M.~F. and {van Dyk}, S. and {Beichman}, C.~A. and {Carpenter}, J.~M. and {Chester}, T. and {Cambresy}, L. and {Evans}, T. and {Fowler}, J. and {Gizis}, J. and {Howard}, E. and {Huchra}, J. and {Jarrett}, T. and {Kopan}, E.~L. and {Kirkpatrick}, J.~D. and {Light}, R.~M. and {Marsh}, K.~A. and {McCallon}, H. and {Schneider}, S. and {Stiening}, R. and {Sykes}, M. and {Weinberg}, M. and {Wheaton}, W.~A. and {Wheelock}, S. and {Zacarias}, N.},
        title = "{VizieR Online Data Catalog: 2MASS All-Sky Catalog of Point Sources (Cutri+ 2003)}",
 howpublished = {VizieR On-line Data Catalog: II/246.  Originally published in: University of Massachusetts and Infrared Processing and Analysis Center, (IPAC/California Institute of Technology) (2003)},
         year = 2003,
        month = jun,
          eid = {II/246},
       adsurl = {https://ui.adsabs.harvard.edu/abs/2003yCat.2246....0C}
}

@dataset{2020yCat.1350....0G,
       author = {{Gaia Collaboration}},
        title = "{VizieR Online Data Catalog: Gaia EDR3 (Gaia Collaboration, 2020)}",
 howpublished = {VizieR On-line Data Catalog: I/350.  Originally published in: 2021A\&A...649A...1G},
         year = 2020,
        month = nov,
          eid = {I/350},
          doi = {10.26093/cds/vizier.1350},
       adsurl = {https://ui.adsabs.harvard.edu/abs/2020yCat.1350....0G}
}

@ARTICLE{jennings2022,
       author = {{Jennings}, Jeff and {Tazzari}, Marco and {Clarke}, Cathie J. and {Booth}, Richard A. and {Rosotti}, Giovanni P.},
        title = "{Superresolution trends in the ALMA Taurus survey: structured inner discs and compact discs}",
      journal = {\mnras},
         year = 2022,
        month = aug,
       volume = {514},
       number = {4},
        pages = {6053-6073},
          doi = {10.1093/mnras/stac1770},
archivePrefix = {arXiv},
       eprint = {2206.11308},
 primaryClass = {astro-ph.EP},
       adsurl = {https://ui.adsabs.harvard.edu/abs/2022MNRAS.514.6053J}
}

@ARTICLE{guerri2011,
       author = {{Guerri}, G{\'e}raldine and {Daban}, Jean-Baptiste and {Robbe-Dubois}, Sylvie and {Douet}, Richard and {Abe}, Lyu and {Baudrand}, Jacques and {Carbillet}, Marcel and {Boccaletti}, Anthony and {Bendjoya}, Philippe and {Gouvret}, Carole and {Vakili}, Farrokh},
        title = "{Apodized Lyot coronagraph for SPHERE/VLT: II. Laboratory tests and performance}",
      journal = {Experimental Astronomy},
         year = 2011,
        month = may,
       volume = {30},
       number = {1},
        pages = {59-81},
          doi = {10.1007/s10686-011-9220-y},
       adsurl = {https://ui.adsabs.harvard.edu/abs/2011ExA....30...59G}
}

@ARTICLE{lawlor2026,
       author = {{Lawlor}, Chloe and {van Capelleveen}, Richelle F. and {Bourdarot}, Guillaume and {Ginski}, Christian and {Kenworthy}, Matthew A. and {Stolker}, Tomas and {Close}, Laird and {Bohn}, Alexander J. and {Eisenhauer}, Frank and {Garcia}, Paulo and {H{\"o}nig}, Sebastian F. and {Kammerer}, Jens and {Kreidberg}, Laura and {Lacour}, Sylvestre and {Le Bouquin}, Jean-Baptiste and {Mamajek}, Eric and {Nowak}, Mathias and {Paumard}, Thibaut and {Straubmeier}, Christian and {van der Marel}, Nienke and {The Exogravity Collaboration}},
        title = "{Direct Spectroscopic Confirmation of the Young Embedded Protoplanet WISPIT 2c}",
      journal = {\apjl},
         year = 2026,
        month = apr,
       volume = {1000},
       number = {2},
          eid = {L38},
        pages = {L38},
          doi = {10.3847/2041-8213/ae4b3b},
archivePrefix = {arXiv},
       eprint = {2603.22085},
 primaryClass = {astro-ph.EP},
       adsurl = {https://ui.adsabs.harvard.edu/abs/2026ApJ..1000L..38L}
}

@ARTICLE{zhang2026,
       author = {{Zhang}, Zixin and {Wang}, Wenqin and {Ma}, Xinyue and {Chen}, Zhangliang and {Wang}, Yonghao and {Yu}, Cong and {Liu}, Shangfei and {Gao}, Yang and {Tang}, Baitian and {Chen}, Dichang and {Ma}, Bo},
        title = "{Constraining exotrojans in hot-Jupiter systems using transit timing variation observations from TESS}",
      journal = {\aap},
         year = 2026,
        month = apr,
       volume = {709},
          eid = {A35},
        pages = {A35},
          doi = {10.1051/0004-6361/202557704},
archivePrefix = {arXiv},
       eprint = {2603.27215},
 primaryClass = {astro-ph.EP},
       adsurl = {https://ui.adsabs.harvard.edu/abs/2026A&A...709A..35Z}
}

@ARTICLE{lyra2009,
       author = {{Lyra}, W. and {Johansen}, A. and {Klahr}, H. and {Piskunov}, N.},
        title = "{Standing on the shoulders of giants. Trojan Earths and vortex trapping in low mass self-gravitating protoplanetary disks of gas and solids}",
      journal = {\aap},
         year = 2009,
        month = jan,
       volume = {493},
       number = {3},
        pages = {1125-1139},
          doi = {10.1051/0004-6361:200810797},
archivePrefix = {arXiv},
       eprint = {0810.3192},
 primaryClass = {astro-ph},
       adsurl = {https://ui.adsabs.harvard.edu/abs/2009A&A...493.1125L}
}

@ARTICLE{garufi2024,
       author = {{Garufi}, A. and {Ginski}, C. and {van Holstein}, R.~G. and {Benisty}, M. and {Manara}, C.~F. and {P{\'e}rez}, S. and {Pinilla}, P. and {Ribas}, {\'A}. and {Weber}, P. and {Williams}, J. and {Cieza}, L. and {Dominik}, C. and {Facchini}, S. and {Huang}, J. and {Zurlo}, A. and {Bae}, J. and {Hagelberg}, J. and {Henning}, Th. and {Hogerheijde}, M.~R. and {Janson}, M. and {M{\'e}nard}, F. and {Messina}, S. and {Meyer}, M.~R. and {Pinte}, C. and {Quanz}, S.~P. and {Rigliaco}, E. and {Roccatagliata}, V. and {Schmid}, H.~M. and {Szul{\'a}gyi}, J. and {van Boekel}, R. and {Wahhaj}, Z. and {Antichi}, J. and {Baruffolo}, A. and {Moulin}, T.},
        title = "{The SPHERE view of the Taurus star-forming region. The full census of planet-forming disks with GTO and DESTINYS programs}",
      journal = {\aap},
         year = 2024,
        month = may,
       volume = {685},
          eid = {A53},
        pages = {A53},
          doi = {10.1051/0004-6361/202347586},
archivePrefix = {arXiv},
       eprint = {2403.02158},
 primaryClass = {astro-ph.GA},
       adsurl = {https://ui.adsabs.harvard.edu/abs/2024A&A...685A..53G}
}

@ARTICLE{close2025pds,
       author = {{Close}, Laird M. and {Males}, Jared R. and {Li}, Jialin and {Haffert}, Sebastiaan Y. and {Long}, Joseph D. and {Hedglen}, Alexander D. and {Weinberger}, Alycia J. and {Follette}, Katherine B. and {Apai}, Daniel and {Doyon}, Rene and {Foster}, Warren and {Gasho}, Victor and {Van Gorkom}, Kyle and {Guyon}, Olivier and {Kautz}, Maggie Y. and {Kueny}, Jay and {Lumbres}, Jennifer and {McLeod}, Avalon and {McEwen}, Eden and {Pavao}, Clarissa and {Pearce}, Logan and {Perez}, Laura and {Schatz}, Lauren and {Szul{\'a}gyi}, Judit and {Wagner}, Kevin and {Wu}, Ya-Lin},
        title = "{Three Years of High-contrast Imaging of the PDS 70 b and c Exoplanets at H{\ensuremath{\alpha}} with MagAO-X: Evidence of Strong Protoplanet H{\ensuremath{\alpha}} Variability and Circumplanetary Dust}",
      journal = {\aj},
         year = 2025,
        month = jan,
       volume = {169},
       number = {1},
          eid = {35},
        pages = {35},
          doi = {10.3847/1538-3881/ad8648},
archivePrefix = {arXiv},
       eprint = {2502.14038},
 primaryClass = {astro-ph.EP},
       adsurl = {https://ui.adsabs.harvard.edu/abs/2025AJ....169...35C}
}

@ARTICLE{izquierdo2026,
       author = {{Izquierdo}, Andr{\'e}s F. and {Bae}, Jaehan and {Galloway-Sprietsma}, Maria and {van Dishoeck}, Ewine F. and {Facchini}, Stefano and {Rosotti}, Giovanni and {Stadler}, Jochen and {Benisty}, Myriam and {Testi}, Leonardo},
        title = "{Circumplanetary Disk Candidate in the Disk of HD 163296 Traced by Localized Emission from Simple Organics}",
      journal = {\apjl},
         year = 2026,
        month = jan,
       volume = {997},
       number = {1},
          eid = {L2},
        pages = {L2},
          doi = {10.3847/2041-8213/ae2f59},
archivePrefix = {arXiv},
       eprint = {2601.10631},
 primaryClass = {astro-ph.EP},
       adsurl = {https://ui.adsabs.harvard.edu/abs/2026ApJ...997L...2I}
}

@ARTICLE{vioque2025,
       author = {{Vioque}, Miguel and {Kurtovic}, Nicol{\'a}s T. and {Trapman}, Leon and {Sierra}, Anibal and {P{\'e}rez}, Laura M. and {Zhang}, Ke and {Curone}, Pietro and {Rosotti}, Giovanni P. and {Carpenter}, John and {Tabone}, Beno{\^\i}t and {Pinilla}, Paola and {Deng}, Dingshan and {Pascucci}, Ilaria and {Miley}, James and {Agurto-Gangas}, Carolina and {Cieza}, Lucas A. and {Anania}, Rossella and {Ruiz-Rodriguez}, Dary A. and {Gonz{\'a}lez-Ruilova}, Camilo and {TorresVillanueva}, Estephani E. and {Kuznetsova}, Aleksandra},
        title = "{The ALMA Survey of Gas Evolution of PROtoplanetary Disks (AGE-PRO). X. Dust Substructures, Disk Geometries, and Dust-disk Radii}",
      journal = {\apj},
         year = 2025,
        month = aug,
       volume = {989},
       number = {1},
          eid = {9},
        pages = {9},
          doi = {10.3847/1538-4357/adc7b0},
archivePrefix = {arXiv},
       eprint = {2506.10746},
 primaryClass = {astro-ph.EP},
       adsurl = {https://ui.adsabs.harvard.edu/abs/2025ApJ...989....9V}
}

\begin{appendix}
\nolinenumbers

\section{IRDIS double-LOCI post-processing}
\label{sec:loci_reduction}

The IRDIS datasets described in Sect.~\ref{sec:irdis_ti_red} were also post-processed using a double-LOCI
approach, implemented as a custom \texttt{Python} version 
from the \texttt{IDL star-hopping} pipeline (\citealt{swastik2025}).
The first LOCI step produced a preliminary image used solely to create an accurate background mask isolating the circumstellar signal. 
For this, we subtracted from all PDS~70 frames a single PSF model built from the linear combination of all reference frames, optimized by minimizing the residuals (with respect to the basic ADI image) in an anchor region dominated by speckles. 
This region consisted of two annuli, one inside the disk cavity and inner to planet b emission, and an outer one covering the image outskirts where the disk emission is negligible. The resulting preliminary image was obtained by median-stacking all derotated PSF-subtracted science frames, and was then used to isolate the disk signal.
In the second LOCI step, we built an independent PSF model for each science frame by selecting its 20 best reference frames, identified via Root Mean Square (\textit{rms}) minimization within the background mask from the previous step. 
This approach typically selected the closest reference frames in time and with similar WDH orientation as the corresponding science frames, which effectively reduced both self- and over-subtraction.
The LOCI coefficients were then derived in the same anchor region as in the first step, composed of two annuli.
We optimized the annulus boundaries through a grid search, finding that the \textit{rms} of the final image was minimized when adopting a wider inner annulus, and an outer annulus placed further away (yet with a less significant effect).
Based on these results, we used an inner ring ranging $0.05\arcsec$-$0.12\arcsec$ (4-11 pixels; or from the center in the case of the non-coronagraphic datasets), and an outer ring of  $0.92\arcsec$-$1.53\arcsec$ (75-125 pixels).

\section{GRAVITY fiber pointings}
\label{sec:grav_point}

We estimate the location of the putative inner planet under the hypothesis that the N and S emissions trace dust trapped at its stable Lagrangian points (see Sect.~\ref{sec:astr_spec} and Sect.~\ref{sec:dis}).
We started from the posterior orbital distributions inferred by \citealt{hammond2025} for the N emission and retained those solutions consistent within 17\,mas of our 2024 IRDIS $H$-band astrometry for N and S (Table~\ref{tab:astr_spec}).
Following the preferred solution by those distributions, we fixed the eccentricity to zero, so that the \Lfour\ and \Lfive\ points are simply shifted by $\pm 60^\circ$ in mean anomaly ($M$) from the putative planet.

We then determined the true anomalies of N and S at the 2024 epoch by finding the solution for the sky-projected coordinates of the Keplerian orbit (e.g., \citealt{meeus1998}). 
Such true anomalies were then converted to mean anomalies using Kepler's equations (e.g., \citealt{murray1999}), and were used to predict the mean anomaly at the GRAVITY epoch according to
\begin{equation}
\label{eq:pred_M}
    M(t_{\rm pred}) = \frac{2\pi}{P}\left(t_{\rm pred} - t_{\rm ref} \right) + M(t_{\rm ref}),
\end{equation}
where $P$ is the orbital period. 

The predicted planet location was finally estimated by shifting the 2025 mean anomalies of N and S by $\pm 60 ^\circ$, respectively (clockwise orbital motion).
To maximize the chance of detection, we split the GRAVITY observations in two pointings, with the most probable location of the putative planet lying in the overlapping area (considering $\sim$30\,mas radius field of view; \citealt{wang2021}).
In total, these two pointings sample mean longitudes between +20$^\circ$ and +100$^\circ$ from N (or equivalently --20$^\circ$ and --100$^\circ$ from S).
We note that these pointings are slightly shifted in radial separation ($\sim$10\,mas) from the star to reduce contamination from the inner regions.
The pointing coordinates are shown in Table~\ref{tab:obs_grav} and Fig.~\ref{fig:astrometry}~and~\ref{fig:chi2_gravity}.

\section{Position angle of the shadow}
\label{sec:shadow_loc}
To isolate the local minima from the azimuthal profiles extracted in Sect.~\ref{fig:az_prof_shadow}, we subtracted a spline model from them fitted to the curve with the region of interest masked out. 
We note that this subtraction may affect the depth of the features, although their azimuthal position remains unchanged.
However, we found that these features exhibit a contrast roughly three times higher in the $J$ and $H$ bands as compared to the $K_1$ and $K_2$ bands, possibly as a result of the different angular resolution.

To infer the azimuthal location of this feature, we used a hierarchical model for these local minima.
This consisted on jointly modeling the eight curves as negative Gaussian profiles, with their centers belonging from a common Gaussian distribution. 
Each curve was described by the three parameters of their Gaussian (amplitude, center, and standard deviation),
and the two hyper-parameters were the population mean ($\mu$) and the population standard deviation ($\sigma$). 
This results in a total of 26 free parameters, whose posterior distributions were obtained using the Markov chain Monte Carlo (MCMC) affine invariant ensemble sampler \texttt{emcee}\footnote{\url{https://github.com/dfm/emcee}} (\citealt{foreman2013}).
We used Uniform priors of all parameters, with the amplitudes within three times the local minimum to 0, the medians (of each curve and the hyper-parameter) covering from 190$^\circ$ to 250$^\circ$, and the standard deviations (also from the curves and from the population) ranging 0.1$^\circ$ to 50$^\circ$.
We ran \texttt{emcee} with five times the number of free parameters as walkers, and 100 000 steps plus other additional 50 000 steps around the solution of the previous burn-in phase.
The convergence was reached, with all the chain lengths above 50 times the autocorrelation time.
Figure~\ref{fig:pa_shadow} shows the seven curves used in the modeling and the posterior distribution for the population mean, being $\mu = 220.74 ^{+0.52^\circ}_{-0.49^\circ}$.
This position is similar (difference of $\sim6^\circ$) to that from the putative planet (see Fig.~\ref{fig:shadow_pl}).

\section{Astrometry and photometry extraction}
\label{app:astro_phot}

To derive the astrometry and photometry of PDS\,70\,b, PDS\,70\,c, and the N \& S objects, 
we proceeded by injecting negative fake companions (\citealt{lagrange2010}) using the \texttt{VIP} package, the so-called NEGFC method.
The PSF from non-saturated frames is injected into the pre-processed cubes at an initial position and a negative flux.
The cubes are then post-processed using PCA, and the optimal position-flux pair is found by minimizing a given figure of merit.
This minimization is carried out using the Nelder-Mead simplex algorithm, which efficiently explores the parameter space of coordinates and flux.
For each epoch and object, we tested six definitions of the figure of merit to identify the most suitable one given the image quality, the extension of the source, and the presence of nearby emission, which is particularly relevant for PDS~70\,b and S due to their proximity. 
These included the absolute determinant of the Hessian matrix computed within 1, 2\,$\times$\,2, and 3\,$\times$\,3 pixel neighbors (\citealt{quanz2015}), and the standard deviation of the residual intensities in circular apertures with radius of 0.5, 1.0, and 1.5\,$\times$\,the full width at half maximum (FWHM) of the PSF (\citealt{wertz2017}).
We selected the final figure of merit for each case by verifying that no over-subtraction or residual emission was present in the post-processed image after injecting the negative companion with the best-fit parameters.

To run the NEGFC, we applied PCA-ARDI and PCA-RDI for the IRDIS and the IFS datasets, respectively. 
For IRDIS, we used a pre-selected range of principal components that maximize the signal of the corresponding object, whereas for IFS we adopted a single optimized value maximizing both the N and S signals. 
We applied the NEGFC to both the wavelength-collapsed IFS cubes ($YJH$-bands) and only collapsing the $H$-band channels, except for the 22 July 2025 epoch, where the $H$-band-only data had insufficient SNR.
For the 18 July 2025 dataset, we additionally extracted the astrometry and flux to each individual spectral channel in order to extract the N and S spectra, as both features are retrieved with a higher significance in this epoch.
The recovered flux was then corrected for the coronagraphic transmission when applicable.
In particular, at a separation of $\sim$100\,mas (corresponding to the locations of N and S), the coronagraphic throughput curves indicate a transmission of approximately 50\,\%.

To estimate the uncertainties associated to the speckle noise, we injected 200 fake companions into the pre-processed cubes also through \texttt{VIP}.
Each injection shared the radius and flux with the best-fit solution, while the azimuth was varied.
Since the N and S features lie at similar separations from the star, we excluded the azimuths corresponding to the opposite feature to avoid contamination.
For PDS~70\,c, the dominant source of uncertainty is the disk rather than the speckle noise. 
To account for this, we injected the companions along the inner edge of the bright side of the outer disk.
The NEGFC procedure was then executed, yielding a distribution of deviations between the injected parameters and the retrieved values.
The systematic uncertainties are also considered  by adding them in quadrature to the speckle noise (\citealt{wertz2017}).

We have calibrated the flux from the 2025 datasets using the spectro-photometric standard star \object{HD~131243}, observed at the end of the star-hopping sequences.
For this, we computed its synthetic photometry with \texttt{synphot}\footnote{\url{https://github.com/spacetelescope/synphot_refactor}} \citep[v1.6.0]{synphot2018}. 
The stellar flux in each band was derived by convolving the SED with the corresponding filter transmission curves.
The calibrator SED was obtained using the Virtual Observatory SED Analyzer (\texttt{VOSA}; \citealt{bayo2008}), which provided the ATLAS9-ODFNEW model (\citealt{castelli2003}) that best fits the multi-wavelength broadband photometry of the calibrator (17 bands from $Tycho$-2 $B_T$ to AllWISE $W4$).
The transmission curves for the IRDIS $K_1K_2$-bands were acquired from the SVO Filter Profile Service (\citealt{rodrigo2020}), while those for the IFS $YJH$-bands were taken from the SPHERE manual\footnote{\url{https://www.eso.org/sci/facilities/paranal/instruments/sphere/doc/}}.
Finally, the conversion factors per band to transform the observed counts into physical flux units were computed as the ratio between the synthetic flux and the aperture-photometry flux measured in the median frames with \texttt{photutils}, with an aperture diameter selected based on the growth curve ($\sim$1\arcsec).

As no standard star was observed for the remaining datasets, we calibrated the IRDIS $H$-band fluxes from 2021 and 2024 epochs using 2MASS $H$-band photometry.
We estimated a conversion factor as the ratio between the 2MASS flux and the aperture-photometry flux measured from the non-coronagraphic PDS 70 observations taken at the beginning of the run. 
Note that we did not calibrate the flux of the archival 2021 and 2022 IFS observations, given the lack of suitable photometric standards.

\section{SPHERE $K_1K_2$-band contrast curves}
\label{app:contr_curve}

The contrast curves were built using an annular variant of the \texttt{4S} algorithm (\citealt{bonse2025}). 
The contrast values were calculated by injecting 20 fake companions at each angular separation and iterating on the injection contrast in order to find the flux at which 19 out of the 20 companions are detected, constituting 95\% completeness limits. 
The detection threshold of a companion was defined using Standardized Trajectory Intensity Mean (STIM) maps (\citealt{pairet2019}), convolved with the FWHM to smooth out outlier pixels.
To be considered detected, a companion must have a STIM value 5$\sigma$ above the average value of the STIM map calculated with the inverse of the parallactic angle vector for the derotation of the data cube.
This 5$\sigma$ STIM threshold also varies with the angular separation as it was calculated in annuli 1.5~$\times$~FWHM wide centered on all separations.
Lastly, PDS~70\,b was subtracted from the data before computing the contrast curves in order to limit its influence on the result.
The resulting curves for the $K_1$ and $K_2$ IRDIS 22 Jul 2025 dataset are shown in Fig.~\ref{fig:contr_curve}.

\section{Self-cast shadow}
\label{sec:self_shadow}

A puffed-up inner disk is also capable of imprinting a shadow on the disk (e.g., \citealt{garufi2014}).
Support for this scenario comes from the dipper events reported in PDS~70 by \cite{gaidos2024}. 
To explain the observed variability, the authors proposed that episodes of enhanced magnetic activity lift dust above the inner disk plane, producing the optical dimming events detected with the Transiting Exoplanet Survey Satellite (TESS) and the associated infrared excess.
If the inner and outer disks are aligned (which remains uncertain, as discussed above), such vertically extended dust structures would be expected to attenuate the stellar illumination over a broad range of azimuths. 
This would likely imprint a relatively wide shadow on the outer disk, in contrast with the narrow feature detected in our data.
In polarized scattered light, \citet{ma2024} reported broad, variable shadows in the outer disk at different PA across epochs, which they attributed to this structural variability in the inner disk. 
This interpretation would not explain the much narrower shadow we detect.

\onecolumn 
\section{Additional Figures}
\FloatBarrier

\begin{figure*}[h!]
\centering\includegraphics[width=0.77\textwidth{}]{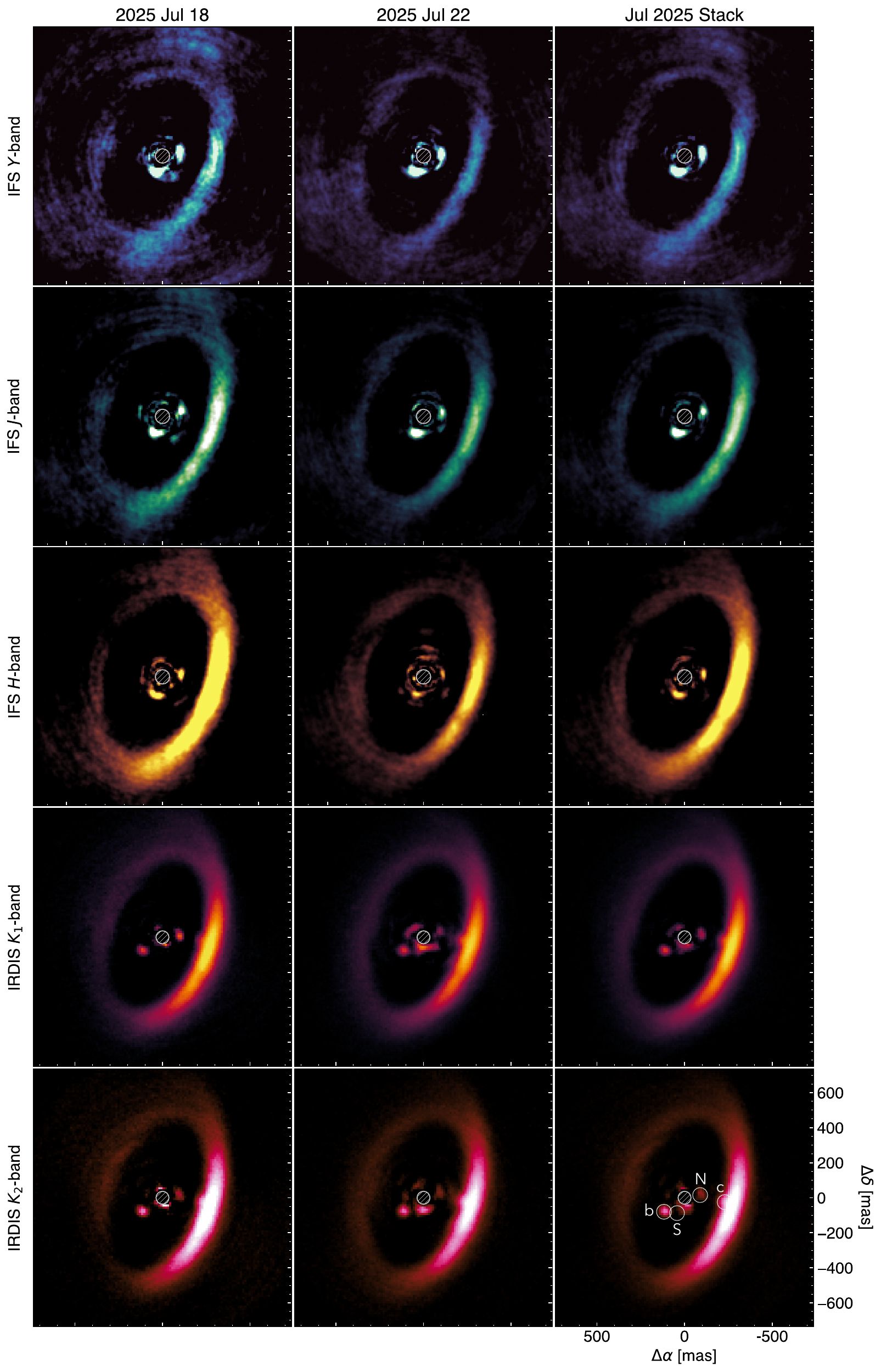}
\caption{Gallery of the simultaneous IRDIS and IFS images from 2025 post-processed using IPCA-ARDI and PCA-RSDI with \texttt{VIP}, respectively. 
The masked region corresponds to saturated pixels.
The detected emissions are labeled in the $K_2$-band combined image (we note S is not detected in $K_1$ and $K_2$ bands).
In all panels, North is up and East is to the left.}
\label{fig:irdifs}
\end{figure*}

\begin{figure}[h!]
\centering
\includegraphics[width=0.85\textwidth{}]{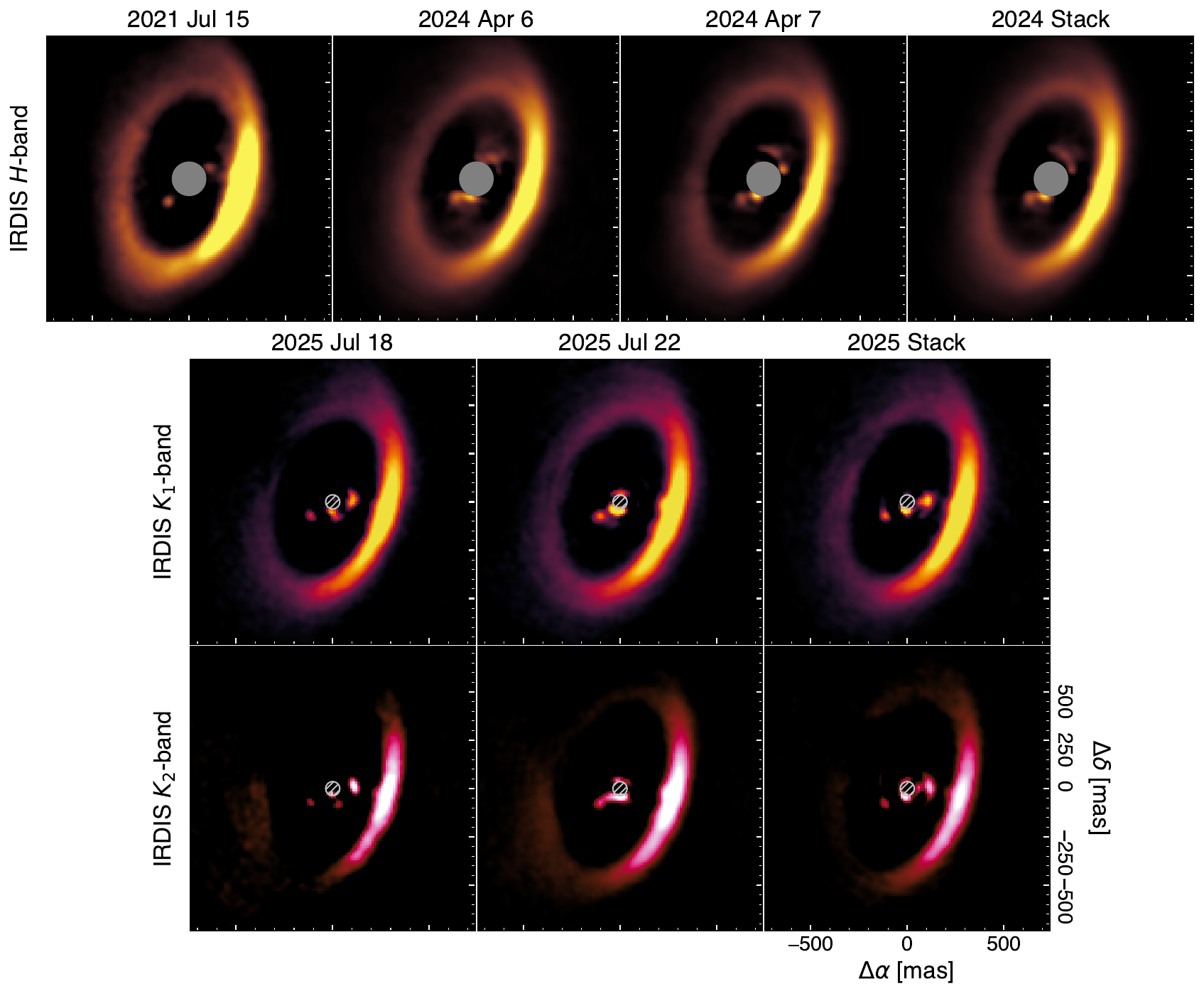}
\caption{IRDIS images post-processed via double LOCI.}
\label{fig:irdis_ti_2loci}
\end{figure}

\begin{figure}[h!]
\centering\includegraphics[width=0.85\textwidth{}]{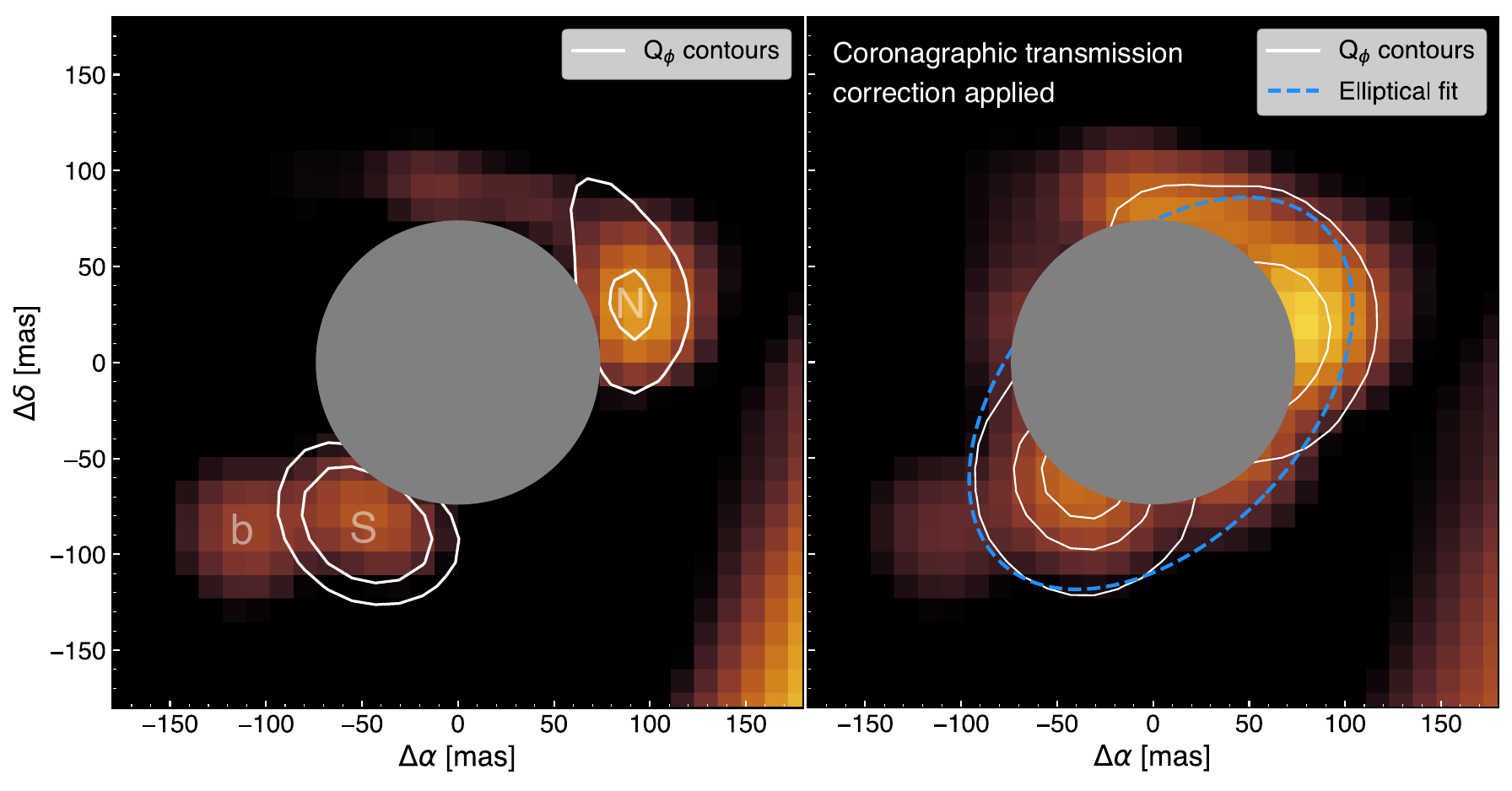}
\caption{\textit{Left:} comparison for the total-intensity (background) and Q$_{\phi}$ (contours) images corresponding to the combined 2024 IRDIS $H$-band dataset.
\textit{Right:} equivalent after applying the coronagraphic transmission correction to both images. Elliptical fit to the Q$_{\phi}$ emission is shown in dashed blue line.
The masked grey region corresponds to the coronagraphic mask.}
\label{fig:qp_ti}
\end{figure}

\begin{figure*}[t]
\centering

\noindent
\begin{minipage}{0.49\textwidth}
\centering
\vspace{1.2cm}
\includegraphics[width=0.85\textwidth{}]{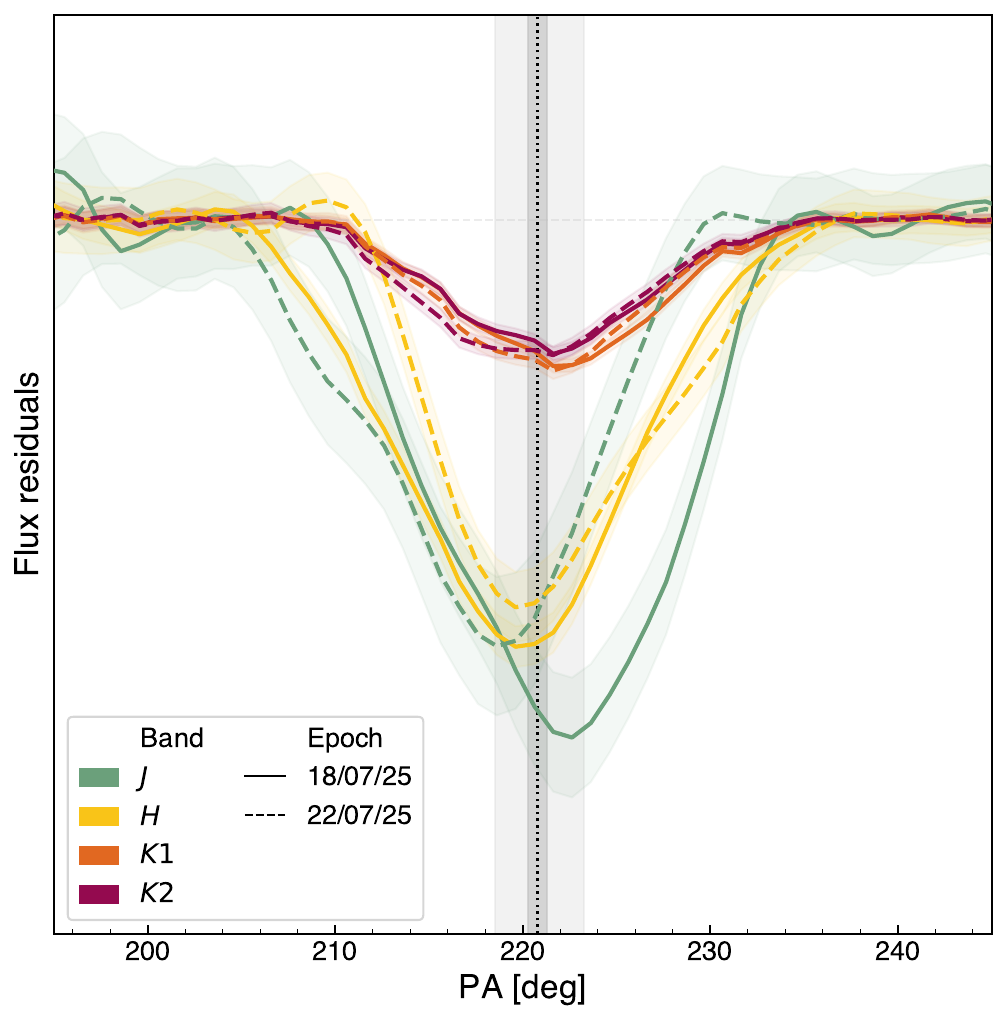}
\vspace{0.55cm}
\caption{July 2025 residual azimuthal profiles of the outer disk after subtracting a spline model, zoomed into the shadow region.
Colors indicate the band, and line styles the epoch as detailed in the legend.
Their associated shaded areas represent the 1-$\sigma$ uncertainties of the curves.
The vertical black-dotted line is the inferred population mean of the hierarchical model, with the grey shaded regions representing 1 (dark) and 2$\sigma$ (light) of its posterior distribution.}
\label{fig:pa_shadow}
\end{minipage}
\hfill
\begin{minipage}{0.49\textwidth}
\centering
\vspace{-0.7cm}
\includegraphics[width=0.9\textwidth{}]{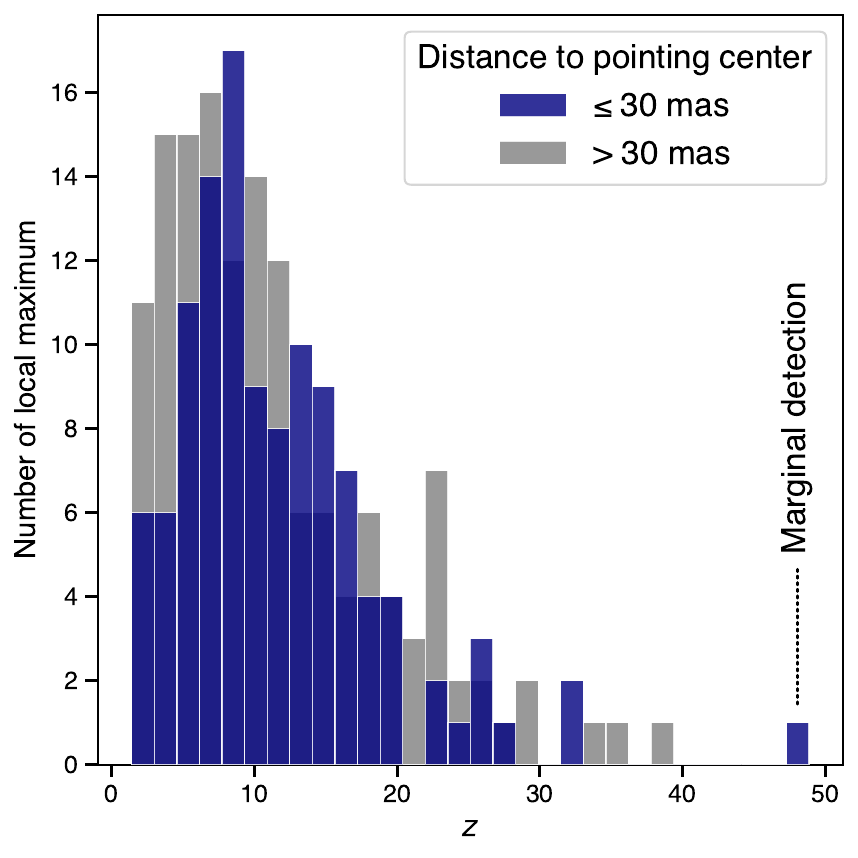}
\vspace{0.6cm}
\caption{Histogram of local maxima in the $z$ map from GRAVITY observations (Fig.~\ref{fig:chi2_gravity}). 
Blue and gray bars indicate maxima within and beyond 30\,mas of the nearest pointing center, respectively.}

\label{fig:distr_grav}
\end{minipage}

\vspace{1cm}

\noindent
\begin{minipage}{0.49\textwidth}
\centering
\includegraphics[width=1\textwidth{}]{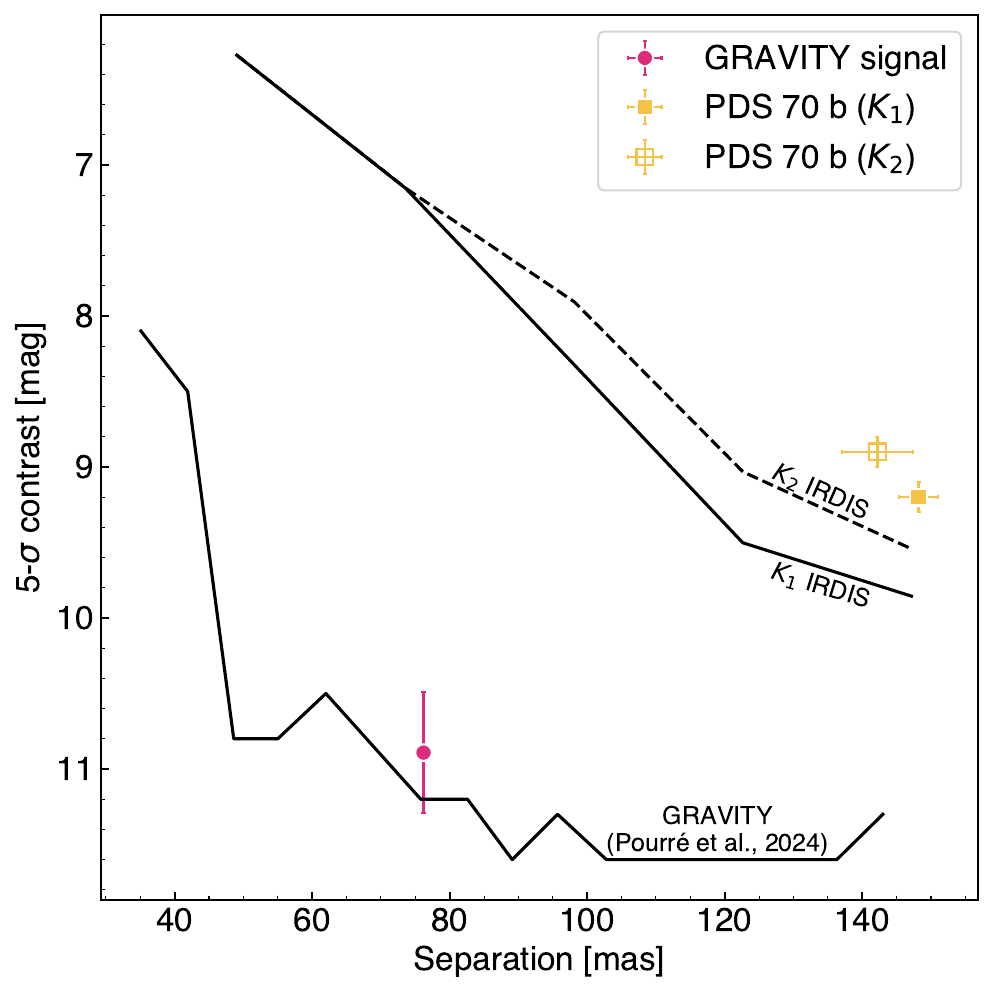}
\caption{Contrast curves extracted for the 22 Jul 2025 IRDIS dataset in $K_1$ and $K_2$-bands. The approximate contrast curve for the GRAVITY observations is also displayed (\citealt{pourre2024}).
}
\label{fig:contr_curve}
\end{minipage}
\hfill
\begin{minipage}{0.49\textwidth}
\centering
\vspace{-0.6cm}
\includegraphics[width=1\textwidth{}]{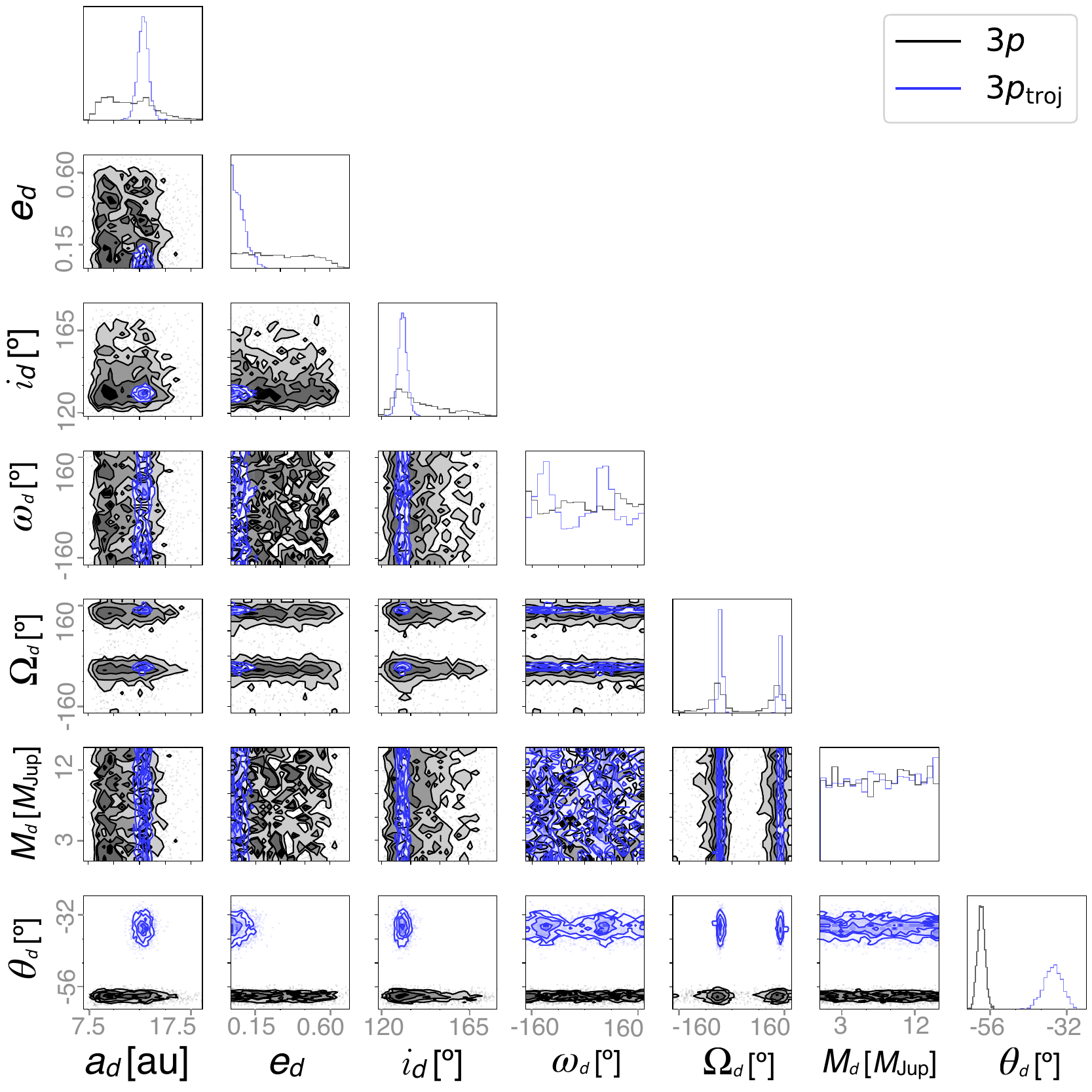}
\caption{Corner plot for the posteriors inferred for the third inner putative planet (here labeled $d$) in PDS~70.
The color identifies the model as shown in the legend.}
\label{fig:corner}
\end{minipage}

\end{figure*}

\FloatBarrier

\section{Tables}

\begin{table*}[h!]
\caption[]{Summary of the VLT/SPHERE observations.}
\label{tab:obs_sph}

\setlength{\tabcolsep}{2.5pt}
\begin{tabular}{lccccccccccc}
\hline 
\hline \noalign{\smallskip}
\multirow{2}{*}{Date\,$^a$} & \multirow{2}{*}{Program ID} & \multirow{2}{*}{Instrument} & \multirow{2}{*}{Strategy} & \multirow{2}{*}{Filter} & DIT\,$^b$ & On-source & On-ref & $\Delta$PA\,$^c$ & Pol. & SR\,$^e$ & $\tau_0\,^f$ \\ 
 & & & & & [s] & time [min] & time [min] & [deg] & cycles\,$^d$ & [\%] & [ms]  \\ \noalign{\smallskip}  
 \hline \noalign{\smallskip} 



 
\multirow{2}{*}{2025 May 28}$^\dagger$ & \multirow{6.6}{*}{115.29EH.002} & IFS & ADI/SDI/RDI & $YJH$ & 4 & 31 & 6 & \multirow{2}{*}{54} & \multirow{2}{*}{-} & \multirow{2}{*}{69\,$\pm$\,5} & \multirow{2}{*}{3.7\,--\,7.4} \\ 

 &  & IRDIS & ADI/RDI & $K_1K_2$ & 0.84 & 15$^\ddag$ & 3 &  &  &  &  \\ 

\noalign{\smallskip}

\multirow{2}{*}{2025 Jul 18} & & IFS & ADI/SDI/RDI & $YJH$ & 4 & 39 & 7 & \multirow{2}{*}{39} & \multirow{2}{*}{-} &  \multirow{2}{*}{65\,$\pm$\,6} & \multirow{2}{*}{3.0\,--\,4.8} \\  

 & & IRDIS & ADI/RDI & $K_1K_2$ & 0.84 & 26 & 3 &  &  & & \\

\noalign{\smallskip}

\multirow{2}{*}{2025 Jul 21} & & IFS & ADI/SDI/RDI & $YJH$ & 4 & 46 & 9 & \multirow{2}{*}{50} & \multirow{2}{*}{-} &  \multirow{2}{*}{74\,$\pm$\,5} & \multirow{2}{*}{5.6\,--\,12.9} \\ 

 & & IRDIS & ADI/RDI & $K_1K_2$ & 0.84 & 31 & 4 & & & & \\ \noalign{\smallskip}

\hline \noalign{\smallskip} 
 
2024 Apr 6 & \multirow{2}{*}{113.26PM.001} & \multirow{2}{*}{IRDIS} & \multirow{2}{*}{ADI/RDI/PDI} & \multirow{2}{*}{$H$} & \multirow{2}{*}{32} & 54 & 32 & 80 & 25$^*$ & 68\,$\pm$\,5 & 4.4\,--\,13.0 \\ 

2024 Apr 7 & & & & & & 58 & 32 & 48 & 27 & 60\,$\pm$\,6 & 5.6\,--\,9.0 \\  

\hline \noalign{\smallskip}

2022 Feb 27 & \multirow{2}{*}{107.22UJ.001} & \multirow{2}{*}{IFS} &  \multirow{2}{*}{ADI/SDI/RDI} & \multirow{2}{*}{$YJH$} & \multirow{2}{*}{32} & 36 & 21 & 20 & \multirow{2}{*}{-} &  76\,$\pm$\,4 & 7.8\,--\,15.2 \\

2021 Aug 21 &  &  &  &  &  & 27 & 16 & 14  &  & 74\,$\pm$\,9 & 4.6\,--\,15.2 \\

\hline \noalign{\smallskip}

2021 Jul 15 & 60.A-9801(S) & IRDIS & ADI/RDI/PDI & $H$ & 16 & 21 & 9 & 31 & 38 &  56\,$\pm$\,11 & 2.8\,--\,7.4 \\ 

\noalign{\smallskip}
\hline 
\end{tabular}







\tablefoot{
\tablefoottext{a}{Date at the beginning of the observing night.}
\tablefoottext{b}{Detector integration time for both science and reference sources.}
\tablefoottext{c}{Total parallactic angle rotation.}
\tablefoottext{d}{Repetitions rotating the polarizer four angles.}
\tablefoottext{e}{Average Strehl ratio measured at 1.6\,$\mu$m using SPARTA real time computer data.}
\tablefoottext{f}{Coherence time.}
\tablefoottext{$^*$}{Two polarimetric cycles removed due to the opening of the AO loop.}
\tablefoottext{$\dagger$}{Uncompleted observations due to worsening atmospheric conditions.}
\tablefoottext{$\ddag$}{One observing block removed as accidentally observed with a neutral density filter.}
} 
\end{table*}

\begin{table*}[h!]
\centering
\caption[]{Summary of the VLTI/GRAVITY observation.}
\label{tab:obs_grav}

\setlength{\tabcolsep}{8.8pt}
\begin{tabular}{lccccccc}
\hline
\hline \noalign{\smallskip}
Date & Program ID & $\Delta\alpha\,^a$ [mas] & $\Delta\delta\,^a$ [mas] & DIT [s] & On-source [min] & Seeing [$\arcsec$] & $\tau_0$ [ms] \\
\noalign{\smallskip} 

\hline \noalign{\smallskip}
\multirow{2}{*}{2025 May 14} & \multirow{2}{*}{115.29EH.001} & $-27.0$ & $-89.4$ & \multirow{2}{*}{30} & \multirow{2}{*}{45} & \multirow{2}{*}{$0.4-0.9$} & \multirow{2}{*}{$4.5-10.9$} \\ 

& & $-54.0$ & $-64.4$ & & & & \\



\hline
\end{tabular}

\tablefoot{
\tablefoottext{a}{Fiber pointing coordinates relative to the star.}}
\end{table*}

\setlength{\tabcolsep}{2pt}
 \begin{table}[]
     \centering
    \caption[]{Astro-photometric data for the target and the reference stars.}\label{tab:stars}
    \begin{tabular}{@{}lcccr@{}}
    \hline \hline     \noalign{\smallskip}
     & \multirow{2.5}{*}{\em PDS~70} & {\em UCAC2} & {\em UCAC2} & \multirow{2.5}{*}{Ref.$^a$}\\    
    \noalign{\smallskip}
    & &  {\em 14412811} &  {\em 14413562} & \\\noalign{\smallskip}
    \hline 
    \noalign{\smallskip}
    \noalign{\smallskip}
    $J$ [mag] & 9.55~$\pm$~0.02 & 9.60~$\pm$~0.02 & 9.69~$\pm$~0.02 & [1] \\ \noalign{\smallskip}
    $H$ [mag] & 8.82~$\pm$~0.04 & 8.92~$\pm$~0.03 & 8.90~$\pm$~0.03 & [1] \\ \noalign{\smallskip}
    $K$ [mag]  & 8.54~$\pm$~0.02 & 8.76~$\pm$~0.02 & 8.73~$\pm$~0.02 & [1] \\ \noalign{\smallskip}
    $\alpha$ (J2016.0)  & 14:08:10.11 & 14:08:43.04 & 14:13:20.96 & [2] \\ \noalign{\smallskip}
    $\delta$ (J2016.0) & $-$41:23:52.96 & $-$40:45:46.17 & $-$40:53:48.55 & [2]\\ 
    \noalign{\smallskip}
    \hline
    \end{tabular}
    \tablefoot{
    \tablefoottext{a}{
    [1]: \cite{2003yCat.2246....0C}}
    [2]: \cite{2020yCat.1350....0G}.}
 \end{table}
\setlength{\tabcolsep}{2pt}
\begin{table*}[h!]
\centering
\caption{Relative astrometry and flux measurements.}
\label{tab:astr_spec}
\begin{tabular}{llcccccccr}
\hline \hline \noalign{\smallskip}																															
\multirow{2}{*}{Date}	&	\multirow{2}{*}{Band}	&	\multirow{2}{*}{Object}	&	\multirow{2}{*}{$\Delta \alpha$ [mas]}			&	\multirow{2}{*}{$\Delta \delta$ [mas]}			&	\multirow{2}{*}{$\rho$ [mas]}			&	\multirow{2}{*}{PA [$^\circ$]}			&	\multirow{2}{*}{$\Delta$mag}			&	Flux\,$\times\,10^{-17}$				&	\multirow{2}{*}{FoM$^a$ / $n_{\rm pc}$}	\\
	&		&		&				&				&				&				&				&	 [W\,m$^{-2}\mu$m$^{-1}$]			&		\\ \noalign{\smallskip}  \hline \noalign{\smallskip}
15 Jul 2021$^b$	&	$H$	&	b	&	115.2	$ \pm $	3.3	&	-123.8	$ \pm $	3.4	&	169.1	$ \pm $	4.3	&	137.1	$ \pm $	0.6	&	8.3	$ \pm $	0.1	&	16.3	$ \pm $	1.6	&	 $\sigma$(1) / 1\,-\,7	\\ \noalign{\smallskip}
	&		&	N	&	-101.0	$ \pm $	16.3	&	50.9	$ \pm $	23.2	&	113.1	$ \pm $	13.2	&	296.8	$ \pm $	12.7	&	8.2	$ \pm $	0.3	&	18.5	$ \pm $	5.9	&	 $\sigma$(1) / 1\,-\,2	\\ \noalign{\smallskip}
	&		&	S	&	42.9	$ \pm $	6.3	&	-96.2	$ \pm $	8.6	&	105.3	$ \pm $	9.0	&	156.0	$ \pm $	3.1	&	7.9	$ \pm $	0.1	&	24.4	$ \pm $	2.4	&	 $\sigma$(1) / 3\,-\,5	\\ \noalign{\smallskip} \hline \noalign{\smallskip}
21 Aug 2021	&	$YJH$	&	N	&	-93.8	$ \pm $	7.0	&	61.1	$ \pm $	7.4	&	111.9	$ \pm $	6.7	&	303.1	$ \pm $	3.9	&			-	&	-			&	 $\sigma$(1) / 10	\\ \noalign{\smallskip}
	&		&	S	&	29.3	$ \pm $	5.9	&	-95.9	$ \pm $	4.8	&	100.3	$ \pm $	4.7	&	163.0	$ \pm $	3.4	&			-	&		-		&	 $\sigma$(1) / 10	\\ \noalign{\smallskip} \cline{3-10} \noalign{\smallskip}
	&	$H$	&	N	&	-91.3	$ \pm $	12.1	&	60.7	$ \pm $	15.9	&	109.6	$ \pm $	8.1	&	303.6	$ \pm $	9.6	&		-		&	-			&	 $\sigma$(1) / 10	\\ \noalign{\smallskip}
	&		&	S	&	29.0	$ \pm $	45.9	&	-89.5	$ \pm $	17.4	&	94.1	$ \pm $	9.6	&	162.0	$ \pm $	29.3	&		-		&	-			&	H(2) / 10	\\ \noalign{\smallskip} \hline \noalign{\smallskip}
27 Feb 2022	&	$YJH$	&	N	&	-98.2	$ \pm $	5.7	&	46.8	$ \pm $	4.0	&	108.8	$ \pm $	6.1	&	295.5	$ \pm $	1.7	&		-		&	-			&	 H(2) / 3	\\ \noalign{\smallskip}
	&		&	S	&	58.4	$ \pm $	14.1	&	-97.9	$ \pm $	15.3	&	114.0	$ \pm $	15.9	&	149.2	$ \pm $	6.7	&		-		&	-			&	 $\sigma$(1) / 3	\\ \noalign{\smallskip} \cline{3-10} \noalign{\smallskip}
	&	$H$	&	N	&	-97.2	$ \pm $	13.0	&	55.9	$ \pm $	10.1	&	112.2	$ \pm $	14.2	&	299.9	$ \pm $	4.3	&		-		&	-			&	 $\sigma$(0.5) / 3	\\ \noalign{\smallskip}
	&		&	S	&	66.7	$ \pm $	6.4	&	-107.2	$ \pm $	5.8	&	126.3	$ \pm $	5.4	&	148.1	$ \pm $	3.1	&		-		&	-			&	 $\sigma$(1) / 1	\\ \noalign{\smallskip} \hline \noalign{\smallskip}
6 Apr 2024$^b$	&	$H$	&	b	&	116.3	$ \pm $	4.8	&	-97.5	$ \pm $	4.7	&	151.7	$ \pm $	5.1	&	130.0	$ \pm $	1.7	&	8.5	$ \pm $	0.1	&	14.1	$ \pm $	0.7	&	H(3) / 8\,-\,16	\\ \noalign{\smallskip}
	&		&	N	&	-93.9	$ \pm $	6.8	&	35.8	$ \pm $	8.7	&	100.5	$ \pm $	6.4	&	290.9	$ \pm $	5.1	&	6.8	$ \pm $	0.2	&	64.9	$ \pm $	13.6	&	H(3) / 1\,-\,2	\\ \noalign{\smallskip}
	&		&	S	&	46.1	$ \pm $	17.4	&	-93.7	$ \pm $	14.5	&	104.5	$ \pm $	13.5	&	153.8	$ \pm $	10.0	&	7.6	$ \pm $	0.3	&	32.5	$ \pm $	9.3	&	 $\sigma$(1) / 2\,-\,4	\\ \noalign{\smallskip} \hline \noalign{\smallskip}
7 Apr 2024$^b$	&	$H$	&	b	&	122.0	$ \pm $	3.5	&	-101.9	$ \pm $	3.4	&	159.0	$ \pm $	3.9	&	129.9	$ \pm $	1.1	&	8.3	$ \pm $	0.2	&	16.0	$ \pm $	2.4	&	 $\sigma$(1) / 3\,-\,5	\\ \noalign{\smallskip}
	&		&	N	&	-85.5	$ \pm $	3.3	&	29.4	$ \pm $	3.5	&	90.4	$ \pm $	3.3	&	289.0	$ \pm $	2.3	&	6.7	$ \pm $	0.2	&	73.1	$ \pm $	11.8	&	H(1) / 1\,-\,8	\\ \noalign{\smallskip}
	&		&	S	&	45.2	$ \pm $	9.4	&	-82.3	$ \pm $	11.4	&	93.9	$ \pm $	12.1	&	151.2	$ \pm $	5.1	&	7.5	$ \pm $	0.4	&	34.8	$ \pm $	11.8	&	 $\sigma$(1) / 3\,-\,6	\\ \noalign{\smallskip} \hline \noalign{\smallskip}
18 Jul 2025$^c$	&	$YJH$	&	N	&	-86.0	$ \pm $	2.6	&	19.8	$ \pm $	4.8	&	88.3	$ \pm $	2.5	&	282.9	$ \pm $	3.2	&	9.5	$ \pm $	0.2	&	8.8	$ \pm $	1.6	&	H(2) / 30	\\ \noalign{\smallskip}
	&		&	S	&	39.4	$ \pm $	3.0	&	-91.8	$ \pm $	1.9	&	99.9	$ \pm $	1.6	&	156.8	$ \pm $	1.8	&	8.9	$ \pm $	0.2	&	14.4	$ \pm $	3.1	&	 $\sigma$(0.5) / 30	\\ \noalign{\smallskip} \cline{3-10} \noalign{\smallskip}
	&	$H$	&	N	&	-87.7	$ \pm $	17.4	&	10.1	$ \pm $	29.6	&	88.3	$ \pm $	17.1	&	276.6	$ \pm $	19.3	&	9.7	$ \pm $	0.8	&	6.7	$ \pm $	5.1	&	 $\sigma$(1) / 55	\\ \noalign{\smallskip}
	&		&	S	&	39.4	$ \pm $	5.8	&	-95.1	$ \pm $	6.3	&	103.0	$ \pm $	6.4	&	157.5	$ \pm $	3.2	&	9.7	$ \pm $	0.5	&	6.3	$ \pm $	3.0	&	 $\sigma$(0.5) / 55	\\ \noalign{\smallskip} \cline{3-10} \noalign{\smallskip}
	&	$K_1$	&	b	&	114.6	$ \pm $	4.0	&	-80.4	$ \pm $	3.2	&	140.1	$ \pm $	4.6	&	125.1	$ \pm $	0.9	&	9.1	$ \pm $	0.1	&	1.9	$ \pm $	0.1	&	 $\sigma$(1) / 2\,-\,20	\\ \noalign{\smallskip}
	&		&	c	&	-220.1	$ \pm $	23.2	&	-26.1	$ \pm $	11.0	&	221.7	$ \pm $	23.4	&	263.2	$ \pm $	2.8	&	10.3	$ \pm $	0.3	&	0.6	$ \pm $	0.2	&	 $\sigma$(1.5) / 8\,-\,20	\\ \noalign{\smallskip}
	&		&	N	&	-83.0	$ \pm $	6.3	&	0.0	$ \pm $	14.9	&	83.0	$ \pm $	6.3	&	270.0	$ \pm $	10.3	&	8.2	$ \pm $	0.2	&	4.1	$ \pm $	0.9	&	H(2) / 2\,-\,3	\\ \noalign{\smallskip} \cline{3-10} \noalign{\smallskip}
	&	$K_2$	&	b	&	114.3	$ \pm $	5.0	&	-79.2	$ \pm $	3.9	&	139.1	$ \pm $	5.8	&	124.7	$ \pm $	1.1	&	8.8	$ \pm $	0.2	&	2.3	$ \pm $	0.3	&	 $\sigma$(1) / 2\,-\,20	\\ \noalign{\smallskip}
	&		&	c	&	-219.2	$ \pm $	20.7	&	-27.9	$ \pm $	11.0	&	221.0	$ \pm $	20.9	&	262.7	$ \pm $	2.8	&	9.9	$ \pm $	0.3	&	0.8	$ \pm $	0.2	&	 $\sigma$(1.5) / 5\,-\,20	\\ \noalign{\smallskip}
	&		&	N	&	-95.4	$ \pm $	12.0	&	0.0	$ \pm $	35.9	&	95.4	$ \pm $	12.0	&	270.0	$ \pm $	21.6	&	10.4	$ \pm $	0.6	&	0.6	$ \pm $	0.3	&	H(2) / 2\,-\,3	\\ \noalign{\smallskip} \hline \noalign{\smallskip}
22 Jul 2025$^c$	&	$YJH$	&	N	&	-89.6	$ \pm $	4.5	&	9.2	$ \pm $	6.6	&	90.1	$ \pm $	4.5	&	275.9	$ \pm $	4.2	&	9.6	$ \pm $	0.3	&	5.0	$ \pm $	1.3	&	H(2) / 45	\\ \noalign{\smallskip}
	&		&	S	&	43.0	$ \pm $	5.3	&	-87.8	$ \pm $	4.3	&	97.8	$ \pm $	4.0	&	153.9	$ \pm $	3.3	&	8.8	$ \pm $	0.3	&	10.3	$ \pm $	2.5	&	 $\sigma$(1) / 45	\\ \noalign{\smallskip} \cline{3-10} \noalign{\smallskip}
	&	$K_1$	&	b	&	122.8	$ \pm $	2.4	&	-83.0	$ \pm $	1.8	&	148.2	$ \pm $	2.8	&	124.0	$ \pm $	0.5	&	9.2	$ \pm $	0.1	&	1.5	$ \pm $	0.1	&	$\sigma$(1.5) / 14\,-\,20	\\ \noalign{\smallskip}
	&		&	c	&	-218.2	$ \pm $	13.3	&	-22.6	$ \pm $	7.0	&	219.4	$ \pm $	13.3	&	264.1	$ \pm $	1.8	&	10.1	$ \pm $	0.2	&	0.6	$ \pm $	0.1	&	$\sigma$(1.5) / 14\,-\,20	\\ \noalign{\smallskip}
	&		&	N	&	-85.7	$ \pm $	7.5	&	27.4	$ \pm $	13.6	&	90.0	$ \pm $	6.4	&	287.7	$ \pm $	9.0	&	9.1	$ \pm $	0.2	&	1.6	$ \pm $	0.3	&	H(2) / 7\,-\,8	\\ \noalign{\smallskip} \cline{3-10} \noalign{\smallskip}
	&	$K_2$	&	b	&	117.6	$ \pm $	4.4	&	-80.0	$ \pm $	3.2	&	142.2	$ \pm $	5.2	&	124.2	$ \pm $	0.7	&	8.9	$ \pm $	0.1	&	1.7	$ \pm $	0.1	&	$\sigma$(1) / 14\,-\,20	\\ \noalign{\smallskip}
	&		&	c	&	-229.3	$ \pm $	9.5	&	-25.0	$ \pm $	9.6	&	230.6	$ \pm $	9.5	&	263.8	$ \pm $	2.4	&	10.1	$ \pm $	0.3	&	0.6	$ \pm $	0.2	&	$\sigma$(1) / 14\,-\,20	\\ \noalign{\smallskip}
	&		&	N	&	-89.1	$ \pm $	7.2	&	15.3	$ \pm $	6.2	&	90.4	$ \pm $	7.2	&	279.7	$ \pm $	3.9	&	8.9	$ \pm $	0.4	&	1.7	$ \pm $	0.6	&	$\sigma$(0.5) / 7\,-\,8	\\
\hline
\end{tabular} \tablefoot{This table is available in electronic format at CDS: \todo{link}. \tablefoottext{a}{Figure of merit, labeled as H for the hessian matrix determinant and $\sigma$ for the standard deviation, with the number in parentheses indicating either the size of the square matrix or the aperture radius in FWHM units, respectively.}
\tablefoottext{b}{Flux calibrated with PDS~70.}
\tablefoottext{c}{Flux calibrated with HD~131243.}}
\end{table*}																															
\setlength{\tabcolsep}{11pt}
\renewcommand{\arraystretch}{0.9}
\begin{table*}[h!]
\centering
\caption{Relative astrometry and flux measurements for N and S emissions per IFS channel from 2025 July 18 observation.}
\label{tab:astr_spec_ch}
\begin{tabular}{llcccccc}
\hline \hline \noalign{\smallskip}
\multirow{2}{*}{$\lambda$ [$\mu$m]} 	&	\multirow{2}{*}{Object}	&	\multirow{2}{*}{$\Delta \alpha$ [mas]}			&	\multirow{2}{*}{$\Delta \delta$ [mas]}			&	\multirow{2}{*}{$\rho$ [mas]}			&	\multirow{2}{*}{PA [$^\circ$]}			&	\multirow{2}{*}{$\Delta$mag}			&	Flux\,$\times\,10^{-17}$				\\
	&		&				&				&				&				&				&	[W\,m$^{-2}\mu$m$^{-1}$]			\\ \noalign{\smallskip}  \hline \noalign{\smallskip}
\multirow{2}{*}{0.95}	&	N	&	-71.9	$ \pm $	4.7	&	9.6	$ \pm $	2.4	&	72.5	$ \pm $	2.3	&	277.6	$ \pm $	3.7	&	8.83	$ \pm $	0.31	&	33.9	$ \pm $	9.6	\\ \noalign{\smallskip}
	&	S	&	39.3	$ \pm $	7.6	&	-104.9	$ \pm $	6.0	&	112.0	$ \pm $	7.8	&	159.5	$ \pm $	2.9	&	10.15	$ \pm $	0.45	&	10.1	$ \pm $	4.2	\\ \noalign{\smallskip} \hline \noalign{\smallskip}
\multirow{2}{*}{0.97}	&	N	&	-71.5	$ \pm $	3.1	&	9.3	$ \pm $	3.5	&	72.1	$ \pm $	3.5	&	277.4	$ \pm $	2.5	&	8.68	$ \pm $	0.21	&	42.3	$ \pm $	8.2	\\ \noalign{\smallskip}
	&	S	&	27.8	$ \pm $	12.1	&	-82.4	$ \pm $	11.4	&	87.0	$ \pm $	12.2	&	161.4	$ \pm $	7.5	&	10.15	$ \pm $	0.81	&	10.9	$ \pm $	8.1	\\ \noalign{\smallskip} \hline \noalign{\smallskip}
\multirow{2}{*}{0.99}	&	N	&	-75.6	$ \pm $	4.3	&	11.9	$ \pm $	2.5	&	76.6	$ \pm $	2.4	&	279.0	$ \pm $	3.3	&	8.93	$ \pm $	0.30	&	32.0	$ \pm $	8.9	\\ \noalign{\smallskip}
	&	S	&	36.8	$ \pm $	10.8	&	-83.6	$ \pm $	10.2	&	91.4	$ \pm $	11.0	&	156.3	$ \pm $	6.3	&	10.52	$ \pm $	2.47	&	7.4	$ \pm $	16.9	\\ \noalign{\smallskip} \hline \noalign{\smallskip}
\multirow{2}{*}{1.01}	&	N	&	-71.9	$ \pm $	4.4	&	9.4	$ \pm $	3.1	&	72.5	$ \pm $	3.0	&	277.4	$ \pm $	3.5	&	8.87	$ \pm $	0.20	&	30.1	$ \pm $	5.4	\\ \noalign{\smallskip}
	&	S	&	32.5	$ \pm $	10.6	&	-84.0	$ \pm $	10.0	&	90.1	$ \pm $	10.8	&	158.8	$ \pm $	6.3	&	10.17	$ \pm $	0.85	&	9.1	$ \pm $	7.1	\\   \hline
    \ldots 
\end{tabular}
 \tablefoot{Only the first four IFS channels are displayed here. The full version of this table is available at CDS: \todo{link}.}
\end{table*}
\setlength{\tabcolsep}{35pt}
\renewcommand{\arraystretch}{0.9}
\begin{table*}[]
\centering
\caption{Prior and posterior distributions for the orbit fitting. Note the third inner planet is here labeled as $d$.}
\label{tab:post_orb}
\begin{tabular}{llll}
\hline
\hline \noalign{\smallskip}
Parameter & Prior$^a$ & Posterior$_{\rm3p}$ & Posterior$_{\rm3p_{\rm troj}}$ \\ \hline \noalign{\smallskip}
$a_b$ [au] & $\mathcal{U}$(1, 40) & $24.5 \pm 1.8$ & $25.3^{+2.3}_{-1.8}$ \\ \noalign{\smallskip}
$e_b$ & $\mathcal{U}$(0, 0.99) & $0.103^{+0.054}_{-0.058}$ & $0.113^{+0.069}_{-0.075}$ \\ \noalign{\smallskip}
$i_b$ [$^\circ$] & $\mathcal{G}$(160.5, 10) & $127.7^{+3.1}_{-2.2}$ & $127.0^{+3.7}_{-2.3}$ \\ \noalign{\smallskip} 
$\omega_b$ [$^\circ$] & $\mathcal{U}$(0, 360) & $97^{+55}_{-61}$ & $85^{+52}_{-43}$ \\ \noalign{\smallskip} 
$\Omega_b$ [$^\circ$] & $\mathcal{U}$(0, 360) & $173.9^{+4.1}_{-350}$ & $173.6^{+3.9}_{-10.0}$ \\ \noalign{\smallskip}
$m_b$ [$M_{\rm Jup}$] & $\mathcal{U}$(0.3, 15) & $8.0^{+4.8}_{-5.3}$ & $7.7^{+5.0}_{-5.0}$ \\ \noalign{\smallskip}
$\theta_b$ [$^\circ$] & $\mathcal{U}$(0, 360) & $160.39^{+0.35}_{-0.33}$ & $160.23 \pm 0.35$ \\[2em]

$a_c$ [au] & $\mathcal{U}$(1, 40) & $35.3^{+3.0}_{-3.5}$ & $35.5^{+2.5}_{-3.6}$ \\ \noalign{\smallskip}
$e_c$ & $\mathcal{U}$(0, 0.99) & $0.054^{+0.071}_{-0.036}$ & $0.058^{+0.067}_{-0.041}$ \\ \noalign{\smallskip}
$i_c$ [$^\circ$] & $\mathcal{G}$(160.5, 10) & $129.7^{+3.4}_{-2.5}$ & $129.7^{+3.3}_{-2.9}$ \\ \noalign{\smallskip} 
$\omega_c$ [$^\circ$] & $\mathcal{U}$(0, 360) & $20^{+79}_{-120}$ & $21^{+91}_{-120}$ \\ \noalign{\smallskip} 
$\Omega_c$ [$^\circ$] & $\mathcal{U}$(0, 360) & $160.5^{+5.5}_{-4.8}$ & $160.9^{+4.8}_{-6.9}$ \\ \noalign{\smallskip}
$m_c$ [$M_{\rm Jup}$] & $\mathcal{U}$(0.3, 15) & $7.6^{+5.0}_{-5.0}$ & $7.7^{+5.0}_{-5.0}$ \\ \noalign{\smallskip}
$\theta_c$ [$^\circ$] & $\mathcal{U}$(0, 360) & $293.23^{+0.25}_{-0.28}$ & $293.32^{+0.24}_{-0.25}$ \\ [2em]

$a_d$ [au] & $\mathcal{U}$(1, 40) & $11.1^{+2.5}_{-2.3}$ & $12.87^{+0.50}_{-0.47}$ \\ \noalign{\smallskip}
$e_d$ & $\mathcal{U}$(0, 0.99) & $0.29^{+0.24}_{-0.20}$ & $0.039^{+0.044}_{-0.028}$ \\ \noalign{\smallskip}
$i_d$ [$^\circ$] & $\mathcal{G}$(160.5, 10) & $136.3^{+21}_{-8.7}$ & $131.0^{+2.2}_{-2.0}$ \\ \noalign{\smallskip} 
$\omega_d$ [$^\circ$] & $\mathcal{U}$(0, 360) & $2 \pm 120$ & $13^{+79}_{-130}$ \\ \noalign{\smallskip} 
$\Omega_d$ [$^\circ$] & $\mathcal{U}$(0, 360) & $15^{+130}_{-70}$ & $142.9^{+5.4}_{-180}$ \\ \noalign{\smallskip}
$m_d$ [$M_{\rm Jup}$] & $\mathcal{U}$(0.3, 15) & $7.8^{+5.0}_{-5.1}$ & $8.0^{+4.8}_{-5.3}$ \\ \noalign{\smallskip}
$\theta_d$ [$^\circ$] & $\mathcal{U}$(0, 360) & $351.0^{+4.1}_{-3.9}$ & $323.8^{+2.9}_{-3.0}$ \\ \noalign{\smallskip}
$\Delta\nu_{\rm N}$ [$^\circ$] & $\mathcal{U}$(--180, 180) & - & $-38.2^{+2.4}_{-2.7}$ \\ \noalign{\smallskip}
$\Delta\nu_{\rm S}$ [$^\circ$] & $\mathcal{U}$(--180, 180) & - & $87.2 \pm 7.6$  \\ \noalign{\smallskip}
$J_{\rm N}$ [mas] & $\mathcal{U}$(0, 50) & - & $2.3^{+2.1}_{-1.6}$ \\ \noalign{\smallskip}
$J_{\rm S}$ [mas] & $\mathcal{U}$(0, 50) & - & $16.9^{+6.2}_{-4.4}$ \\ [2em]

$M_\star$ [$M_{\odot}$] &  $\mathcal{G}$(0.88, 0.09) & $0.898^{+0.072}_{-0.063}$ & $0.889^{+0.081}_{-0.080}$ \\ \noalign{\smallskip}
Parallax [mas] & $\mathcal{G}$(8.897, 0.019) & $8.898 \pm 0.019$ & $8.897 \pm 0.019$ \\ \noalign{\smallskip}
\hline
\end{tabular}
\end{table*}

\FloatBarrier

\end{appendix}
\end{document}